\documentclass[fleqn,usenatbib]{mnras}
\usepackage{newtxtext,newtxmath}
\usepackage[T1]{fontenc}
\DeclareRobustCommand{\VAN}[3]{#2}
\let\VANthebibliography\thebibliography
\def\thebibliography{\DeclareRobustCommand{\VAN}[3]{##3}\VANthebibliography}

\usepackage{graphicx}	
\usepackage{amsmath}	
  
\usepackage{amssymb}	
\usepackage{threeparttable}
\usepackage{longtable}
\usepackage{float}

\title[ATOMS VII]{ATOMS: ALMA Three-millimeter Observations of Massive Star-forming regions – VII. A catalogue of SiO clumps from ACA observations}

\author[Rong Liu et al.]
{Rong Liu,$^{1,2}$\thanks{E-mail: liu\underline{~~}rong@bao.ac.cn}
Tie Liu,\thanks{E-mail: liutie@shao.ac.cn }$^{3,4}$
Gang Chen,\thanks{E-mail: ddwhcg@cug.edu.cn}$^{5}$
Hong-Li Liu,$^{6}$
Ke Wang,$^{7}$
Jin-Zeng Li,$^{1}$ 
Xun-Chuan Liu,$^{3,10}$
\newauthor
Chang Won Lee,$^{8,9}$
Paul F. Goldsmith,$^{20}$
Mika Juvela,$^{11}$
Guido Garay,$^{12}$
Leonardo Bronfman,$^{12}$
Tapas Baug,$^{22}$
\newauthor
Jinhua He,$^{15,21,12}$
Si-Ju Zhang,$^{7}$
Yong Zhang,$^{17}$
Feng-Wei Xu,$^{10}$
Archana Soam,$^{14}$
Zhi-Qiang Shen,$^{3}$
\newauthor
Shanghuo Li,$^{8}$
Lokesh Dewangan,$^{13}$
Chakali Eswaraiah,$^{16}$
Yue-Fang Wu,$^{10}$
Sheng-Li Qin,$^{6}$
L. Viktor Tóth,$^{18}$
\newauthor
Zhiyuan Ren,$^{1}$
Guoyin Zhang$^{1}$
Anandmayee Tej,$^{19}$ 
Qiuyi Luo,$^{3}$
Jianwen Zhou$^{1,2}$
Chang Zhang$^{1,2}$
\\
$^{1}$National Astronomical Observatories of China, Chinese Academy of Sciences, Beijing, 100012, China\\
$^{2}$University of Chinese Academy of Sciences, Beijing 100049, Peoples Republic of China  \\
$^{3}$Shanghai Astronomical Observatory, Chinese Academy of Sciences, 80 Nandan Road, Shanghai 200030, Peoples Republic of China \\
$^{4}$Key Laboratory for Research in Galaxies and Cosmology, Chinese Academy of Sciences, 80 Nandan Road, Shanghai 200030, Peoples Republic of China \\
$^{5}$China University of Geosciences, Wuhan 430074, China\\
$^{6}$Department of Astronomy, Yunnan University, Kunming, 650091, PR China \\
$^{7}$Kavli Institute for Astronomy and Astrophysics, Peking University, 5 Yiheyuan Road, Haidian District, Beijing 100871, China\\
$^{8}$Korea Astronomy and Space Science Institute, 776 Daedeokdaero, Yuseong-gu, Daejeon 34055, Republic of Korea\\
$^{9}$University of Science and Technology, Korea (UST), 217 Gajeong-ro, Yuseong-gu, Daejeon 34113, Republic of Korea\\
$^{10}$Department of Astronomy, Peking University, 100871, Beijing,
People’s Republic of China\\
$^{11}$Department of Physics, P.O. Box 64, FI-00014, University of Helsinki, Finland\\
$^{12}$Departamento de Astronomía, Universidad de Chile, Las Condes, Santiago, Chile\\
$^{13}$Physical Research Laboratory, Navrangpura, Ahmedabad—380 009, India \\
$^{14}$SOFIA Science Centre, USRA, NASA Ames Research Centre, MS-12, N232, Moffett Field, CA 94035, USA\\
$^{15}$Yunnan Observatories, Chinese Academy of Sciences, 396 Yangfangwang, Guandu District, Kunming, 650216, China\\
$^{16}$Indian Institute of Science Education and Research (IISER) Tirupati, Rami Reddy Nagar, Karakambadi Road, Mangalam (P.O.), Tirupati 517 507, India\\
$^{17}$School of Physics and Astronomy, Sun Yat-Sen University Zhuhai Campus, Tangjia, Zhuhai 519082, China\\
$^{18}$ E\"otv\"os Lor\'and University, Department of Astronomy, P\'azm\'any P\'eter s\'et\'any 1/A, H-1117, Budapest, Hungary\\
$^{19}$Indian Institute of Space Science and Technology, Thiruvanan-
thapuram 695 547, Kerala, India\\
$^{20}$Jet Propulsion Laboratory, California Institute ofTechnology, 4800 Oak Grove Drive, Pasadena, CA 91109, USA\\
$^{21}$Chinese Academy of Sciences South America Center for Astronomy, National Astronomical Observatories, CAS, Beijing 100101, China \\ 
$^{22}$S. N. Bose National Centre for Basic Sciences, Block-JD, Sector-III, Salt Lake City, Kolkata 700106, India\\
}
\date{Accepted XXX. Received YYY; in original form ZZZ}

\pubyear{2021}

\begin{document}
\label{firstpage}
\pagerange{\pageref{firstpage}--\pageref{lastpage}}
\maketitle

\begin{abstract}
To understand the nature of SiO emission, we conducted ACA observations of the SiO (2-1) lines toward 146 massive star-forming regions, as part of the ALMA Three-millimeter Observations of Massive Star-forming regions (ATOMS) survey. We detected SiO emission in 128 (87.7$\%$) sources and identified 171 SiO clumps, 105 of which are spatially separated from 3 mm continuum emission. A large amount of the SiO line profiles (60$\%$) are non-Gaussian. The velocity dispersion of the SiO lines ranges from 0.3 to 5.43 km s$^{-1}$. In 63 sources the SiO clumps are associated with H{\sc ii} regions characterized by H40$\alpha$ emission. We find that 68$\%$ (116) of the SiO clumps are associated with strong outflows. The median velocity dispersion of the SiO line for outflow sources and non-outflow sources is 1.91 km s$^{-1}$ and 0.99 km s$^{-1}$, respectively. These results indicate that outflow activities could be connected to strongly shocked gas. The velocity dispersion and [SiO]/[H$^{13}$CO$^+$] intensity ratio do not show any correlation with the dust temperature and particle number density of clumps. We find a positive correlation between the SiO line luminosity and the bolometric luminosity, implying stronger shock activities are associated with more luminous proto-clusters. The SiO clumps in associations with H{\sc ii} regions were found to show a steeper feature in $L_\textup{sio}$/$L_\textup{bol}$. The SiO line luminosity and the fraction of shocked gas have no apparent evidence of correlation with the evolutionary stages traced by luminosity to mass ratio ($L_\textup{bol}/M$).

\end{abstract}

\begin{keywords}
stars: formation - ISM: clouds - ISM: molecules 
\end{keywords}

\section{Introduction}
As the main contributors of heavy elements and UV-photon radiation, massive stars play an important role in the evolution of interstellar medium (ISM). In the past decades, studies of massive star formation through multi-wavelengths and high-resolution observations have made great achievements, but there are still difficulties in understanding the formation process because of large distances, short timescales, high extinction, and clustered environments \citep{zinnecker2007toward, motte2018high}. The process of massive star formation itself can also provide strong feedback to the parent clouds and ISM by energetic jets and induced outflows. The jets and outflows interacting with the surrounding medium are ubiquitous at different evolutionary stages in high-mass star formation, such as infrared quiet, infrared bright, and Ultra-Compact (UC) H{\sc ii} stage \citep{churchwell2002formation,bally2016protostellar,li2019sio}. Thus studies of such shock activities can help us deepen understanding of the massive star formation processes.

A deep understanding of the properties of jets and outflows is crucial in answering the problems in massive star formation \citep{motte2018high}.  Molecular lines are powerful tools to study the physical and chemical conditions and the feedback within massive star-forming regions (SFRs). Various molecular emissions trace different layers in the internal structure of massive clumps embedding high-mass stars \citep{csengeri2016atlasgal}. Rotational transitions of different molecules provide us temperatures, densities, UV fields, chemical abundances, and gas kinematics of the massive clumps. Molecules such as CO, SiO, SO, HCO$^+$, and CS are good tracers of jets and outflows \citep{bally2016protostellar}. In particular, due to little contamination from the quiescent gas, SiO emissions have been widely used to study shock activities induced by outflows. 

Shocks are a universal phenomenon in SFRs. High velocity shocks ($v_\textup{s}$ ${\ge}$ 20 km s$^{^{-1}}$) are caused by powerful outflows and jets from massive young proto-stars \citep{qiu2007high,duarte2014sio,li2019formation,li2020alma}, while low velocity shocks ($v_\textup{s}$ < 10 km s$^{^{-1}}$) could be associated with less powerful outflows \citep{lefloch1998widespread}, cloud-cloud collisions \citep{louvet2016tracing} or large-scale gas inflows \citep{jimenez2010parsec}. 
SiO is an excellent tracer of shocks. In the shocked regions, grain material is destructed via sputtering or vaporization, leading the Si atoms and Si-bearing molecules to be injected into the gas phase and subsequently oxidized to SiO \citep{schilke1997sio,gusdorf2008siob}. In active star-forming regions, SiO abundance is enhanced by up to six orders of magnitudes compared to quiescent regions \citep{martin1992sio,codella1999low}. Observations of the SiO emission lines toward different sources exhibit a variety of profiles. Previous studies reported that the spectra of SiO usually shows two Gaussian components both corresponding to similar central velocities but different line widths, namely, the broad component and the narrow one \citep{martin1992sio, lefloch1998widespread}. These two SiO components trace different emission regions and may have different origins \citep{schilke1997sio}. The broad components are thought to be related to high-velocity shocks from protostellar outflows, and the narrow components are attributed to low-velocity shocks linked to low-velocity motions (see above). 

Based on studies targeting the variation of SiO emission toward different evolutionary stages of star formation, several studies found the brighter SiO emission are associated with higher luminosity sources \citep{codella1999low,Liu2020a}, which would be consistent with the trend that the SiO abundance in warmer sources is higher than the one in infrared-quiet sources \citep{miettinen2014malt90,gerner2014chemical}. \citet{motte2007earliest} and \citet{sakai2010survey} found that younger sources show brighter SiO emission than mid-infrared bright sources. The studies of \citet{miettinen2006sio}, \citet{lopez2011sio}, and \citet{sanchez2013evolution} are consistent with a decrease in SiO abundance in more evolved sources. 
The results that the SiO abundance have no trend among different evolutionary stages also have been reported by \citet{sanhueza2012chemistry}, \citet{leurini2014sio},
\citet{csengeri2016atlasgal}, and \citet{li2019sio}. Recently, \citet{liu2021sio} reported an increasing trend of the SiO line luminosity with bolometric luminosity, but they do not see the relation between the SiO line luminosity and the bolometric luminosity-to-mass ratio. So far, there is no consensus on SiO abundance variation in different evolutionary stages.

During the past decades, a large of SiO studies toward massive star-forming regions have been conducted. The SiO abundance and excitation variations have been studied by searching for evolutionary trends in different samples \citep{codella1999low,miettinen2006sio,motte2007earliest,sakai2010survey,gerner2014chemical,csengeri2016atlasgal,li2019sio,Liu2020a}. Moreover, \citet{jimenez2010parsec}, \citet{duarte2014sio}, \citet{louvet2016tracing}, and \citet{cosentino2020sio} investigated the mechanisms responsible for the broad and narrow line profile of SiO. However, previous studies have been focused on either individual sources through interferometer arrays or large samples with single-dish observations. The origin of the SiO emission is still an enigma for astronomers without a high resolution survey of SiO emission toward a large sample of sources with interferometers.

The ALMA Three-millimeter Observations of Massive Star-forming regions (ATOMS) survey employed the Band 3 and covered eight spectral windows (two wide spectral windows in upper side band and six spectral windows in lower side band) toward 146 massive star-forming sources \citep{liu2020atoms}. The survey provides us larger samples and high resolution to systemically study various molecular gas involved  in star formation. In this paper, we statistically analyze SiO emission of 146 sources using the Atacama Compact 7 m Array observations (ACA) to better understand shocked gas under different conditions and in different evolutionary trends. We establish a catalogue of the SiO clumps for future studies. We present the level of shock activities and estimate the fraction of the shocked gas in the clumps. The paper is organized as follows: the sample selection and observations are presented in Sect.~\ref{sec2}. Sect.~\ref{sec3} gives our results. We put discussion in Sect.~\ref{sec4} and summary in Sect.~\ref{sec5}.

\begin{figure*}
   \centerline{\includegraphics[width=1.16\linewidth]{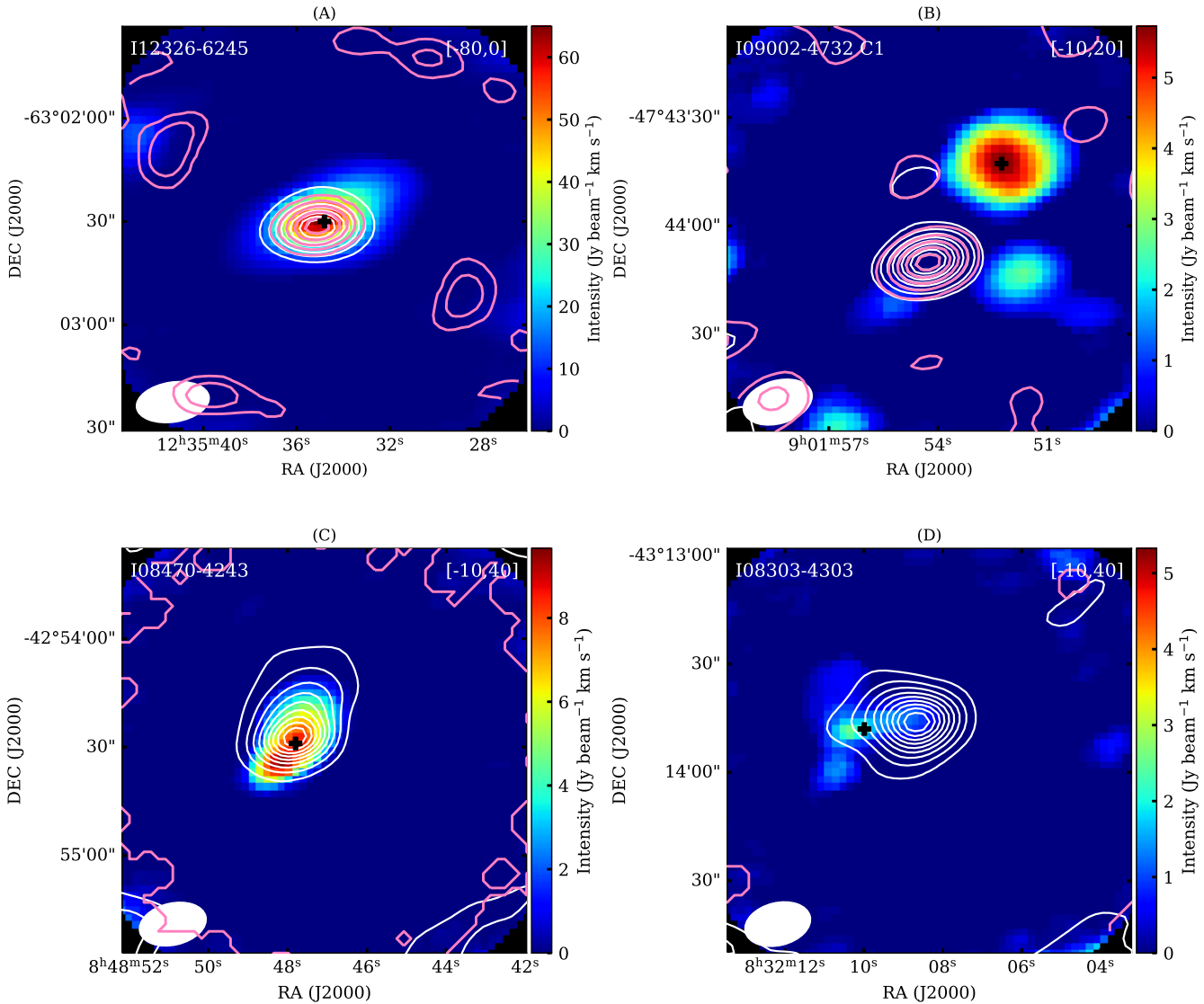}}
   \caption{Example sources. The background is SiO (2-1) integrated intensity maps. The white contours are 3 mm continuum emission, and contours are from 10$\%$ to 100$\%$ in the step of 10$\%$ of peak values. The pink contours represent H40$\alpha$ emission, contours from 10$\%$ to 100$\%$ in the step of 20$\%$ of peak values. 
   The source name is shown on the upper left. The black cross is the central position extracted spectra and the beam size is presented at the lower left corner. The integration ranges are shown in the upper right corner.
   (A) Panel presents source associated with 3 mm emission and coincided with H40$\alpha$ emission. 
   (B) Panel shows source separated with 3 mm emission and coincided with H40$\alpha$ emission. 
   (C) Panel presents source associated with 3mm emission undetected H40$\alpha$ emission. 
   (D) Panel shows source separated with 3mm emission undetected H40$\alpha$ emission. 
   All images have a field of view of 2$^{\prime}$× 2$^{\prime}$.} 
   \label{fig1}
\end{figure*}

\section{Observations}
\label{sec2}
\subsection{Sample}
A sample of 146 massive SFRs from the ATOMS survey were investigated in this paper. The sample sources were selected from the CS $J$=2-1 line survey of UC H{\sc ii} regions in the Galactic Plane \citep{bronfman1996cs}. 
The sample sources are complete for protoclusters characterized by bright CS $J$=2-1 emission ($T_\textup{b}$ > 2 K), implying reasonably dense gas. The basic parameters of the sample were taken from \citet{liu2020atoms}, \citet{liu2020atomsII}, and \citet{liu2021atomsIII} ( paper I, paper II, and paper III of the series). 
The distance of the sample sources ranges from 0.4 kpc to 13.0 kpc with a mean value of 4.5 kpc. The clump mass ranges from 5.6 to 2.5×10$^5$ $M_{\odot}$, with a median value of 1.4×10$^3$ $M_{\odot}$. The bolometric luminosity ranges from 16 to 8.1×10$^6$ $L_{\odot}$, with a median value of 5.7×10$^4$ $L_{\odot}$. The radii ranges from 0.06 to 4.26 pc, with a median value of 0.86 pc. The dust temperature ranges from 18 to 46 K, with a median value of 29 K.

\begin{figure*}
   \begin{minipage}[t]{0.44\linewidth}

     \centerline{\includegraphics[width=1.12\linewidth]{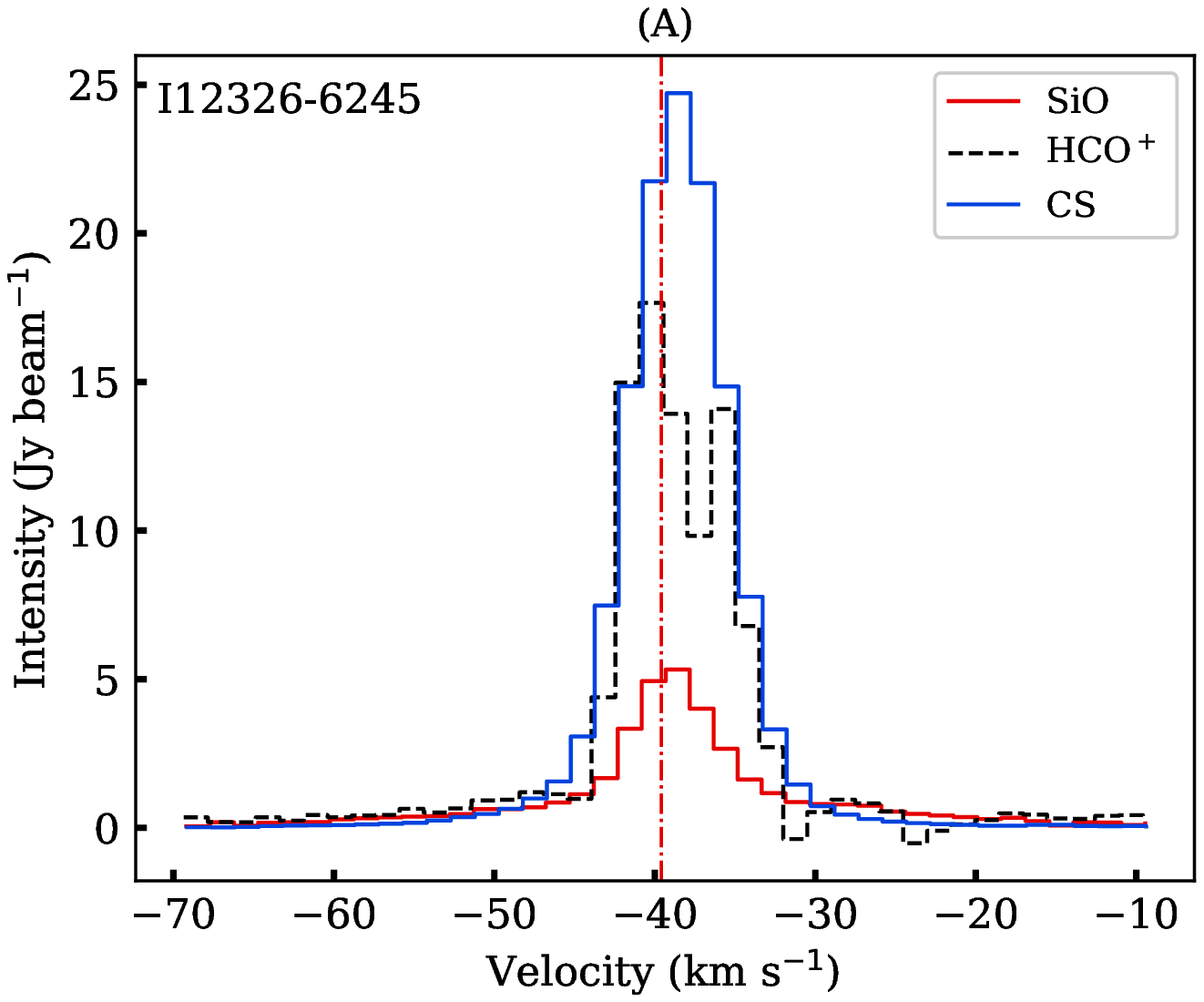}}
     \centerline{\includegraphics[width=1.12\linewidth]{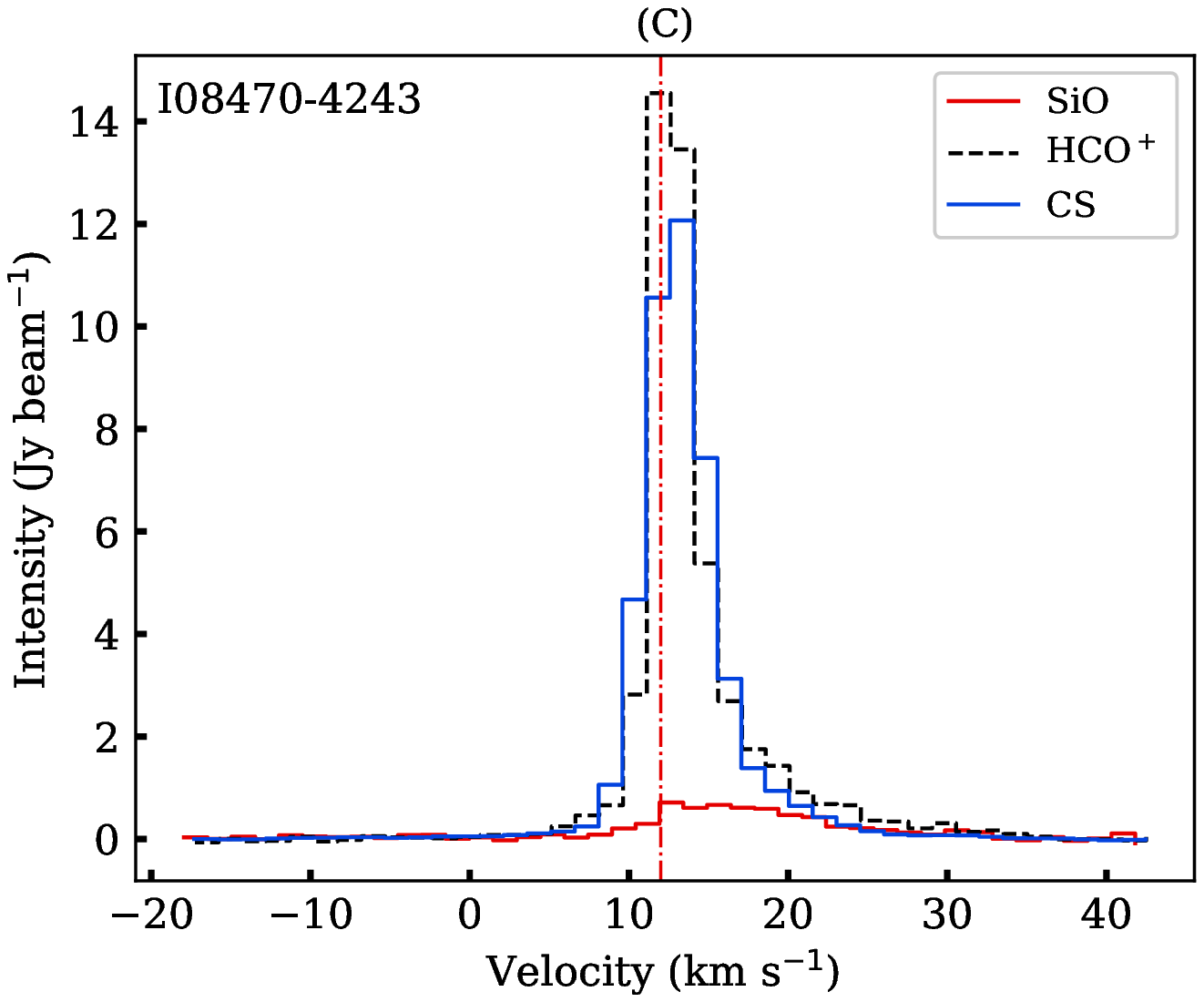}}
      \end{minipage}
  \begin{minipage}[t]{0.44\linewidth}

     \centerline{\includegraphics[width=1.12\linewidth]{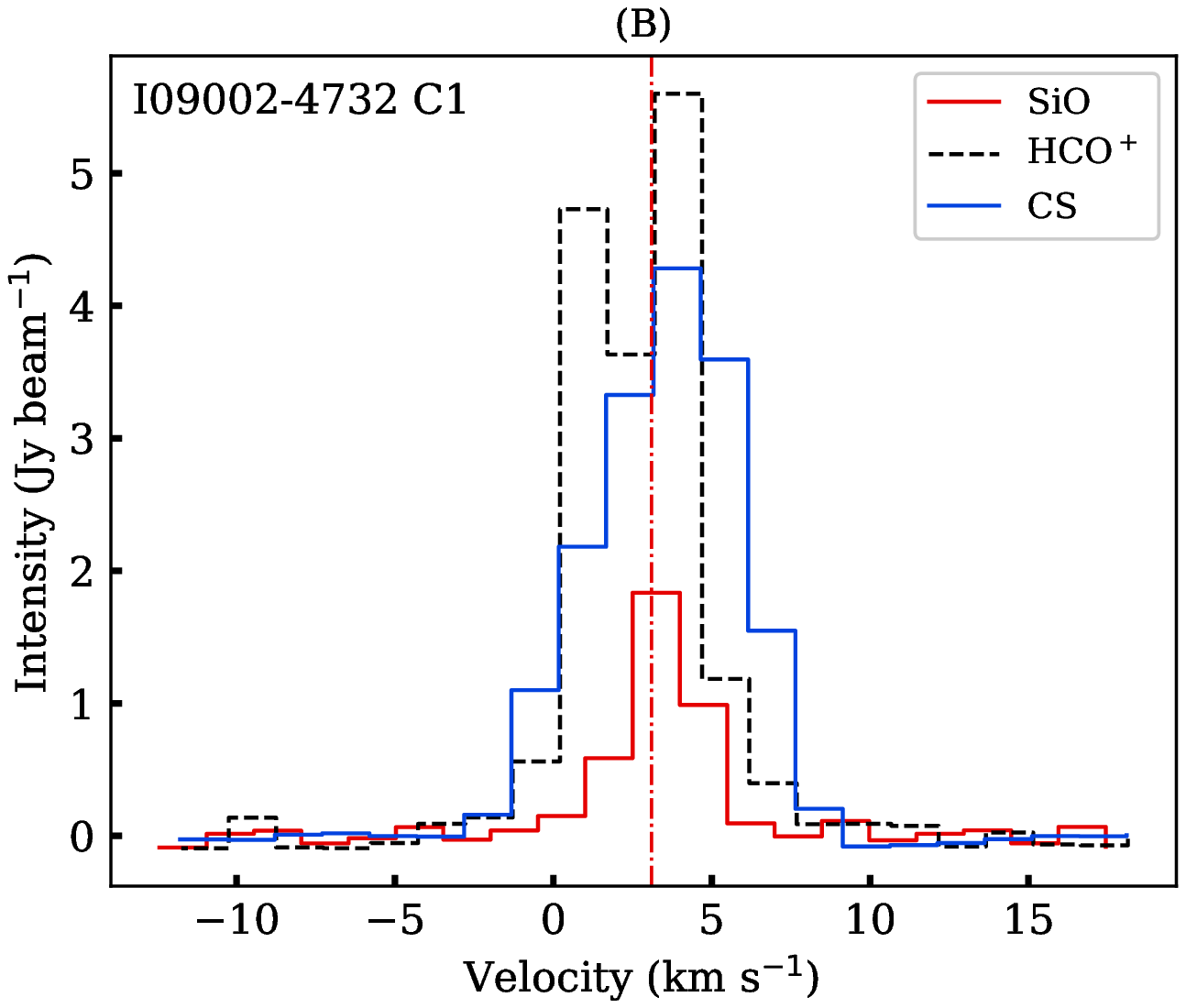}}
     \centerline{\includegraphics[width=1.12\linewidth]{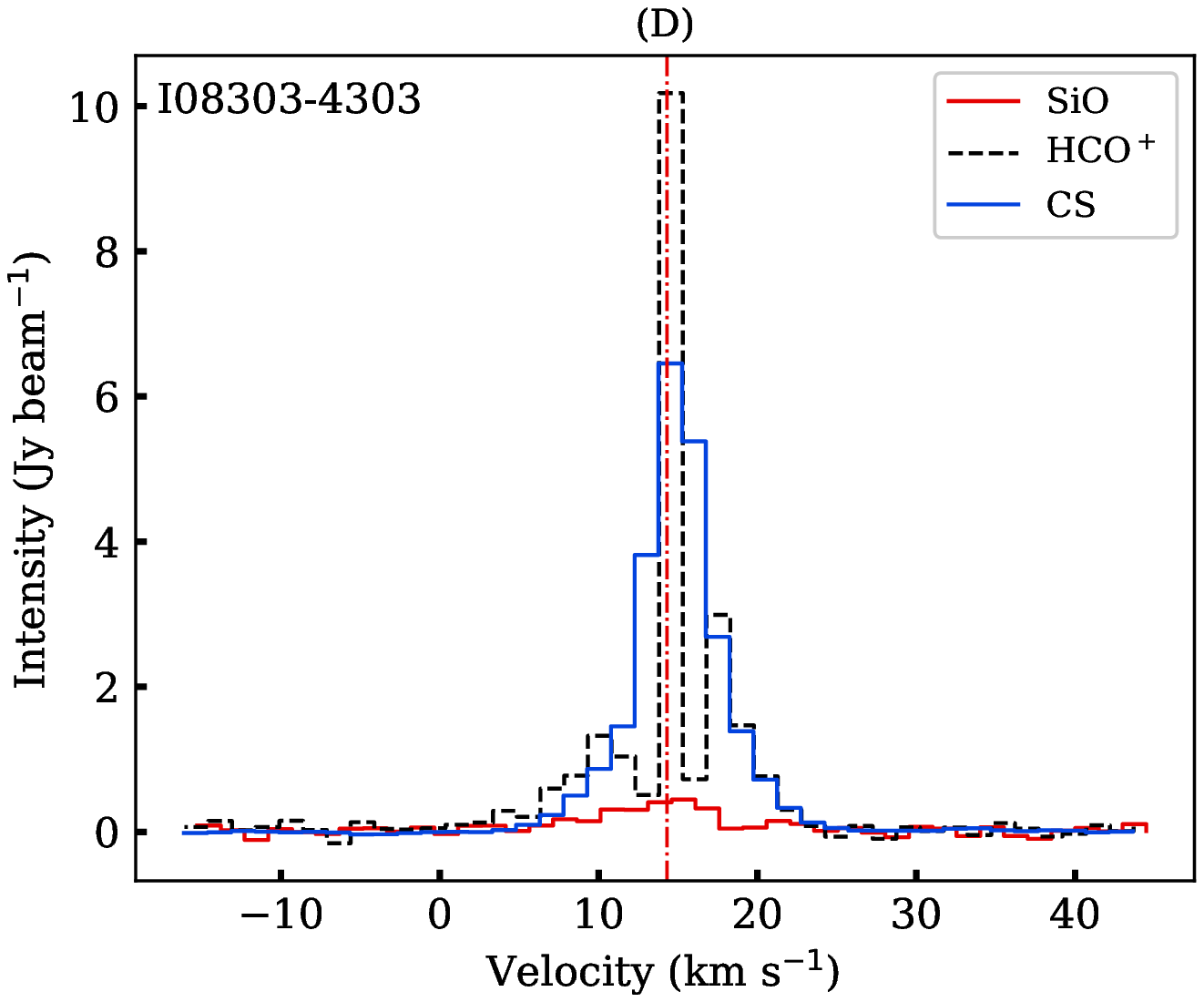}}
    \end{minipage}
   \caption{Examples of sources. The red spectrum represents the SiO average spectral extracted over the aperture of SiO clumps or beam. The HCO$^+$ and CS spectra are presented as black dashed and blue solid lines. The dash-dot line indicates the systemic velocity of the source.} 
   \label{fig2}
\end{figure*}

\subsection{ALMA Observations}
The present study is based on the ATOMS survey data (Project ID: 2019.1.00685.S; PI: Tie Liu). The survey has collected both 12m-array and ACA data. We focus the latter in this work with an aim at compiling a complete catalog of SiO emission clumps owing to the benefit from the large field of view of the ACA observations. The Atacama Compact 7 m Array (ACA) observations were conducted from September to mid-November 2019. 
The typical ACA observing time is ${\sim}$ 8 minutes. 
The angular resolution of ACA observations is ${\sim}$ 13.1$^{\prime \prime}$ $-$ 13.8$^{\prime \prime}$ and the maximum recovered angular scale is ${\sim}$ 53.8$^{\prime \prime}$ $-$ 76.2$^{\prime \prime}$. The SiO J=2-1, HCO$^+$ J=1-0, H$^{13}$CO$^+$ $J$=1-0 lines are included in SPWs 1${\sim}$6  at the lower side band with the spectral resolutions of  0.21, 0.11 and 0.21 km s$^{-1}$, respectively \citep{liu2021atomsIII, Liu2021b}. The H40$\alpha$ and CS $J$=2-1 lines are included in SPWs 7 at the upper side band with the spectral resolutions of 1.49 km s$^{-1}$. The ACA data were calibrated and imaged by CASA software package version 5.6 \citep{mcmullin2007casa}, and more details about data reduction can be found in Paper I.

\section{Results}
\label{sec3}
For the identification of SiO clumps, we firstly generate the integrated intensity (Moment 0) maps of SiO emission using channels with signals higher than 3${\sigma}$. The rms level ${\sigma}$ for each source are listed in Table~\ref{tab:TableA1} and Table~\ref{tab:TableA2}. 
In Moment-0 maps, we have identified 171 SiO clumps in 128 sources by eye. We identified SiO clumps by eye because there are not so many SiO clumps within individual sources and the emission peaks of most SiO clumps are clearly separated from each other in these low resolution ACA images. Therefore, we did not apply any algorithm in clump identification in order to avoid fake detection close to the edges of images where the signal-to-noise levels are low. Then we used the two-dimensional fitting tools ($imfit$) in CASA to fit these SiO clumps one by one. We get deconvolved sizes (the full width at half maximum (FWHM) and position angle (PA)), center positions and integrated intensity for these SiO clumps. SiO clumps in the marginal region are excluded. These results are shown in Table~\ref{tab:TableA1} and Table~\ref{tab:TableA2}. Next we extract the spectra of SiO lines toward the peak positions of SiO clumps in all detected sources. The spectra were extracted over the aperture of clumps size or beam size (clumps smaller than the beam size). Examples of the Moment-0 maps and spectra are shown in Appendix~\ref{figA1}.

Figure~\ref{fig1} shows four examples of detected sources. A Moment-0 map of SiO emission overlaid by contours of 3 mm continuum emission and H40$\alpha$ emission is shown. We classify the sources into four groups based on their different conditions in line emission. The classification criteria are as follows:

(A) 27 SiO clumps in sources containing H{\sc ii} regions: SiO emission associated with both H40$\alpha$ emission and 3 mm continuum emission.

(B) 59 SiO clumps in sources containing H{\sc ii} regions: SiO emission separated from both H40$\alpha$ emission and 3 mm continuum emission. 

(C) 39 SiO clumps in sources containing non H{\sc ii} regions: SiO emission associated with 3mm continuum emission.

(D) 46 SiO clumps in sources containing non H{\sc ii} regions: SiO emission partly separated from 3mm continuum emission.

Figure~\ref{fig2} present the SiO (2-1), HCO$^+$ (1-0), and CS (2-1) spectra extracted toward four exemplar SiO clumps. We smooth the spectral resolution of SiO and HCO$^+$ to the spectral resolution of CS (1.49 km s$^{-1}$) for comparison. 
In I12326-6245, I08470-4243, and I08303-4303, the spectra of SiO show high-velocity wings, and we can see two distinct components: a narrow component and a broad component. The broad SiO components especially with red- and/or blue-shifted line wings are likely related to molecular outflows. In I12326-6245, I08470-4243, and I08303-4303, the spectra of HCO$^+$ and CS also show high-velocity wings. In I09002-4732 C1, there is no high-velocity wing emission in all lines, indicating that this source does not have outflows. The spectra of other SiO clumps are shown in Appendix~\ref{figA1}.

In most detected SiO clumps, we can see the peak velocity of the narrow components is similar to the systemic velocity (marked by the red dash-dot line). We thus consider these components coincided with the ambient cloud. In very rare cases such as I08076-3556 (Appendix~\ref{figA1}), however, the SiO velocity ($\sim$40 km s$^{-1}$) deviated clearly from the systemic velocity (5.9 km s$^{-1}$) and SiO line emission mainly comes from red-shifted line wings.  

\subsection{Detection rates of SiO emission}

Toward the 146 sources, SiO (2-1) emission was detected in 128 (87.7$\%$) sources, indicating that the presence of shocks is very common in high-mass star forming regions. In particular, when we inspect the spectra of the sources that do not show strong SiO emission in their Moment-0 maps, we find additional 18 SiO clumps showing weak SiO emission, with peak intensity close to 2${\sigma}$. We list these weak SiO clumps in Table~\ref{tab:TableA2}. 

Because of the complicated and non-Gaussian SiO line profiles, we instead use their Moment-0 maps, the intensity-weighted velocity (Moment 1) maps, and the intensity-weighted velocity dispersion (Moment 2) maps to derive their velocity ($\nu$), and velocity dispersion ($\sigma_\textup{v}$). The derived parameters of SiO clumps are summarized in Table~\ref{tab:TableA1}. Notably, the velocity dispersion of SiO emission is mostly larger than the velocity resolution (0.21 km s$^{-1}$).
For the 18 weaker clumps, we use one component Gaussian fit of their averaged spectra to get the line parameters (Intensity, velocity, and velocity dispersion). The derived line parameters are shown in Table~\ref{tab:TableA2}. 
  
\subsection{The velocity dispersion of SiO and the line wings of SiO, HCO$^{+}$, and CS}
\label{sec3.2}
\begin{figure}
   \centerline{\includegraphics[width=1.14\linewidth]{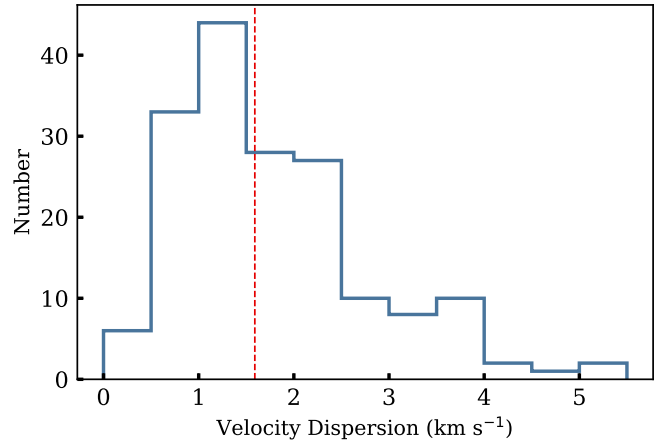}}
   \caption{Histogram of the velocity dispersion of the SiO (2-1) lines. The vertical red dashed line represents the median velocity dispersion.} 
   \label{fig3}
\end{figure}
The velocity dispersions of the detected SiO lines are listed in Table~\ref{tab:TableA1} and Table~\ref{tab:TableA2}. For the SiO lines, the range of the velocity dispersion values is 0.3${\sim}$5.43 km s$^{-1}$. The mean values is 1.77 km s$^{-1}$, and the median values is 1.59 km s$^{-1}$. The histogram of the velocity dispersion is presented in Figure~\ref{fig3}.

To identify strong outflows in these sources, we searched the line wings in SiO, HCO$^{+}$, and CS emission lines. These results are listed in Table~\ref{tabA3}. Because of relatively high abundance, HCO$^{+}$ emission is a very good tracer of outflows \citep{myers1996simple}. 
Therefore, in outflow regions, HCO$^{+}$ emission is detectable even if SiO emission is not detected. However, in some sources the line profile of HCO$^{+}$ emission is self-absorbed. We inspect the spectra of CS and find its spectra to be less self-absorbed. As a consequence, we use the line wings of SiO, HCO$^{+}$, and CS together to identify strong outflows. The spectra of SiO, HCO$^+$, and CS are shown in Appendix~\ref{figA1}. 
According to the high velocity line wings, the SiO clumps are divided into two groups: clumps with strong outflows and without strong outflows. We find 116 clumps are associated with outflows showing wing emission in at least one of the three lines and 50 clumps are not associated with outflows. However, we cannot exclude these clumps without outflows that are not associated with weak outflows.
We exclude I17441-2822 spectra in further analysis because this source is close to the galactic center and shows very complicated line profiles. 

\begin{figure*}
   \centerline{\includegraphics[width=1.13\linewidth]{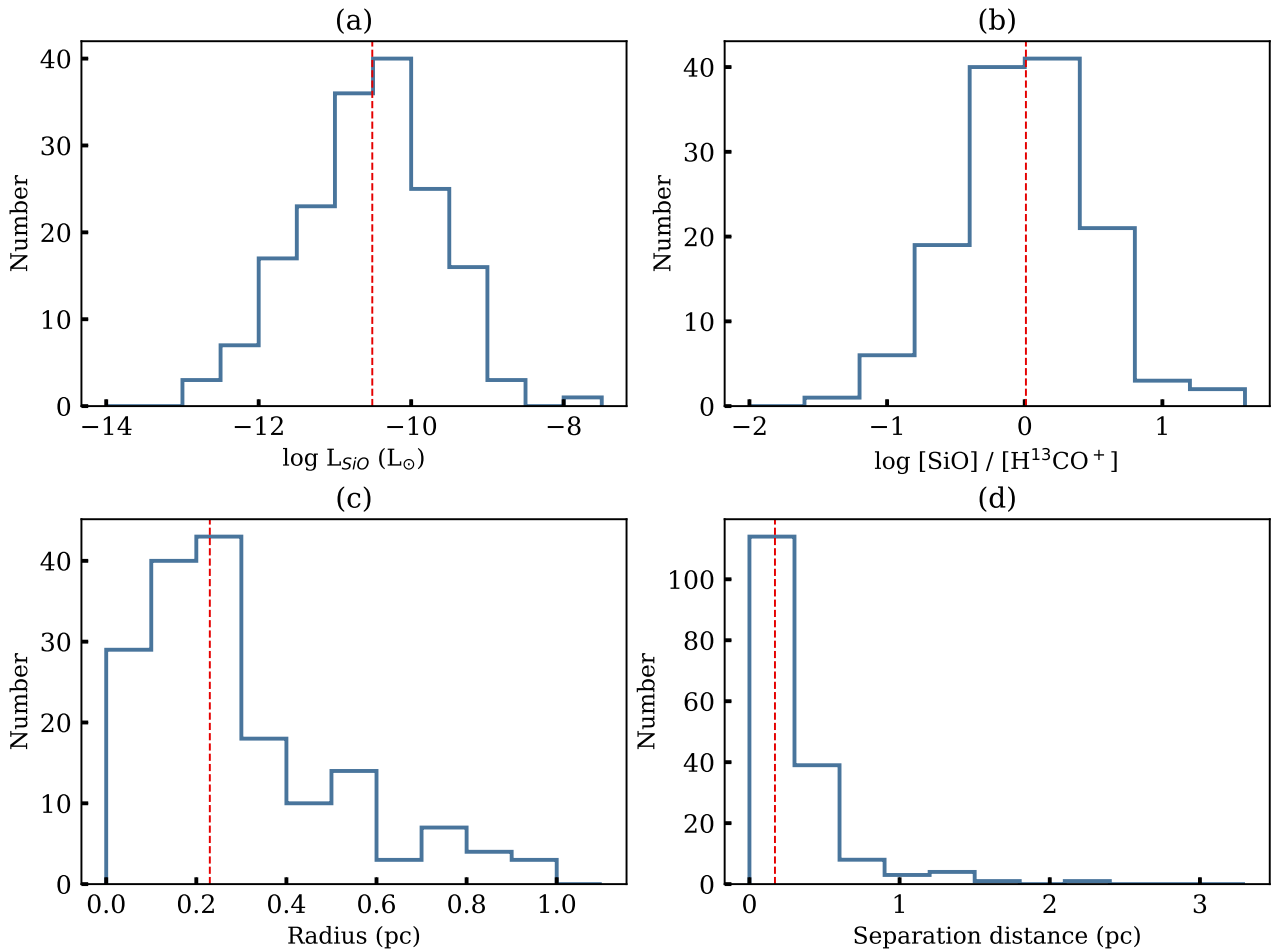}}
   \vspace{1pt}
   \caption{ (a) Histogram of the SiO luminosity. (b) Histogram of the [SiO]/[H$^{13}$CO$^+$] ratio. (c) Histogram of separation distance. (d) Histogram of the SiO clump radius. The vertical red dashed line represents the median values of the parameter.} 
   \label{fig4}
\end{figure*}

\subsection{SiO clumps properties}

Using the Moment-0 maps of SiO emission, we have identified 171 SiO clumps. We calculated the geometric mean of the major axis FWHM and the minor axis FWHM and got the SiO clumps linear radius. The histogram of the SiO clump radii is presented in Figure~\ref{fig4}. The range of the SiO clump radii is 0.03${\sim}$0.99 pc. The mean value is 0.3 pc, and the median value is 0.23 pc. 
In addition, we failed to fit 116 clumps because the shape of SiO clumps is irregular and the size of these SiO clumps is smaller than or close to the beam size. Instead, we give the beam size as these clump sizes, and we marked these clumps in Table~\ref{tab:TableA1}. The velocities of SiO clumps are approximately consistent with the systemic velocities of the clumps, except for I08076-3556 (Sect.~\ref{sec3}). 

In Moment-0 maps, we found that 105 SiO clumps are not spatially coincided with the peak of 3 mm continuum emission with separation larger than half of beam size. There are 66 SiO clumps associated with 3 mm continuum emission. We derived the distances between the central positions of SiO clumps and the peak positions of the 3 mm continuum emission. These results are listed in Table~\ref{tabA3}. The histogram of the separation distance is presented in Figure~\ref{fig4}. The range of the separation is 0${\sim}$3.35 pc. The mean value is 0.3 pc, and the median value is 0.17 pc. An evident tail is presented in the separation histogram which represents these SiO clumps away from the 3 mm continuum emission may not be caused by outflows. This separation between the thermal dust and SiO molecular gas is similar to the results of \citet{lopez2016role} and \citet{li2020alma}.

For the 128 sources with SiO emission, 63 sources have H40$\alpha$ emission above 3${\sigma}$. The H40$\alpha$ emission shows a compact structure and is coincident with the 3 mm continuum emission, indicating the existence of compact H{\sc ii} regions. For some of these sources, we find that SiO clumps are surrounding the H40$\alpha$ emission, indicating the possible existence of shocks induced by interaction between HII regions and their surrounding molecular clouds \citep{cosentino2020sio,liu2020atoms}. We will investigate the properties of shocked gas caused by H{\sc ii} regions using the high resolution 12-m array data in a forthcoming work.

We calculate the SiO (2-1) luminosity ($L_\textup{SiO}$) of the SiO clumps using the integrated intensity and the source distance. $L_\textup{SiO}$ can be derived from the formula:
\begin{equation}
    {L_\textup{SiO}}=4\pi\times{d^2}\times{\int{F_{\rm{\nu}}d{\nu}}}, 
	\label{eq:eq1}
\end{equation}
where d is the distance to the source and $\nu$ represents the velocity. $\int{F_{\nu} d{\nu}}$ is the integrated integrated intensity of the SiO clump.

$L_\textup{SiO}$ from 1.74 $\times{10^{-13}}$ to 1.07 $\times {10^{-8}}$ $L_{\odot}$. The mean value is 2.69 $\times{10^{-11}}$ $L_{\odot}$, and the median value is 3.1 $\times{10^{-11}}$ $L_{\odot}$. Figure~\ref{fig4} (a) shows the number distribution of the SiO luminosity. We used the sum $L_\textup{SiO}$ in one source for statistic analysis. More details on $L_\textup{SiO}$ will be discussed in Sect.~\ref{sec4.1.2}.

\begin{figure*}
   \centerline{\includegraphics[width=1.15\linewidth]{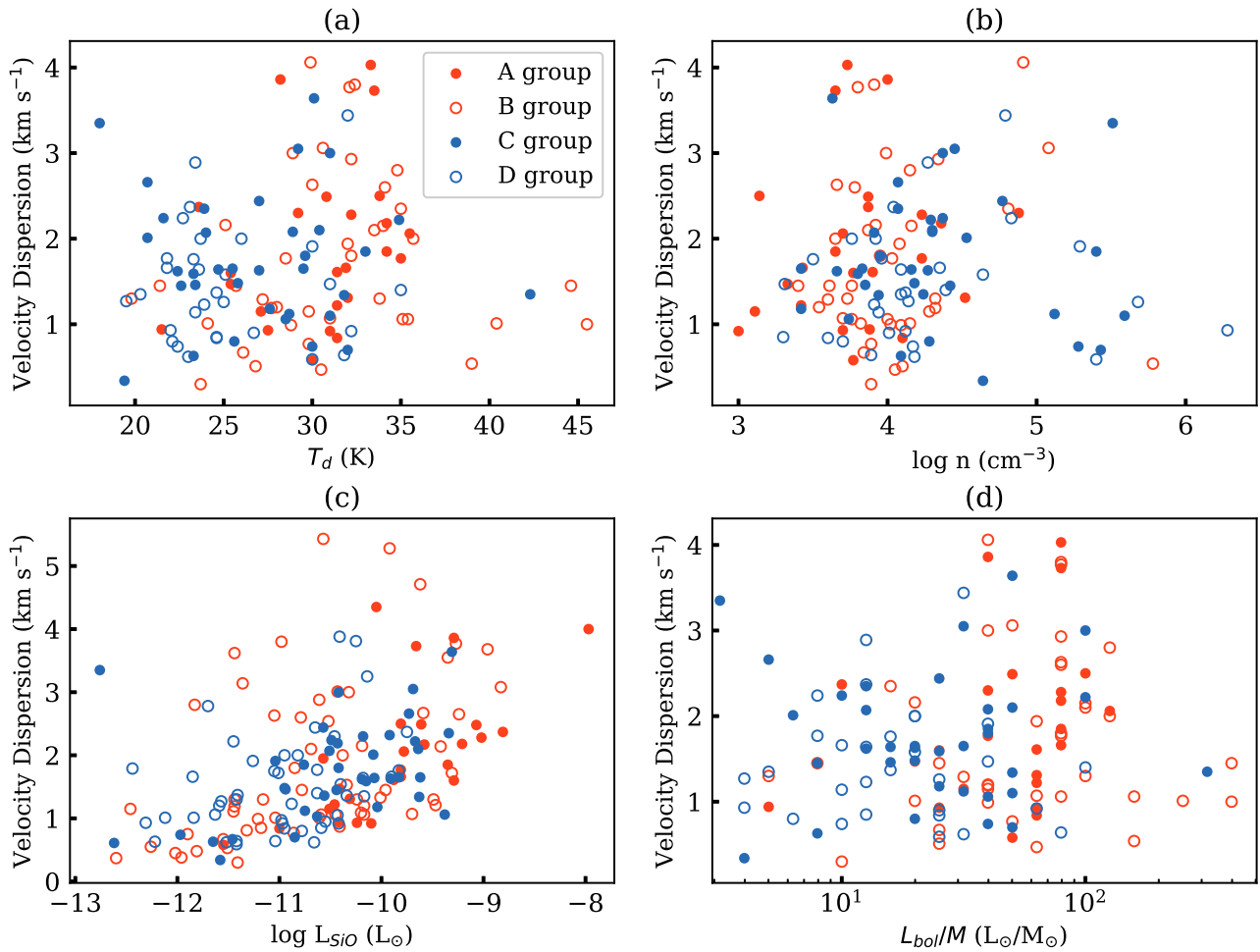}}
   \caption{ (a) The velocity dispersion of SiO emission against dust temperature ($T_\textup{d}$). (b) The velocity dispersion of SiO emission versus the particle number density ($n$) of the clumps. (c) The velocity dispersion of SiO emission against the SiO clump radius. (d) The velocity dispersion of SiO emission vs. the bolometric luminosity to mass ratio ($L_\textup{bol}/M$). The filled red circles (A group) represent the SiO clumps containing H{\sc ii} regions are associated with 3 mm continuum emission. The empty red circles (B group) represent the SiO clumps containing H{\sc ii} regions are separated with 3 mm continuum emission.
   The filled blue circles (C group) represent the SiO clumps containing non H{\sc ii} regions are associated with 3 mm continuum emission. The empty blue circles (D group) represent the SiO clumps containing non H{\sc ii} regions are separated with 3 mm continuum emission.} 
   \label{fig5}
\end{figure*}

\subsection{The fraction of shocked gas}

The H$^{13}$CO$^+$ emission is a good tracer of relatively quiescent gas. The H$^{13}$CO$^+$ abundance does not vary substantially with time \citep{nomura2004physical}. Therefore, column density of H$^{13}$CO$^+$ could reflect the dense gas of the clump. \citet{sakai2010survey} used SiO column density against H$^{13}$CO$^+$ column density to represent the fraction of shocked gas in a dense clump. Assuming both SiO and H$^{13}$CO$^+$ emission are optically thin and have constant excitation temperatures, the integrated intensity ratios of SiO emission and H$^{13}$CO$^+$ emission can approximately reflect the relative fraction of the shocked gas in the clump, and later we use [SiO]/[H$^{13}$CO$^+$] to represent this ratio. The derived ratios are shown in Table~\ref{tabA3}.  We inspected H$^{13}$CO$^+$ spectra and found that 12 sources have absorption features in their spectra. As a result, in the latter analysis, we ignore these 12 sources. For sources containing multiple SiO clumps, we calculate an average [SiO]/[H$^{13}$CO$^+$] ratio. The range of [SiO]/[H$^{13}$CO$^+$] is 0.04${\sim}$42.79. The mean values is 2.41, and the median values is 1.02. The histogram of [SiO]/[H$^{13}$CO$^+$] of all sources is presented in Figure~\ref{fig4} (b).

\begin{figure}
   \centerline{\includegraphics[width=1.14\linewidth]{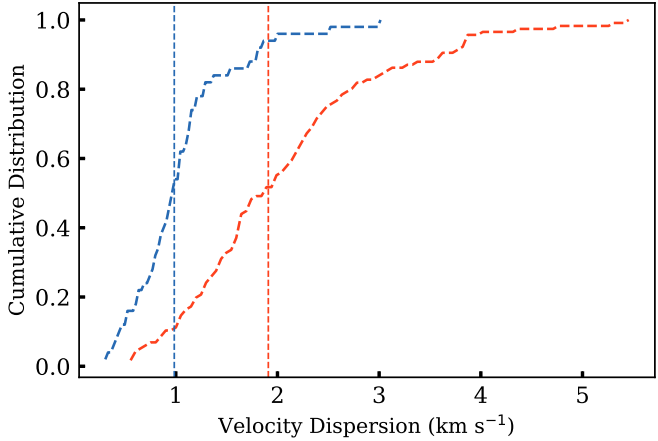}}
   \caption{The velocity dispersion distributions for SiO clumps with outflows and without outflows (the red and blue lines). The vertical dashed lines represent the median velocity dispersion of SiO clumps in the two groups: 1.91 km~s$^{-1}$ (red line) and 0.99 km~s$^{-1}$ (blue line) for clumps with and without strong outflows, respectively.} 
   \label{fig6}
\end{figure}
\section{Discussion}
\label{sec4}

\begin{figure}
   \centerline{\includegraphics[width=1.12\linewidth]{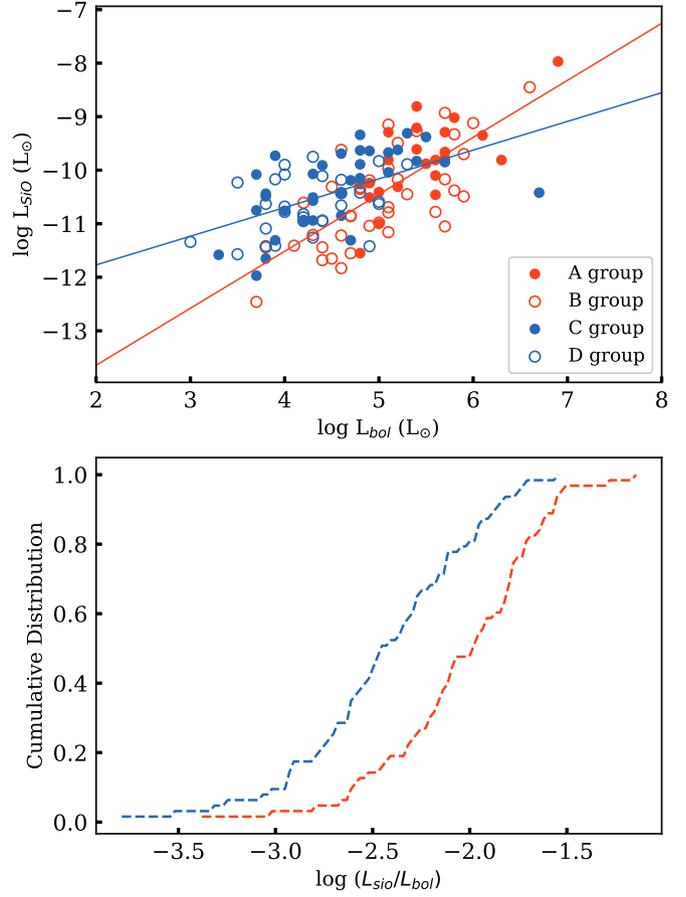}}
   \caption{Upper panel: SiO (2-1) line luminosity $L_\textup{SiO}$ versus bolometric luminosity $L_\textup{bol}$. The red line shows a linear fit of $f(x) = (1.06\pm0.12)x - 15.77$ and the blue line a linear fit of $f(x) = (0.54\pm0.11)x - 12.84$. The symbols are the same with Figure~\ref{fig5}. Lower panel: The $L_\textup{SiO}$/$L_\textup{bol}$ distributions for SiO clumps with H{\sc ii} regions and without H{\sc ii} regions (the red and blue lines). }
   \label{fig7}
\end{figure}

\begin{figure*}
   \centerline{\includegraphics[width=1.14\linewidth]{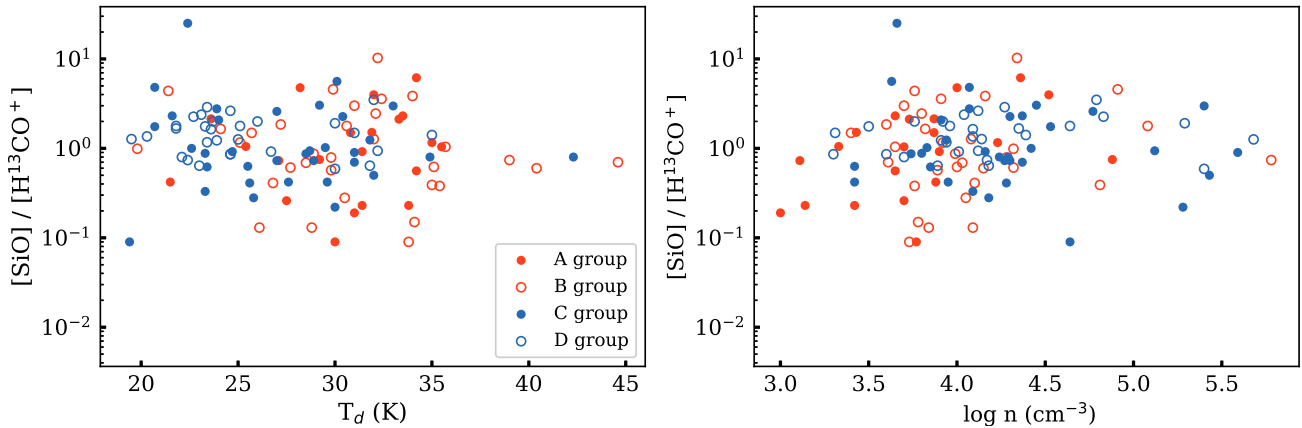}}
   \caption{Left: [SiO]/[H$^{13}$CO$^+$] against dust temperature ($T_\textup{d}$). Right: [SiO]/[H$^{13}$CO$^+$] vs. particle number density ($n$). The symbols are the same as in Figure~\ref{fig5}.} 
   \label{fig8}
\end{figure*}

\begin{figure*}
   \centerline{\includegraphics[width=1.14\linewidth]{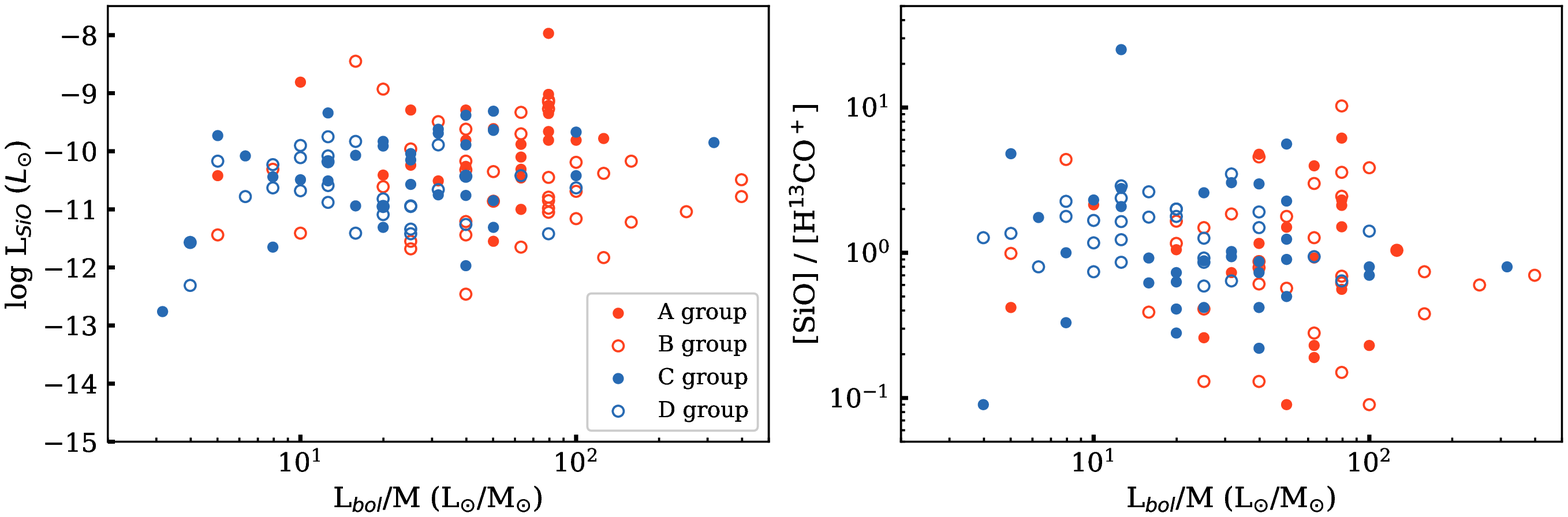}}
   \caption{Left: The SiO line luminosity versus $L_\textup{bol}/M$ in the clumps. Right: [SiO]/[H$^{13}$CO$^+$] vs. $L_\textup{bol}/M$. The symbols are the same with Figure~\ref{fig5}} 
   \label{fig9}
\end{figure*}
\subsection{The line broadening of SiO emission}
\label{sec4.1}

The broad components of SiO emission in our sample are likely caused by the high-velocity gas driven by energetic outflows. The high detection rate of the velocity wings of SiO emission (60$\%$) suggests the presence of outflows. Our finding divides the SiO clumps into two groups (Sect.~\ref{sec3.2}). In the group with outflows, the median velocity dispersion of the SiO emission is as large as 1.91 km s$^{-1}$. 
In the other group without outflows, the narrow components of SiO emission are common and the median velocity dispersion of the SiO emission is 0.99 km s$^{-1}$. This narrow SiO emission could be created by unresolved low mass outflows, cloud-cloud collision, or gas inflows. 

We plot the cumulative distribution of the SiO velocity dispersion for these two groups in Figure~\ref{fig6}. The SiO clumps with outflows are presented as red curve, while the clumps without outflows are shown as blue curve.
For the SiO clumps with strong outflows, the velocity dispersion values range from 0.58 to 5.43 km s$^{-1}$ with a mean value of 2.08 km s$^{-1}$ and a standard deviation of 1.00 km s$^{-1}$.
In contrast, for the SiO clumps without strong outflows, the velocity dispersion values are from 0.32 to 3 km s$^{-1}$, with a mean value of 1.06 km s$^{-1}$ and a standard deviation of 0.53 km s$^{-1}$.
In addition, we use the Kolmogorov-Smirnov (KS) test to compare these distributions. The P-value returned by the KS test is the probability that the two samples were drawn from the same distributions. If the P-value is smaller than 5$\%$, we conclude that the two samples were drawn from different distributions. The KS test gives a P-value of about 10$^{-12}$, indicating that the velocity dispersion distribution of these two groups is very different. The clumps associated with outflows show significantly larger velocity dispersion in SiO emission on average.

In shocked gas, the SiO velocity dispersion is attributed to non-thermal broadening. \citet{gusdorf2008sioa} found that the SiO intensity and line width as the shock dissipates. In Figure~\ref{fig5}, we plot the velocity dispersion of the four groups (Sect.~\ref{sec3}) against dust temperature ($T_\textup{d}$), the particle number density ($n$) of the clumps, the SiO luminosity ($L_\textup{sio}$), and the bolometric luminosity to mass ratio ($L_\textup{bol}/M$) of clumps from \citet{liu2020atoms}, respectively.  The particle number density can be derived as
\begin{equation}
    n = \frac{3 M_{gas}}{4 \pi R^3 m_{H}\mu},
	\label{eq:eq2}
\end{equation}
$M$ and $R$ the clump mass and effective radii compiled in \citet{liu2020atoms}. $n$ is the particle number density. $m{_\textup{H}}$ is the mass of a hydrogen atom. ${\mu}$ = 2.37 is the mean molecular weight per "free particle". $T_\textup{d}$ and $L_\textup{bol}/M$ are potential evolutionary tracers for high-mass protostars and their natal clumps \citep{Molinari2008,Elia2021}. More evolved sources show higher $T_\textup{d}$ and $L_\textup{bol}/M$.

As shown in Figure~\ref{fig5}, there is no obvious trend between the velocity dispersion of all groups SiO emission and either $T_\textup{d}$ or $n$. There is also no statistically significant correlation between the velocity dispersion of all groups and $L_\textup{bol}/M$. We use the Spearman-rank correlation test between those quantities, and the correlation coefficient is 0.15, 0.06, and 0.13, respectively. In our samples, we found no obvious variations in the velocity dispersion of the SiO emission against $T_\textup{d}$, $n$, and $L_\textup{bol}/M$, which indicates that the strengths of shocks are not so different under different physical conditions or at various evolutionary stages. Our results are consistent with the results of \citet{csengeri2016atlasgal} and \citet{li2019sio}, who also found the line width of SiO emission is nearly constant at different evolutionary stages of clumps. 
In A, B, C groups, the $L_\textup{sio}$ has a positive correlation with the velocity dispersion of SiO emission, with a correlation coefficient of 0.72, 0.55, and 0.44, respectively. These results suggest that the more intense SiO sources are associated with more active outflows. Whereas in the D group, the $L_\textup{sio}$ does not correlate with the velocity dispersion of SiO emission. It means it presence of relatively strong SiO emission sources has smaller velocity dispersion. One would expect these SiO clumps may be due to the aging of outflows as it moves away from its driving source. An alternative explanation is to consider some of these SiO sources caused by unresolved lower mass protostars, cloud-cloud collision, or gas inflows. We will investigate these scenarios in the following work with high resolution 12-m array data.

\subsection{The excitation condition for SiO emission}
\label{sec4.1.2}

We examine the relationship between the bolometric luminosity and SiO emission line luminosity in all groups. Figure~\ref{fig7} shows the SiO line luminosity versus the bolometric luminosity in the upper panel. We exclude two low-mass star forming sources \citep[I08076-3556 and I11590-6452;][]{liu2020atoms}. We can see an increasing trend between these two quantities in all groups. But in the A and B groups (with H{\sc ii} regions), the increasing trend is much steeper. 
Thus we get two linear fits. In the groups with H{\sc ii} regions, a linear fit is $f(x) = (1.06\pm0.12)x - 15.77$ and in the groups without H{\sc ii} regions, another linear fit is $f(x) = (0.54\pm0.11)x - 12.84$. The Spearman rank correlation coefficient (p) is 0.7 and 0.58, respectively. This implies that higher luminosity sources could have brighter SiO emission, indicating stronger shock activity in more luminous proto-clusters. These results are consistent with the results of \citet{codella1999low} and \citet{liu2021sio}, who both found a trend of brighter SiO emission in higher luminosity sources. 
As for the different slope of $L_\textup{sio}$/$L_\textup{bol}$, this could be due to the different evolutionary stages of these sources. Next, we plot the cumulative distribution of the $L_\textup{sio}$/$L_\textup{bol}$ for the SiO clumps with H{\sc ii} regions and without H{\sc ii} regions in the lower panel. The KS test reveals that the p-value is 10$^{-5}$, indicating the $L_\textup{sio}$/$L_\textup{bol}$ for these two groups is from different distributions. Sources containing H{\sc ii} regions show relatively lower $L_\textup{sio}$/$L_\textup{bol}$ ratios, indicating that H{\sc ii} regions may have negative feedback on surrounding gas and may suppress further star formation, leading to less energetic outflows and thus weaker shocks.

Figure~\ref{fig8} shows [SiO]/[H$^{13}$CO$^+$] intensity ratio as a function of $T_\textup{d}$ and $n$ in all groups (p = -0.13, 0.12, respectively). However, we find the fraction of shocked gas increase as the $n$ increase in A group (p = 0.5). Summarizing, in a large fraction of sources (except for the A group), the fraction of shocked gas shows no essential dependence on the $T_\textup{d}$ and $n$. This indicates that SiO emission is not affected by thermal condition but is more likely affected by shock activities. However, H$^{13}$CO$^+$ abundance may not be constant with time \citep{sanhueza2012chemistry}, which may affect the interpretation of [SiO]/[H$^{13}$CO$^+$] ratios.

The bolometric luminosity of a molecular clump will increase as the high mass star evolves, while its mass will decrease. Thus the bolometric luminosity to mass ratio ($L_\textup{bol}/M$) can be a good tracer of the evolutionary stage of star formation \citep{Molinari2008,molinari2016calibration,Elia2021}. $L_\textup{bol}/M$ values can distinguish between the young and evolved sources, and its low values are related to young sources. 
In the left panel of Figure~\ref{fig9}, we plot the SiO luminosity ($L_\textup{SiO}$) against $L_\textup{bol}/M$ in all groups (p = 0.3, -0.04, 0.32, and -0.12, respectively). The $L_\textup{SiO}$ does not vary with $L_\textup{bol}/M$, suggesting that SiO luminosity has no relationship with evolutionary stages, which is similar to the results of \citet{liu2021sio}.
The right panel shows [SiO]/[H$^{13}$CO$^+$] versus $L_\textup{bol}/M$. There is also no correlation between [SiO]/[H$^{13}$CO$^+$] and $L_\textup{bol}/M$. This implies that the fraction of shocked gas in high-mass star forming clumps does not change obviously in dense gas at various evolutionary stages. This is consistent with the results reported by \citet{csengeri2016atlasgal} and \citet{li2019sio}. 

\section{Summary}
\label{sec5}
In this work, we used ALMA ACA observational data for a statistical study of shocked gas toward 146 massive star forming regions. We analyze the variation of SiO emission under different physical conditions and evolutionary stages. The main results are summarized as follows:

(1) Among the entire sample, we have detected SiO emission in 128 sources, which contain 171 SiO clumps. SiO (2-1) emission has a high detection rate of 87.7$\%$, above 3${\sigma}$. 

(2) The velocity dispersion of SiO line emission ranges from 0.3 to 5.43 km s$^{-1}$, with a median velocity dispersion of 1.59 km s$^{-1}$. Based on the high-velocity emission wings in SiO, HCO$^+$, and CS lines, we divided the clumps into two groups. There are 116 SiO clumps associated with strong outflows, which show high-velocity wing emission in at least one line, while the other 50 SiO clumps show no wing emission in the three lines and seem to be not associated with energetic outflows. The two groups have an obvious difference in the velocity dispersion of SiO emission. The median velocity dispersion of outflow sources is 1.91 km s$^{-1}$, which is significantly larger than that (0.99 km s$^{-1}$) of the non-outflow sources, indicating that outflow activities have a great influence on the strongly shocked gas. In particular, SiO emission clumps with small velocity dispersion could be formed by low-velocity shocks that are induced by either H{\sc ii} regions or other large-scale compression flows (e.g., cloud-cloud collision), which will be investigated thoroughly in forthcoming works with higher resolution ALMA 12-m array data.

(3) We find a positive correlation between the SiO line luminosity and the bolometric luminosity, implying stronger shock activities associated with more luminous proto-clusters. We also find the SiO clumps with H{\sc ii} regions show a lower $L_\textup{sio}$/$L_\textup{bol}$ ratio than the SiO clumps without H{\sc ii} regions. In most sources, the velocity dispersion of SiO emission and [SiO]/[H$^{13}$CO$^+$] show no obvious correlations with dust temperature ($T_\textup{d}$) and particle number density ($n$). These results indicated that the SiO emission is not likely affected thermal conditions but is more likely affected by shock activities. In addition, we do not see clear correlations between the SiO line luminosity and $L_\textup{bol}/M$. There is also no robust trend in the [SiO]/[H$^{13}$CO$^+$] and $L_\textup{bol}/M$ relation. This implies that the fraction of shocked gas in dense gas does not change obviously at various evolutionary stages.

\section*{Acknowledgements}
Tie Liu acknowledges the supports by National Natural Science Foundation of China (NSFC) through grants No.12073061 and No.12122307, the international partnership program of Chinese academy of sciences through grant No.114231KYSB20200009, and Shanghai Pujiang Program 20PJ1415500.
This work is supported by the Ministry of Science and Technology of China through grant 2010DFA02710, the Key Project of Interntional 
Cooperation, and by the National Natural Science Foundation of China (NSFC) through grants 11503035, 11573036.
H.-L. Liu is supported by NSFC through the grant No.12103045.
K. W. acknowledges support by the National Science Foundation of China (12041305, 11973013) and the High-Performance Computing Platform of Peking University through the instrumental analysis fund of Peking University (0000057511).
This research was carried out in part at the Jet Propulsion Laboratory, which is operated by the California Institute of Technology under a contract with the National Aeronautics and Space Administration (80NM0018D0004).
G.G. and L.B. acknowledge support by the ANID BASAL project FB210003.
S.-L. Qin is supported by NSFC under grant No. 12033005.
C. W. L. is supported by the Basic Science Research Program through the National Research Foundation of Korea (NRF) funded by the Ministry of Education, Science and Technology (NRF-2019R1A2C1010851).
Y. Zhang acknowledges financial support from NSFC (grant No. 11973099).
JHH thanks the National Natural Science Foundation of China under grant No. 11873086. This work is sponsored (in part) by the Chinese Academy of Sciences (CAS), through a grant to the CAS South America Center for Astronomy (CASSACA) in Santiago, Chile.
Guo-Yin Zhang acknowledges support by China Postdoctoral Science 
Foundation (No. 2021T140672).
This paper makes use of the following ALMA data: ADS/JAO.ALMA\#2019.1.00685.S. ALMA is a partnership of ESO (representing its member states), NSF (USA), and NINS (Japan), together with NRC (Canada), MOST and ASIAA (Taiwan), and KASI (Republic of Korea), in cooperation with the Republic of Chile.

\section*{Data availability}
The data underlying this article are available in the article and in
ALMA archive.

\input{ATOMS_liurong.bbl}
\bibliographystyle{mnras}
\bibliography{ATOMS_liurong}



\appendix

\section{}
\begin{figure*}
\begin{minipage}[t]{0.24\linewidth}
   \vspace{5pt}
   \centerline{\includegraphics[width=1.13\linewidth]{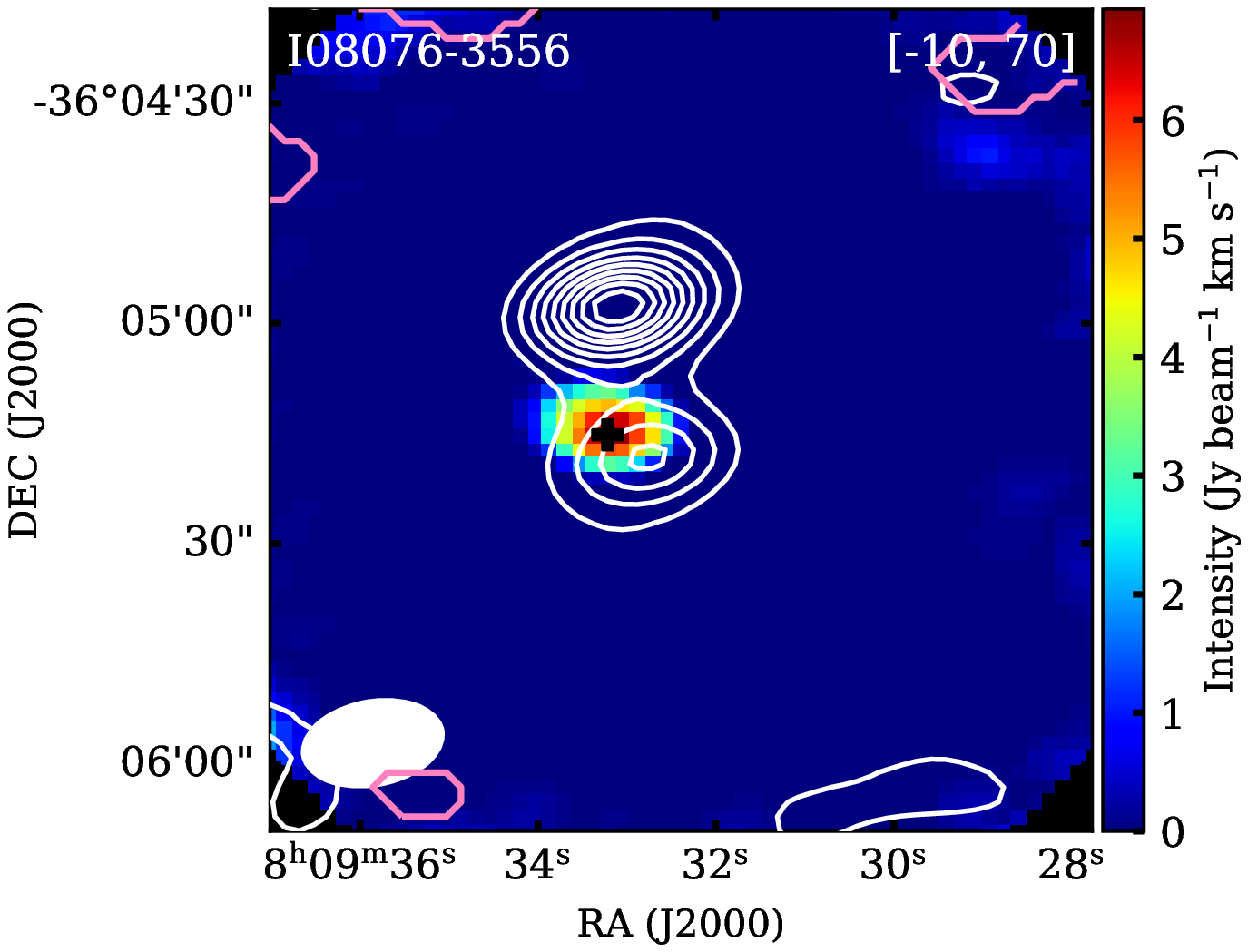}}
   \vspace{5pt}
   \centerline{\includegraphics[width=1.13\linewidth]{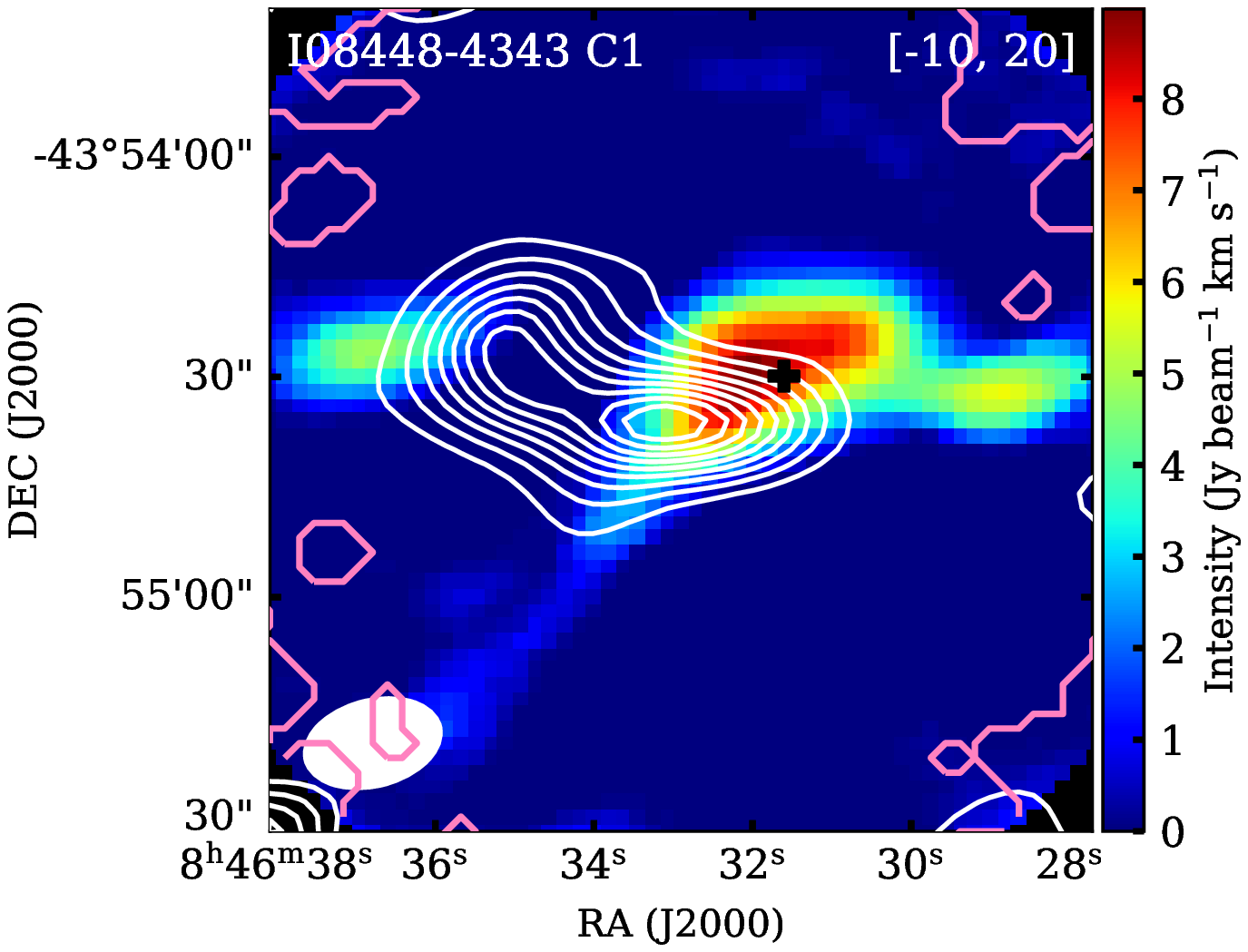}}
   \vspace{5pt}
   \centerline{\includegraphics[width=1.13\linewidth]{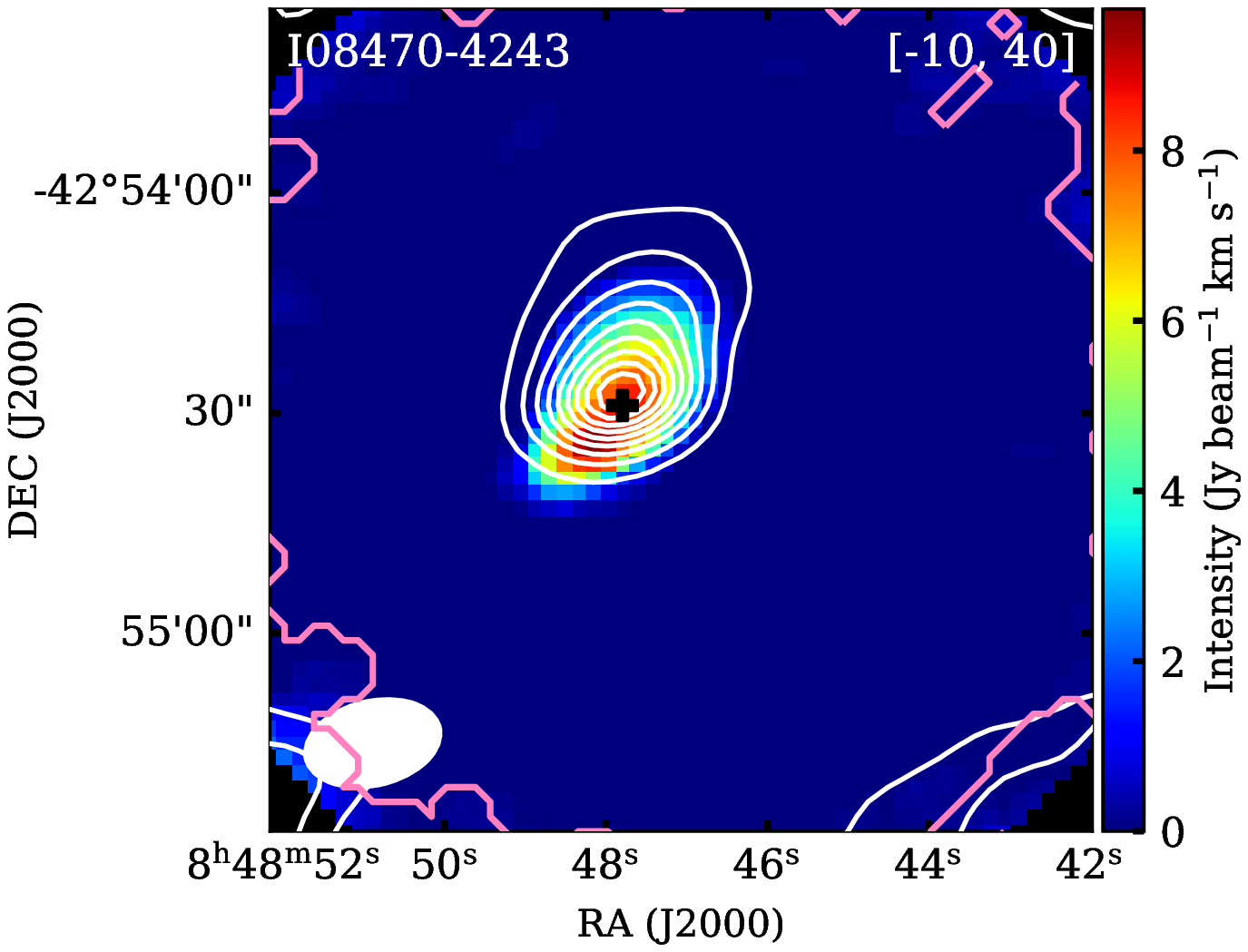}}
   \vspace{5pt}
   \centerline{\includegraphics[width=1.13\linewidth]{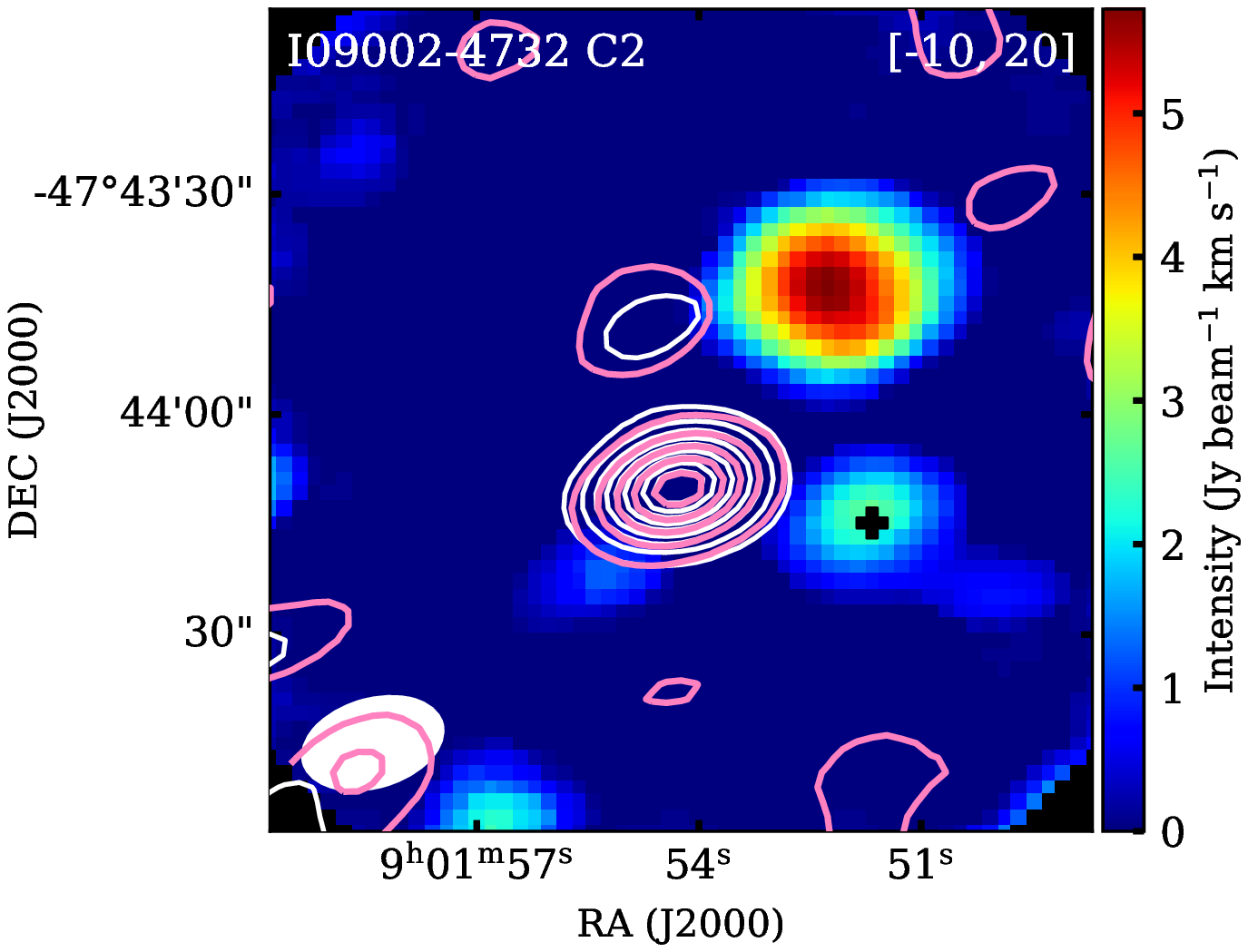}}
   \vspace{5pt}
   \centerline{\includegraphics[width=1.13\linewidth]{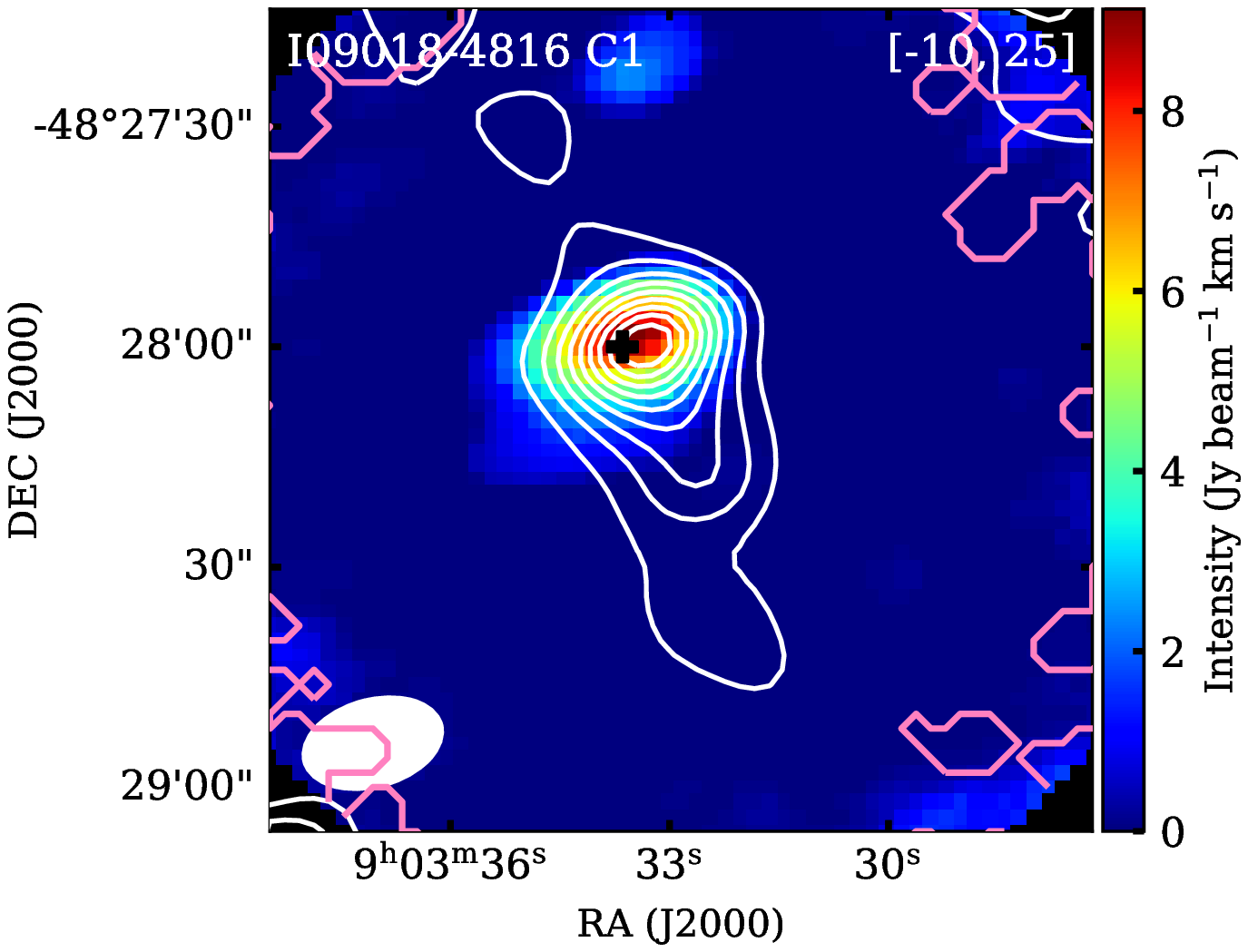}}
\end{minipage}
\begin{minipage}[t]{0.25\linewidth}
   \vspace{5pt}
   \centerline{\includegraphics[width=1.085\linewidth]{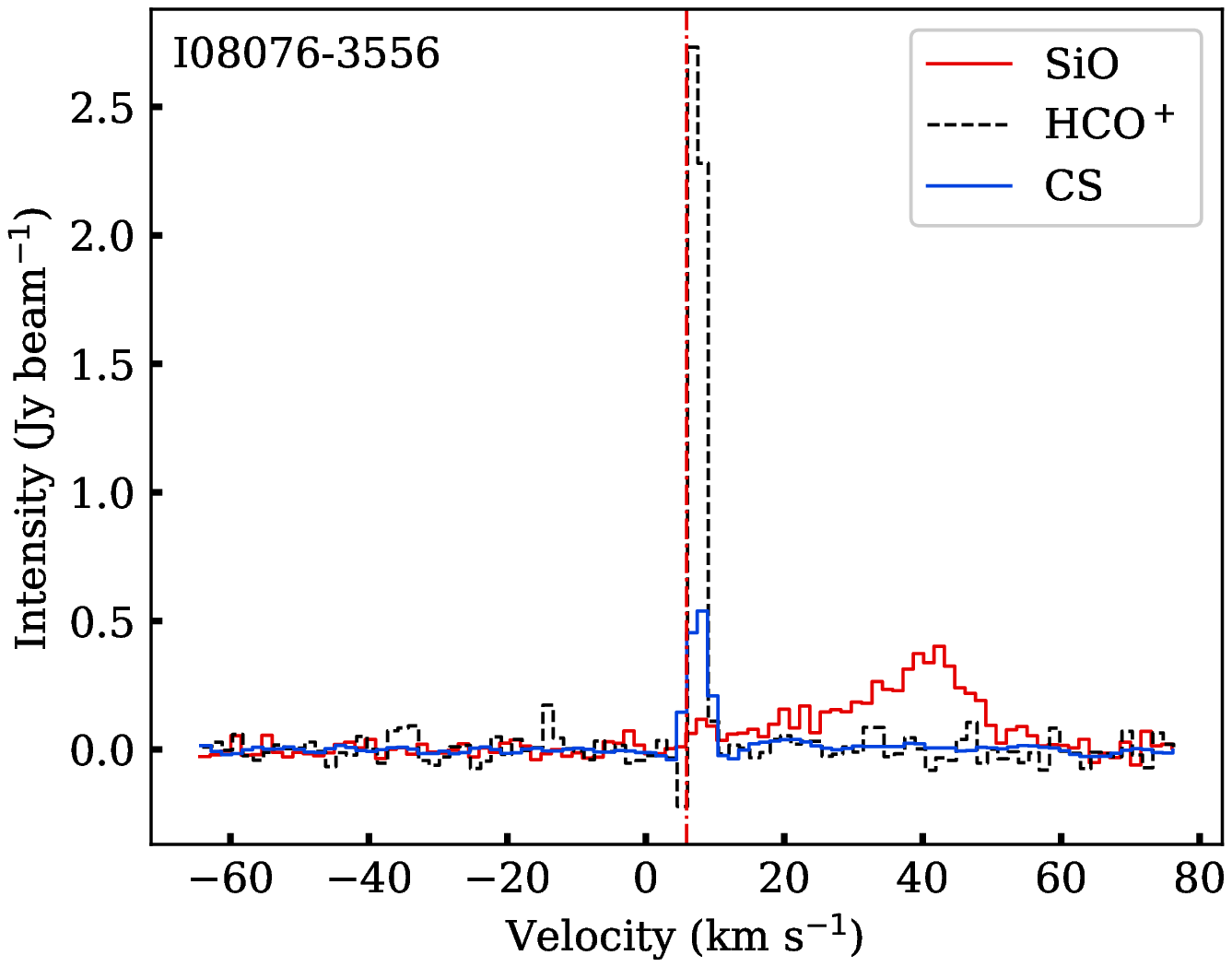}}
   \vspace{5pt}
   \centerline{\includegraphics[width=1.085\linewidth]{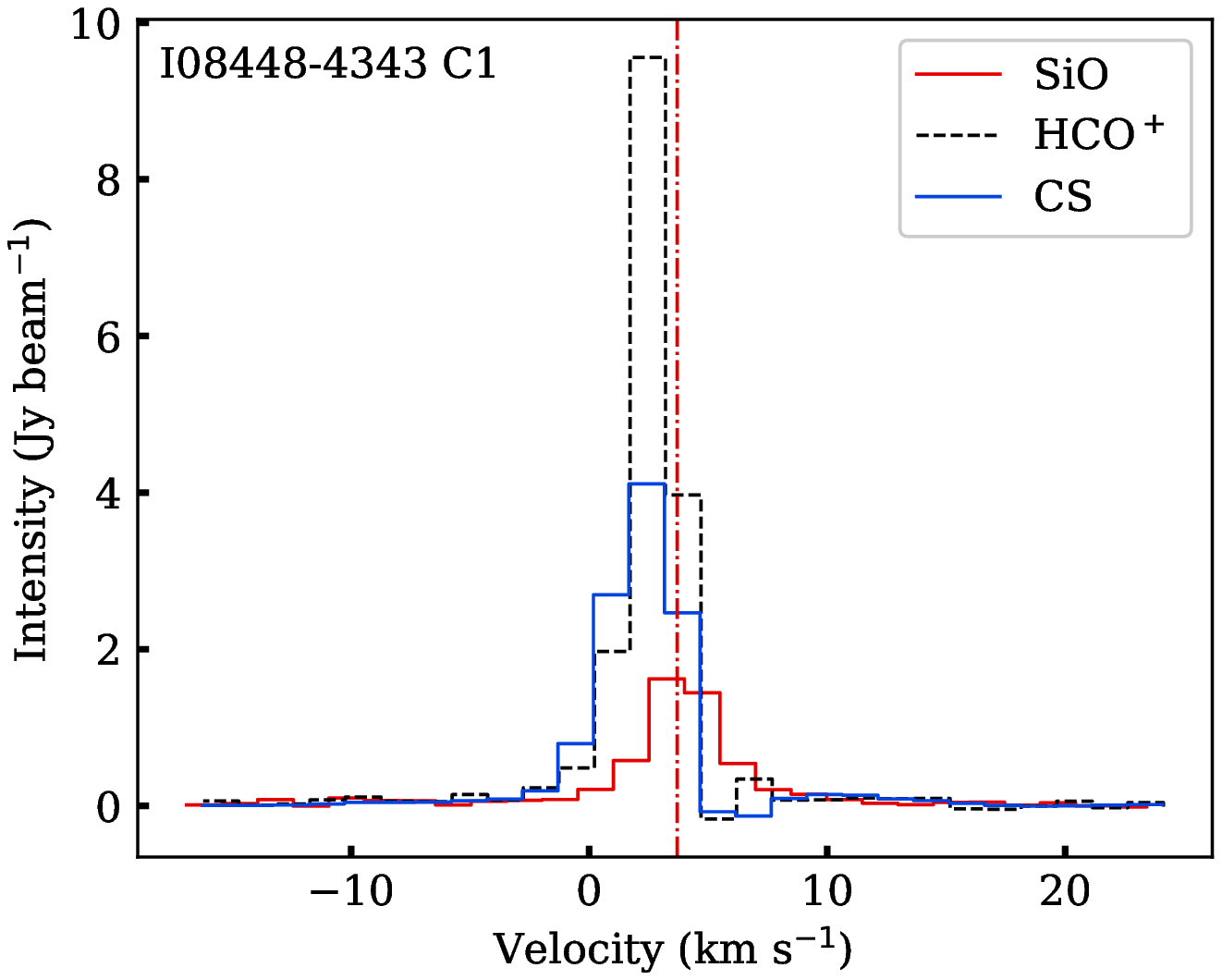}}
   \vspace{5pt}
   \centerline{\includegraphics[width=1.085\linewidth]{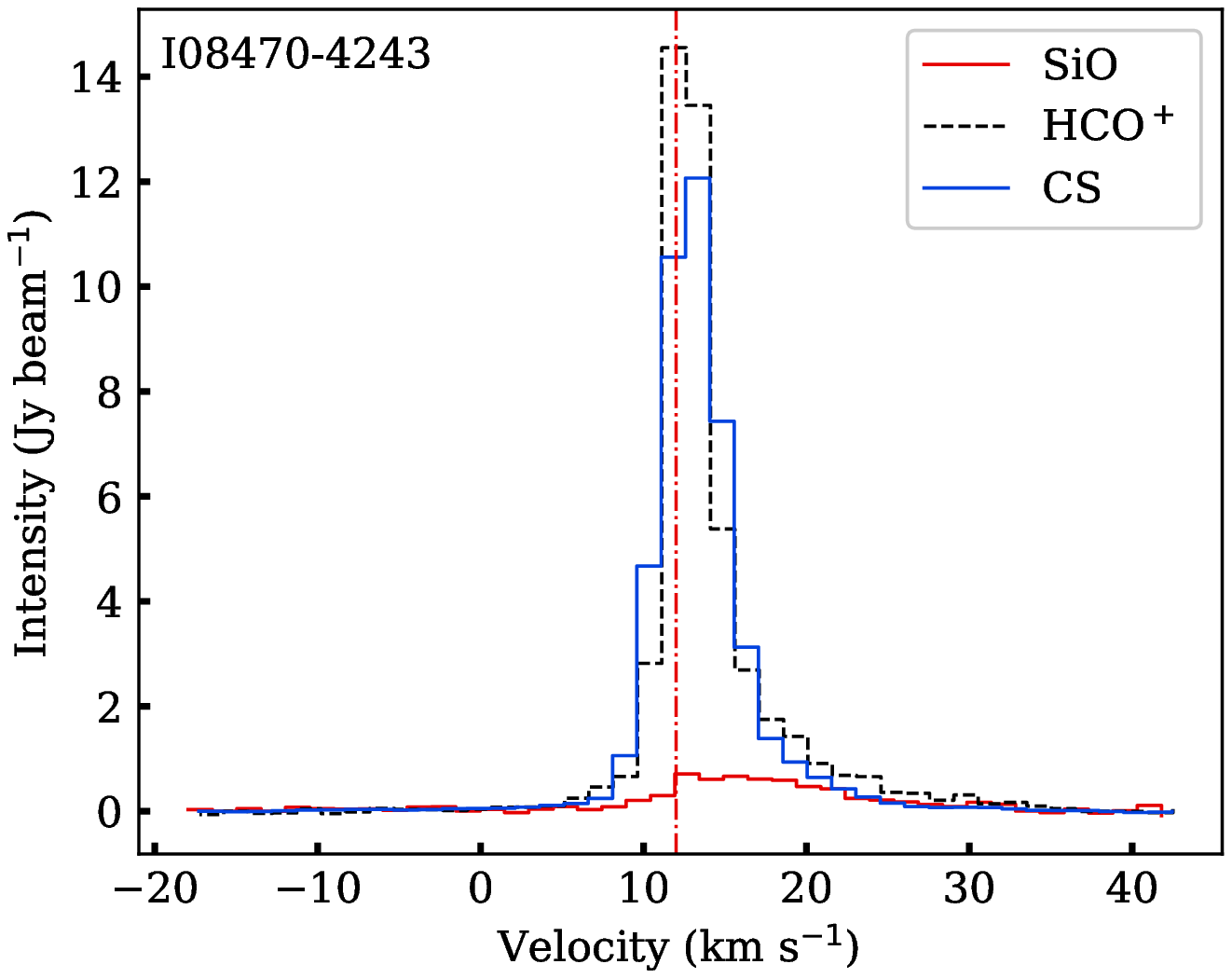}}
   \vspace{5pt}
   \centerline{\includegraphics[width=1.085\linewidth]{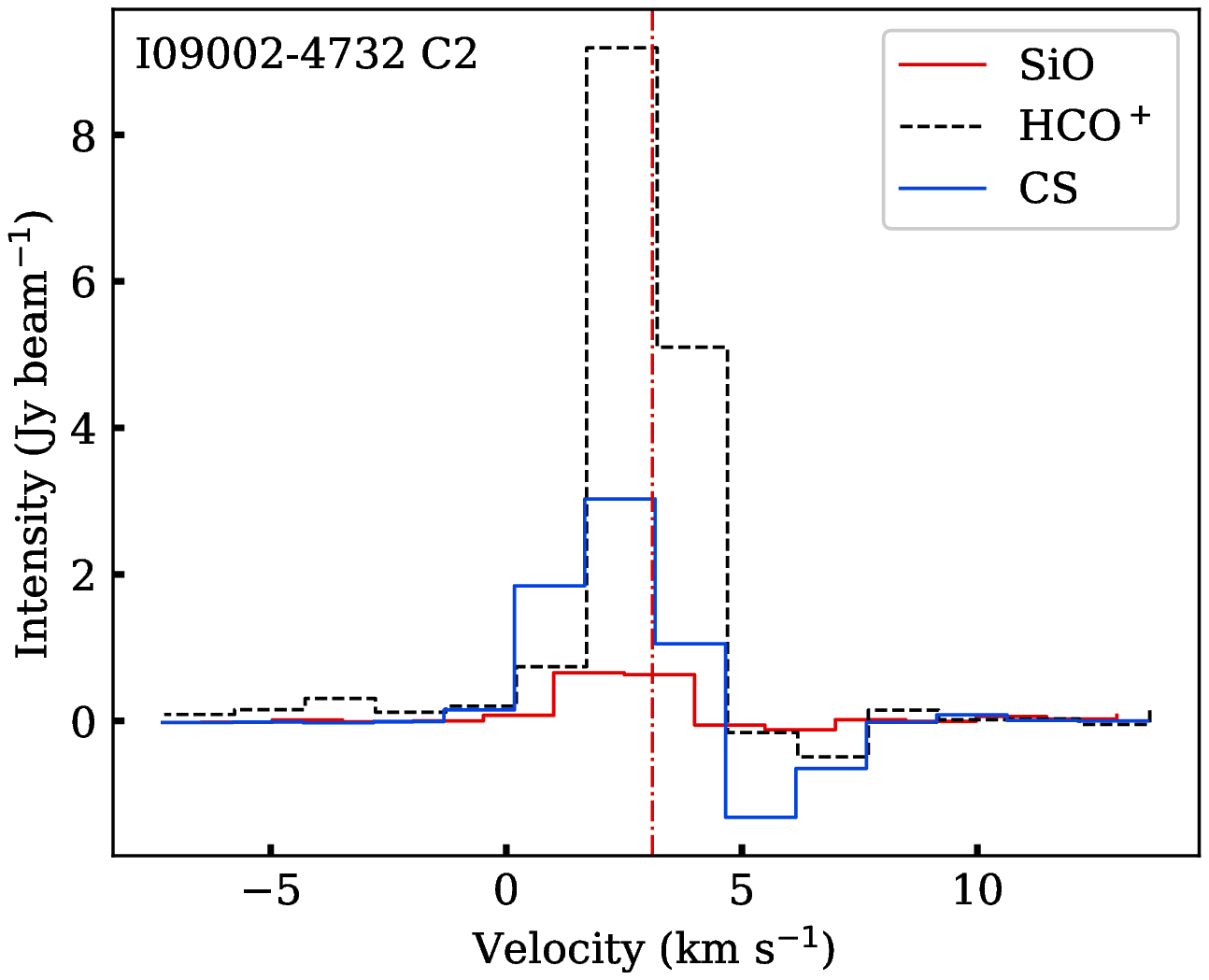}}
   \vspace{5pt}
   \centerline{\includegraphics[width=1.085\linewidth]{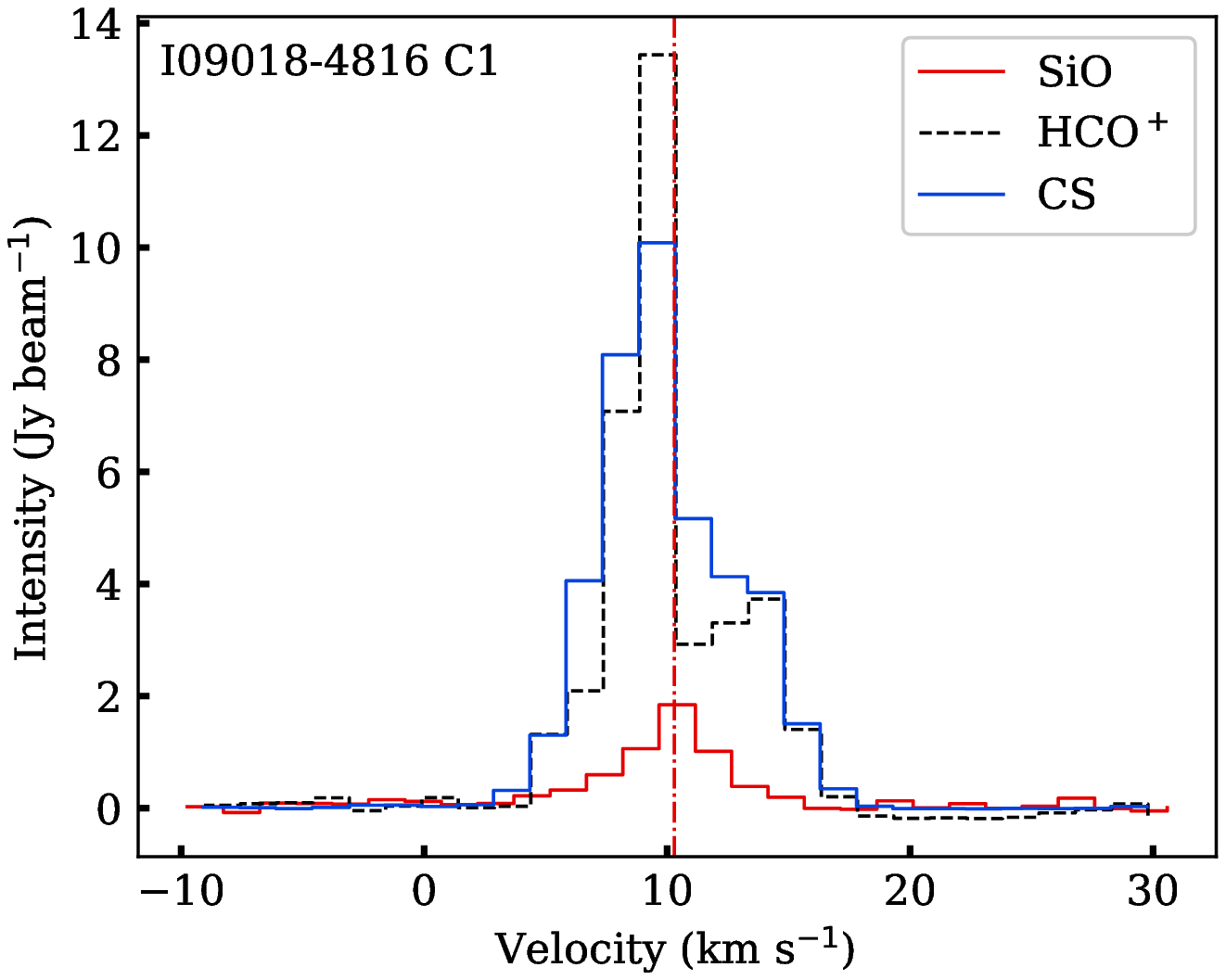}}
\end{minipage}
\begin{minipage}[t]{0.24\linewidth}
   \vspace{5pt}
   \centerline{\includegraphics[width=1.13\linewidth]{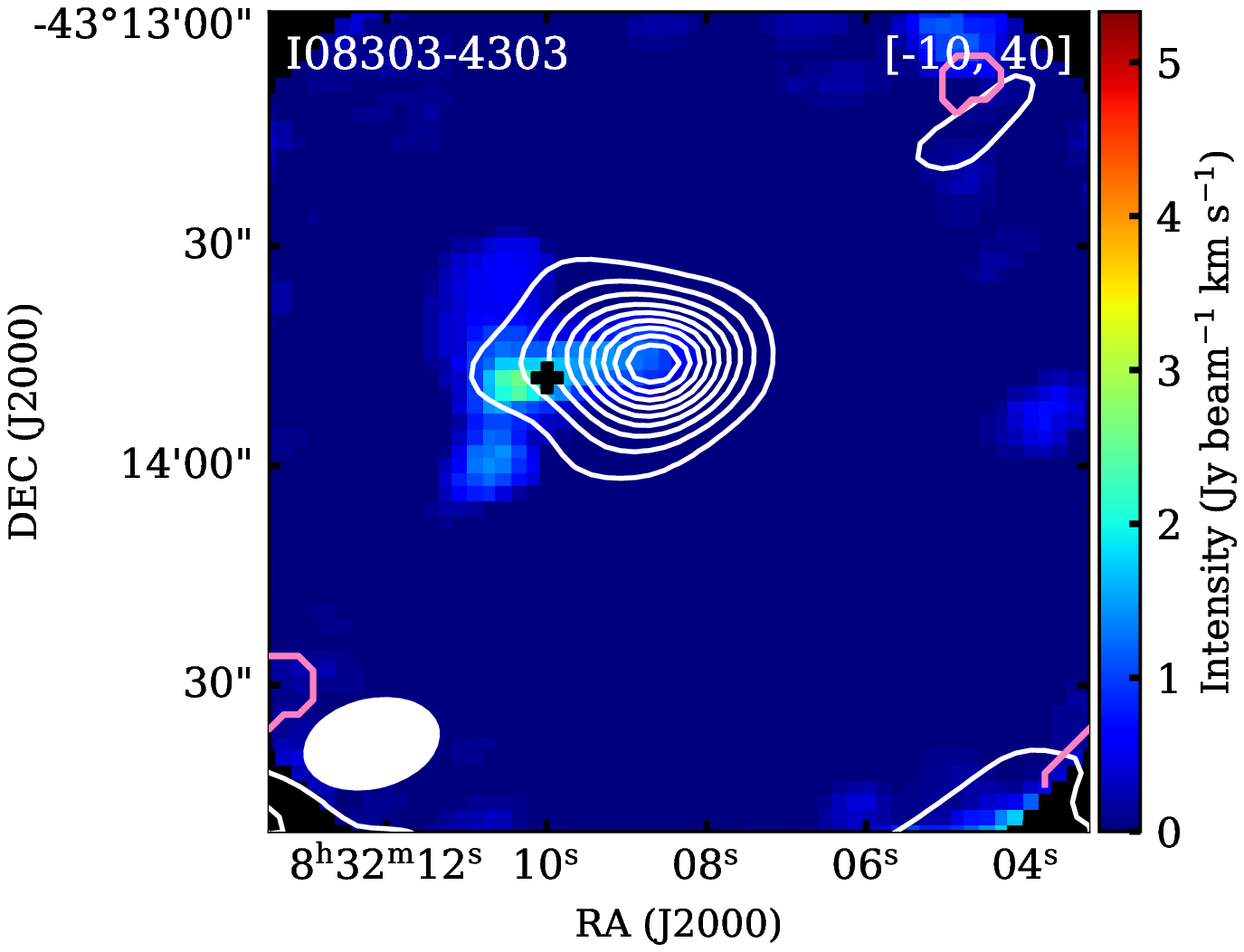}}
   \vspace{5pt}
   \centerline{\includegraphics[width=1.13\linewidth]{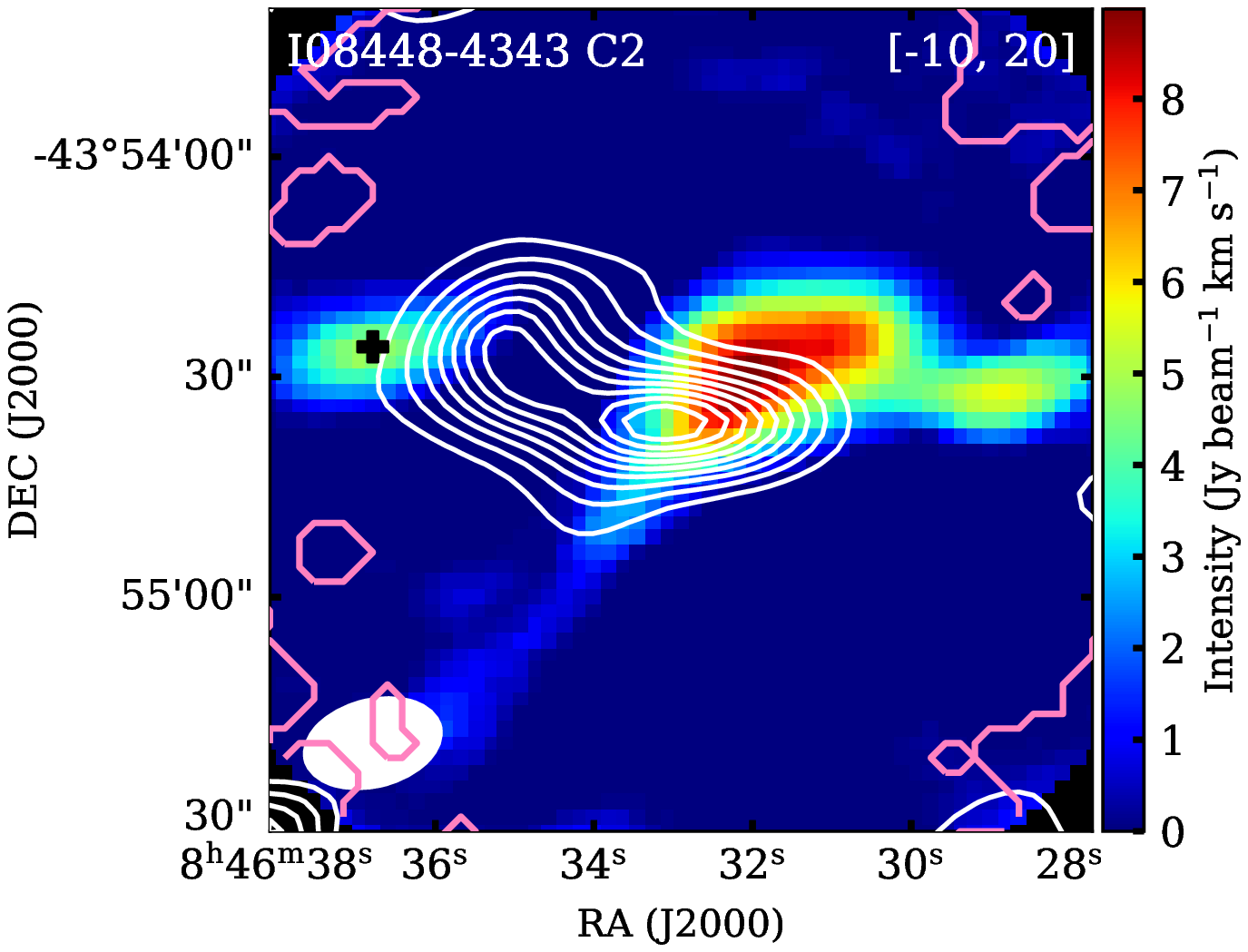}}
   \vspace{5pt}
   \centerline{\includegraphics[width=1.13\linewidth]{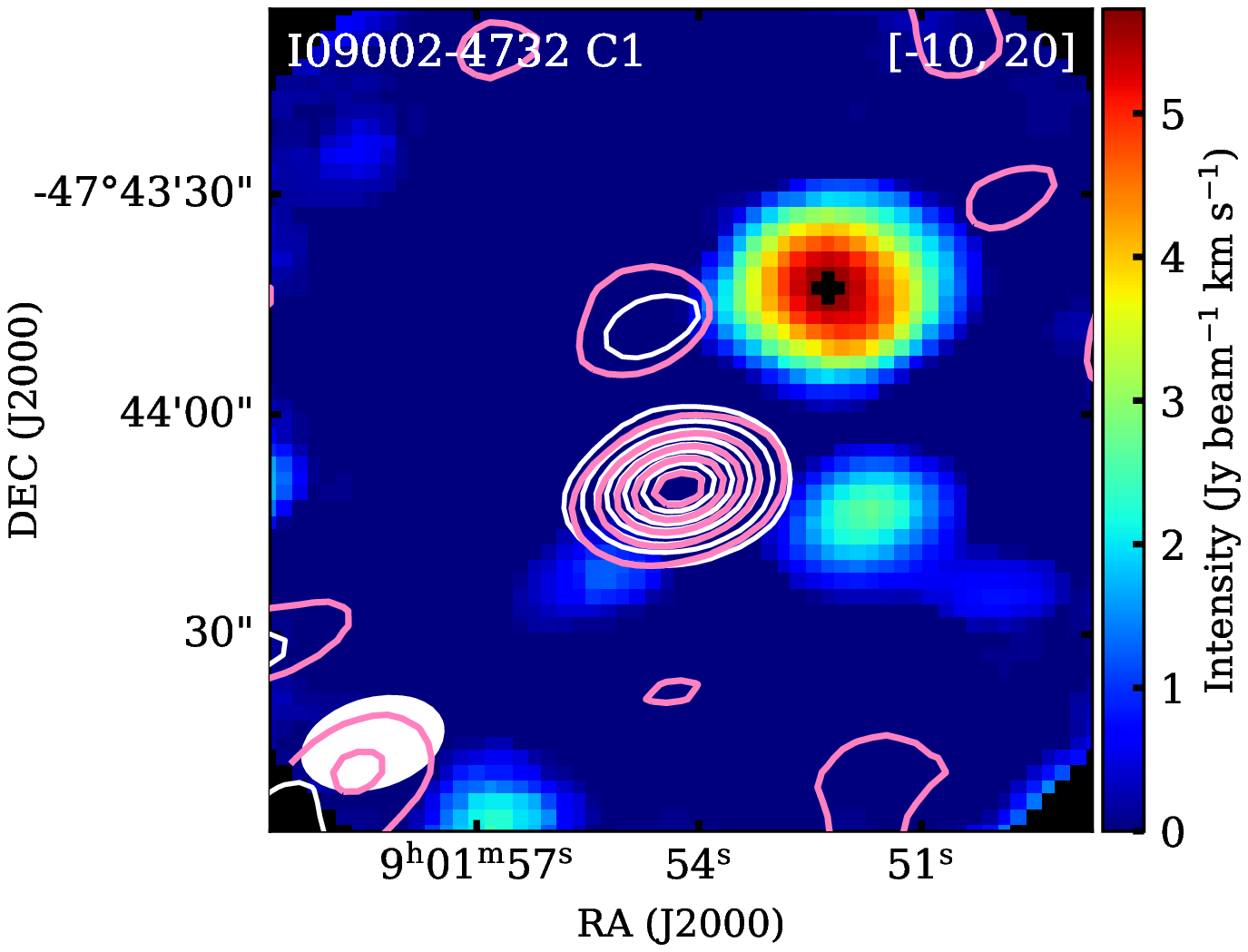}}
   \vspace{5pt}
   \centerline{\includegraphics[width=1.13\linewidth]{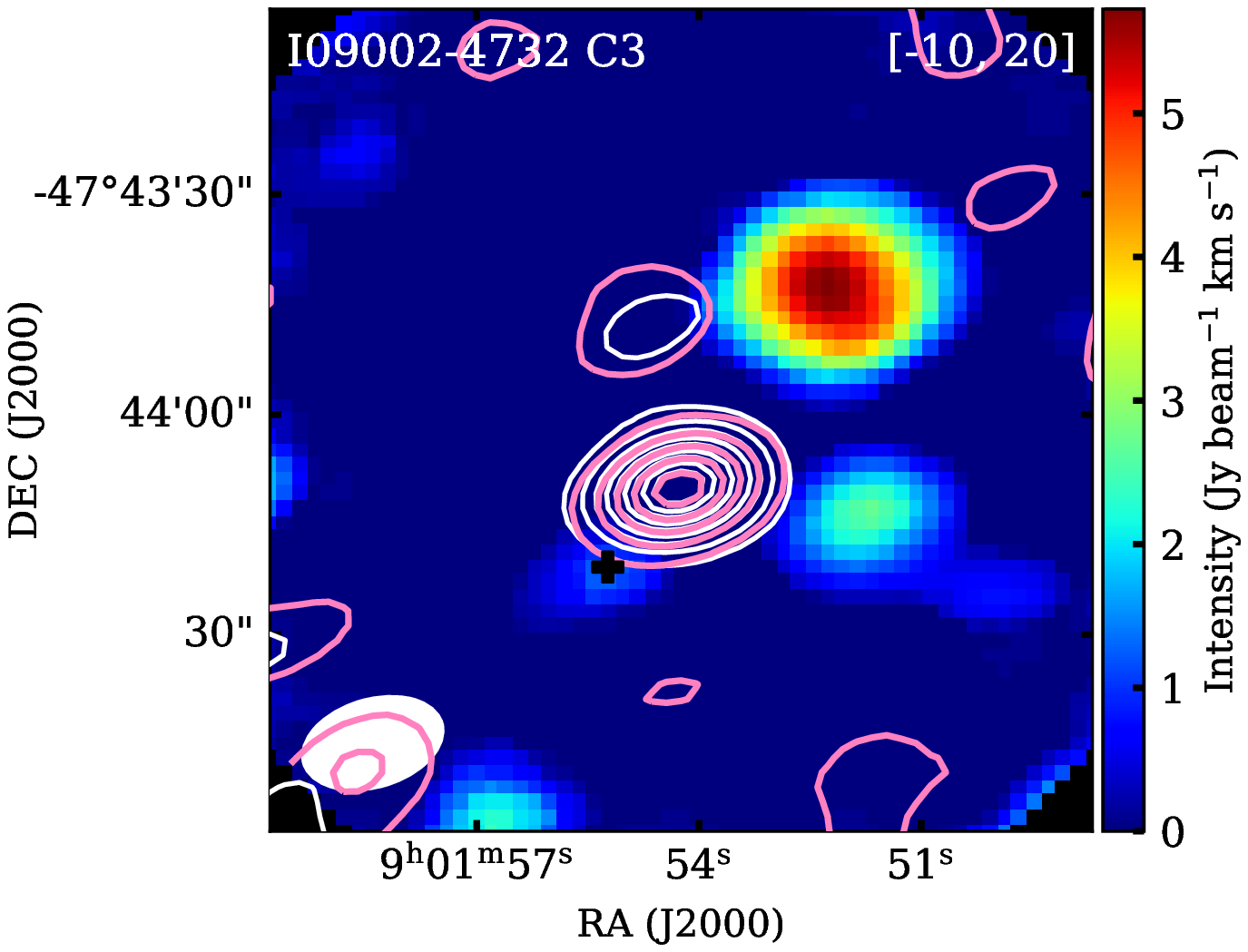}}
   \vspace{5pt}
   \centerline{\includegraphics[width=1.13\linewidth]{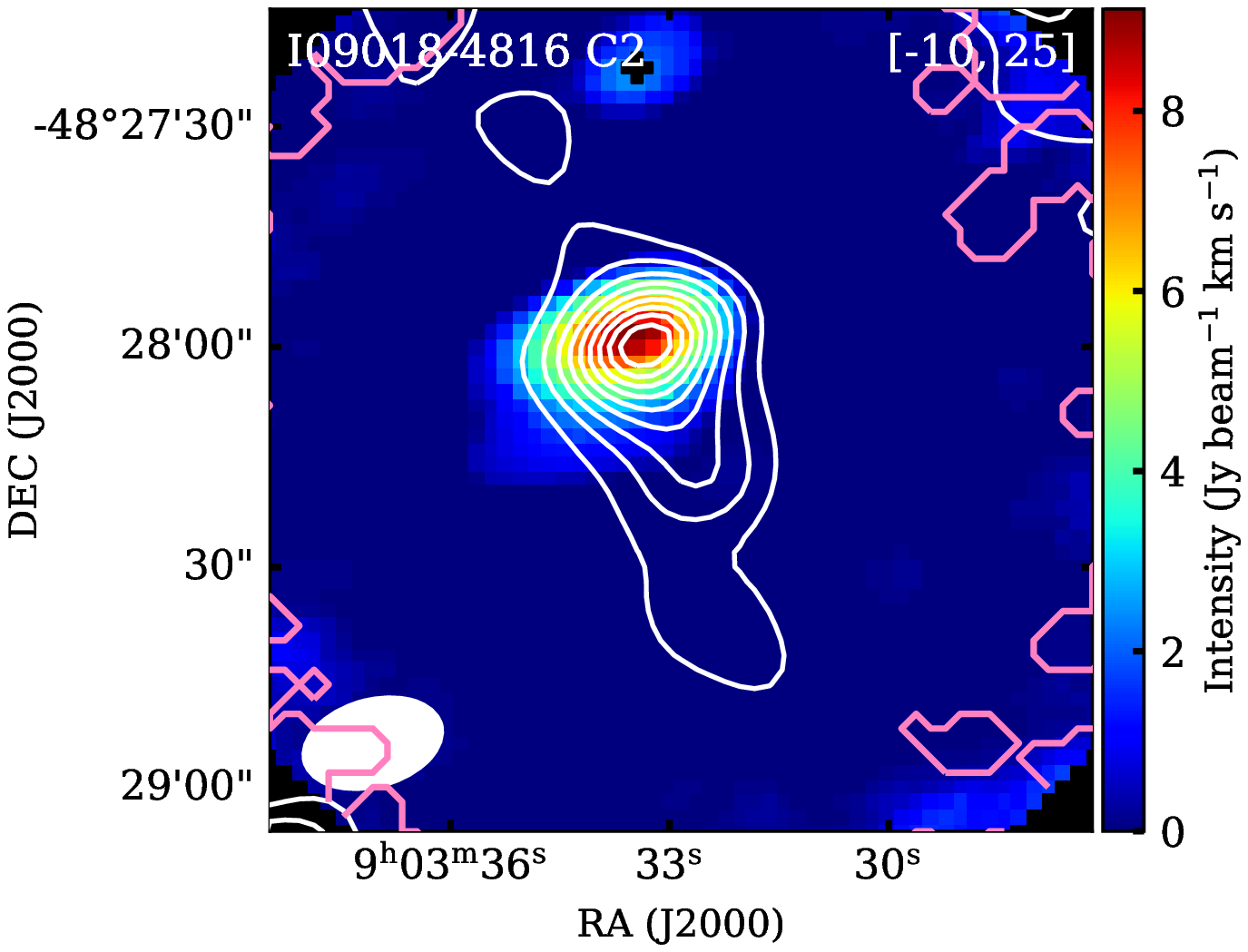}}
\end{minipage}
\begin{minipage}[t]{0.25\linewidth}
   \vspace{5pt}
   \centerline{\includegraphics[width=1.085\linewidth]{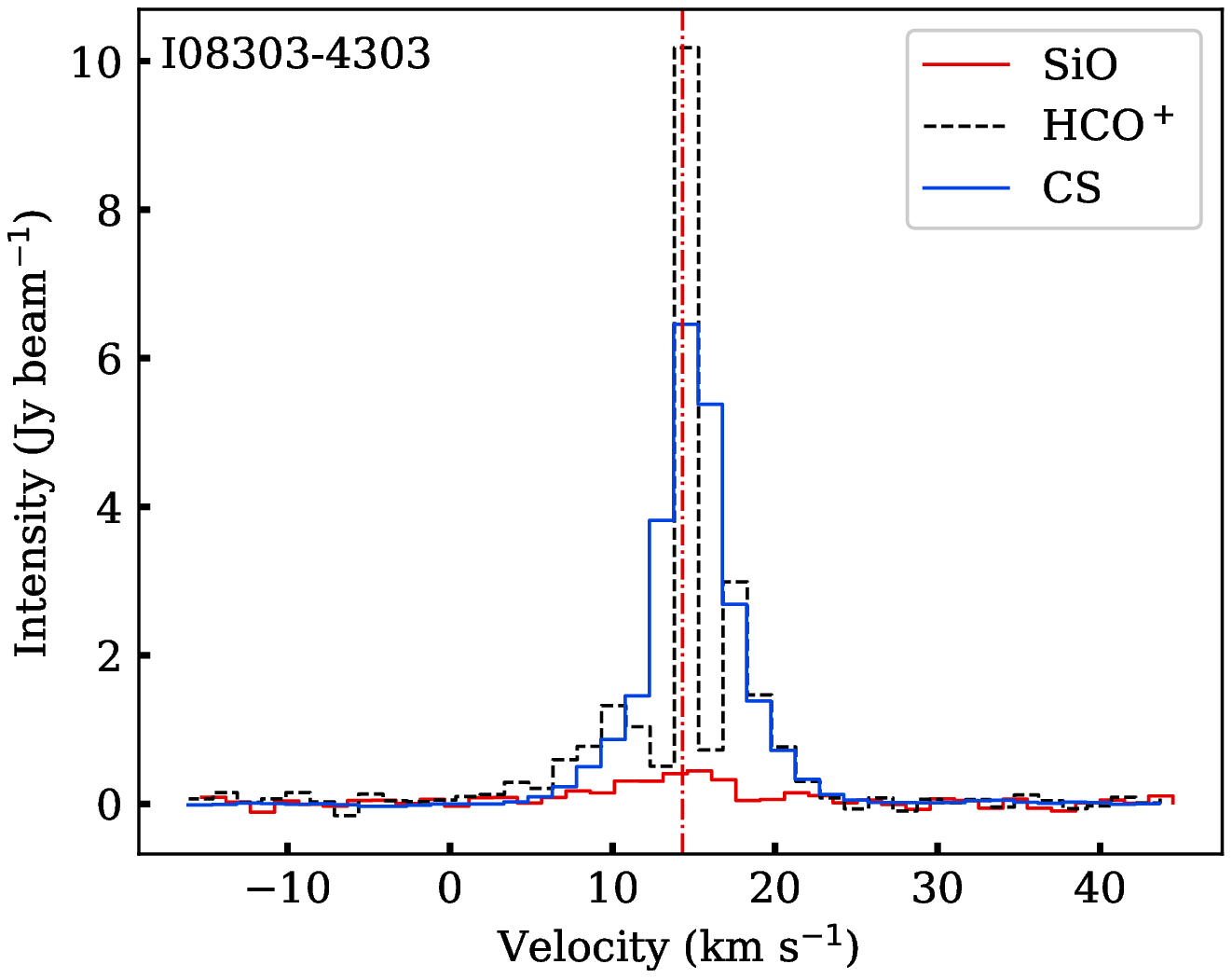}}
   \vspace{5pt}
   \centerline{\includegraphics[width=1.085\linewidth]{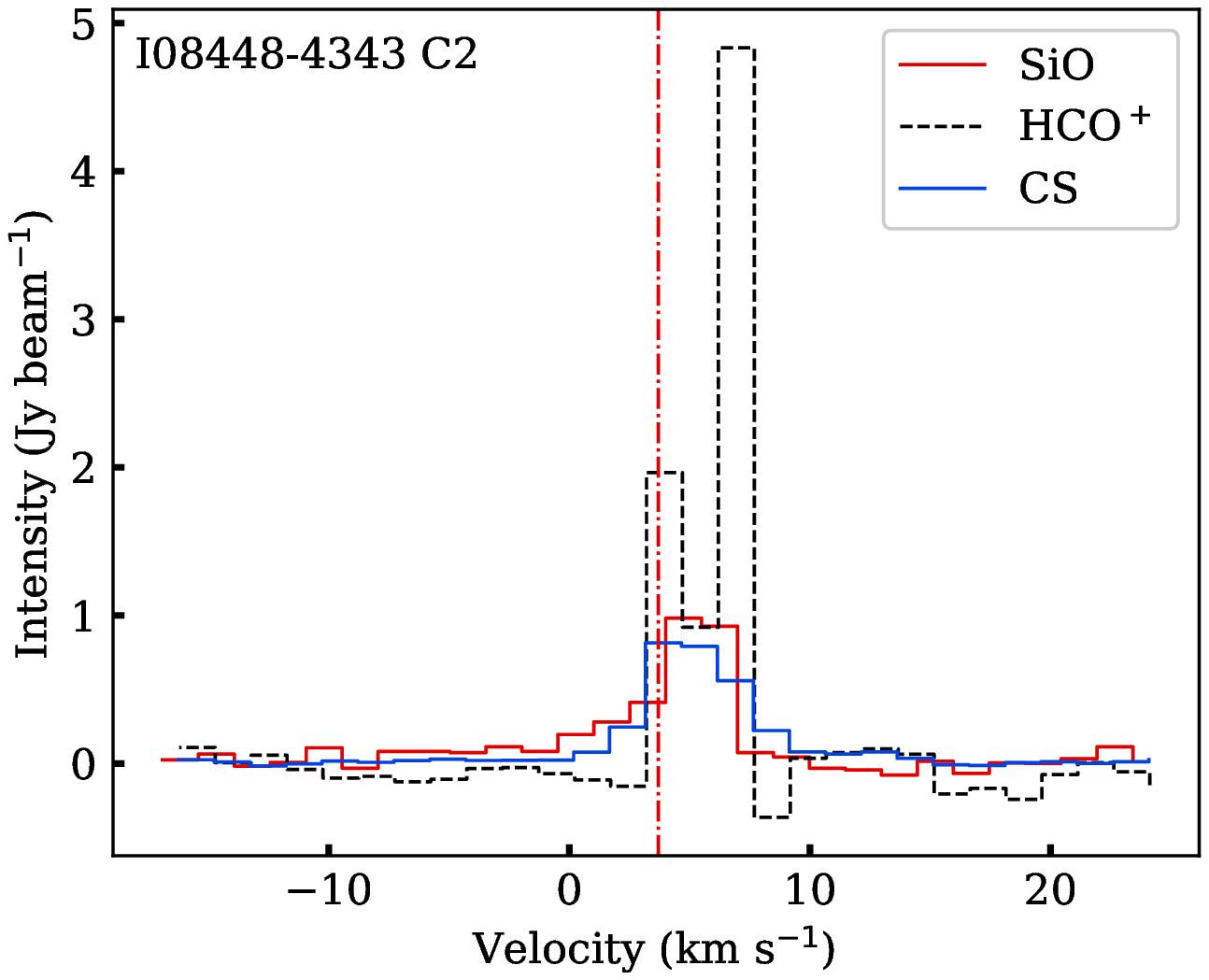}}
   \vspace{5pt}
   \centerline{\includegraphics[width=1.085\linewidth]{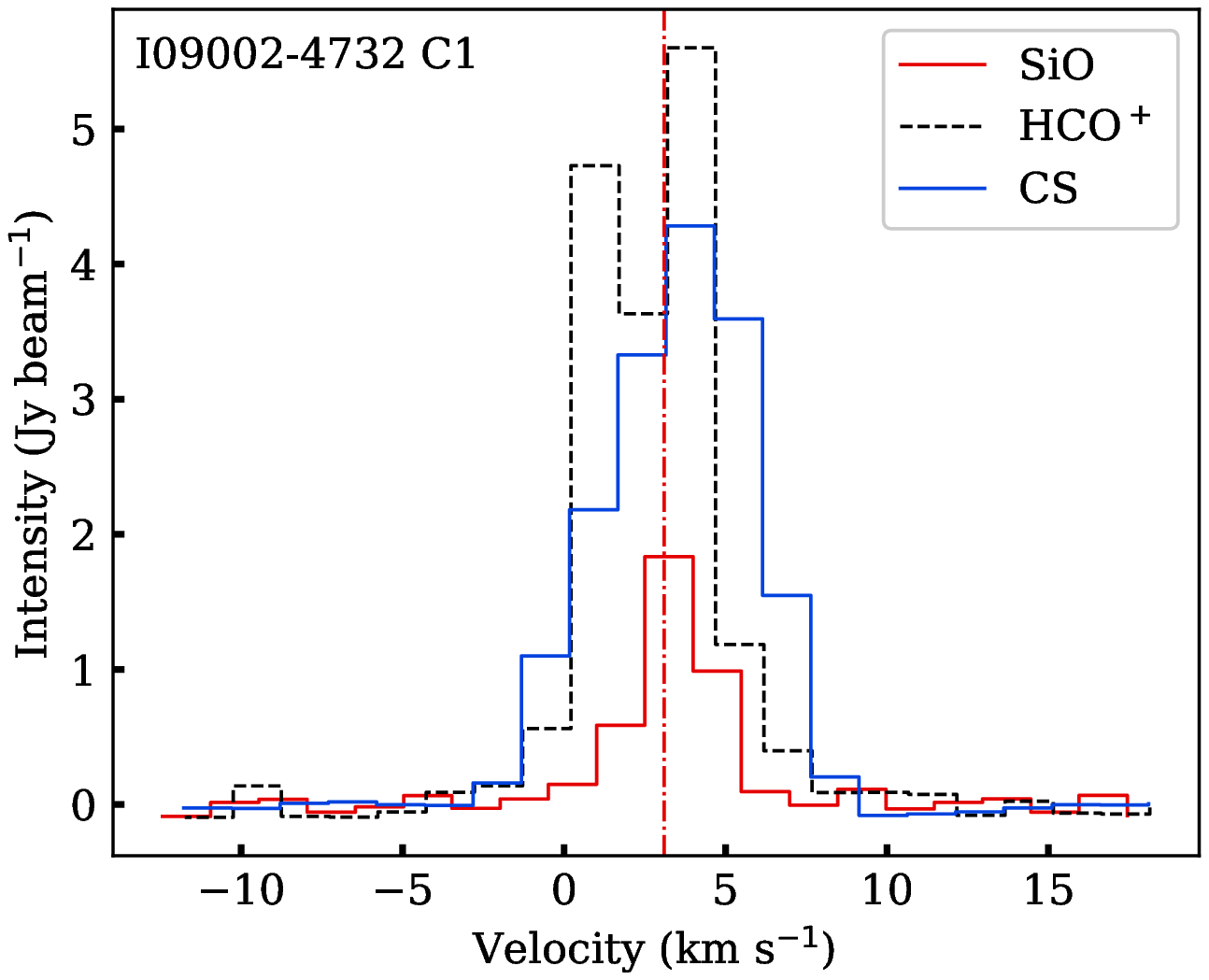}}
   \vspace{5pt}
   \centerline{\includegraphics[width=1.085\linewidth]{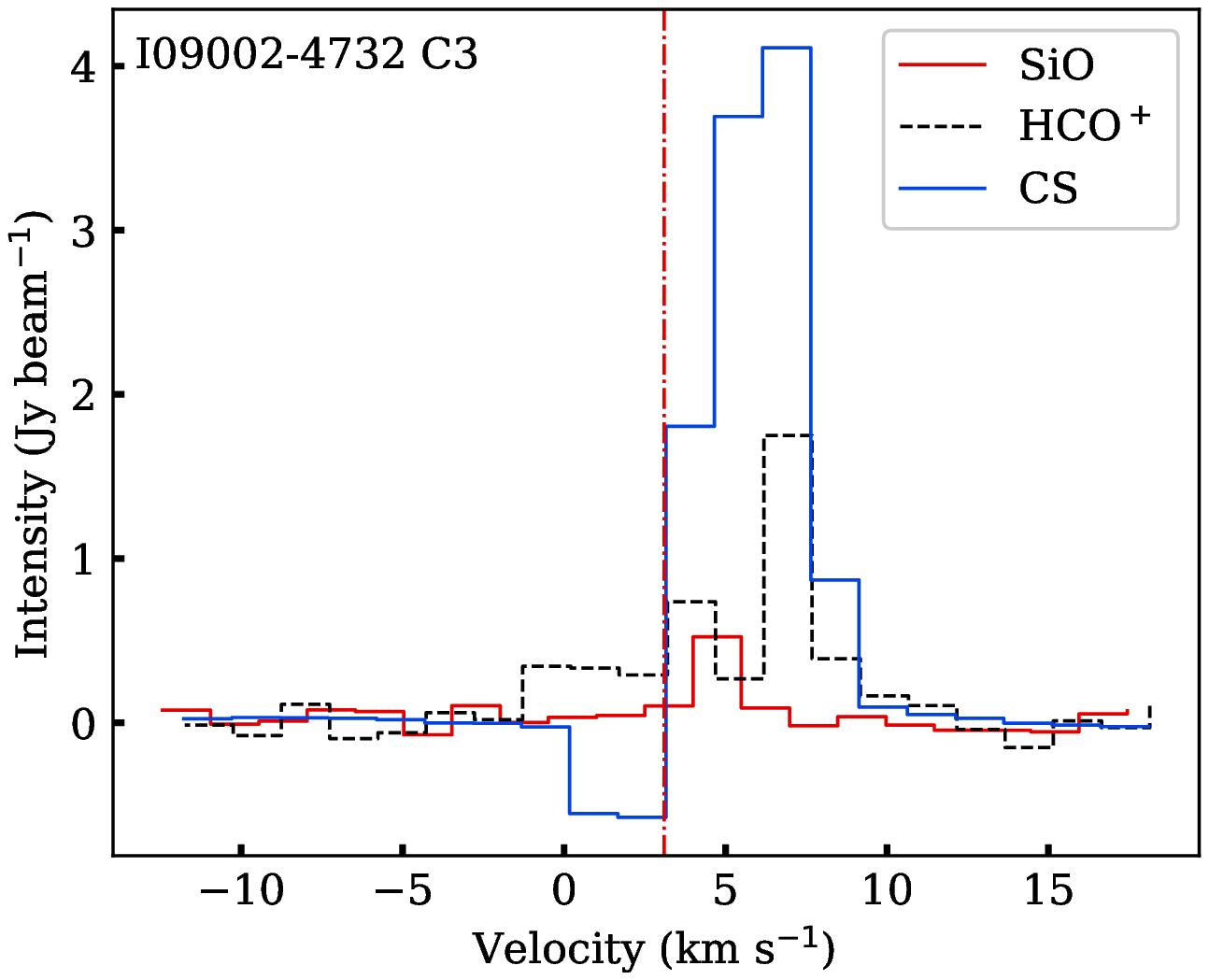}}
   \vspace{5pt}
   \centerline{\includegraphics[width=1.085\linewidth]{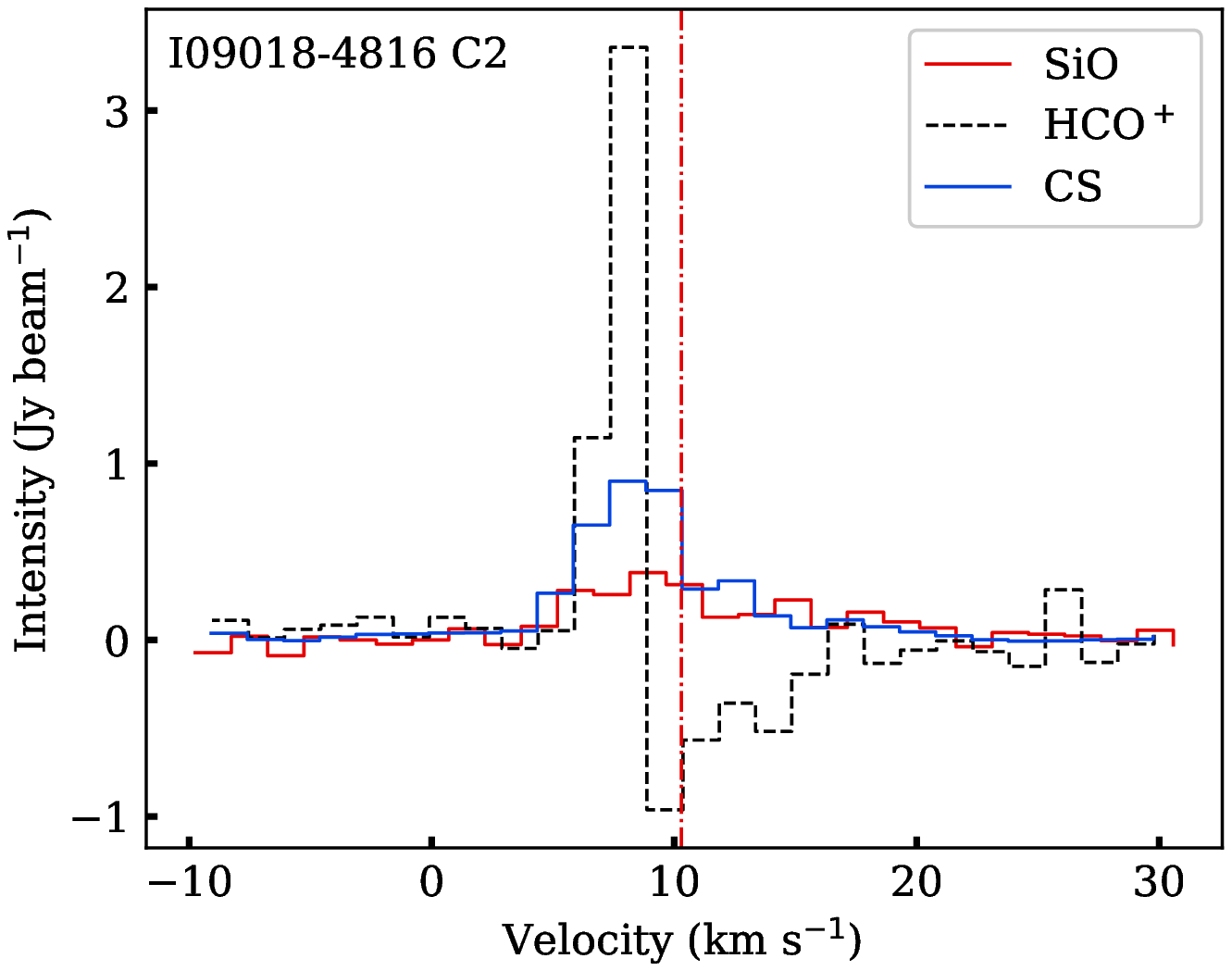}}
   \vspace{7pt}
\end{minipage}
   \caption{Examples of sources. Left: The background is SiO (2-1) integrated intensity maps. The white contours are 3mm continuum emission, and contours are from 10$\%$ to 100$\%$ in the step of 10$\%$ of peak values. The pink contours represent H40$\alpha$ emission, and contours are from 10$\%$ to 100$\%$ in the step of 20$\%$ of peak values. The black cross is the position extracted spectra. The source name is shown on the upper left. The integration velocity ranges are shown in the upper right corner. Right: The clump averaged SiO, HCO$^+$ and CS spectra are shown in red, black dashed and blue lines, respectively. The vertical dash-dot line indicates the systemic velocity of the source. The full images are available as supplementary material.}
\label{figA1}
\end{figure*}

Figure \ref{figA1} present the moment maps and spectra for some exemplar sources. The images for all sources are available as on-line supplementary material. Tables \ref{tab:TableA1}, \ref{tab:TableA2},and \ref{tabA3} show the derived parameters for SiO clumps.

\begin{table*}
    \centering
	\caption{ SiO (2-1) Line Parameters and Clump Properties in the Region with Clear Detections}
	\renewcommand\tabcolsep{2.8pt}
	\label{tab:TableA1}
    \begin{tabular}{lcclllccccclcc}
	    \hline
	    \hline
\multicolumn{1}{c}{IRAS} & \multicolumn{1}{c}{RA} & \multicolumn{1}{c}{DEC} & \multicolumn{1}{c}{$R_\textup{a}$} & \multicolumn{1}{c}{$\int{F_{\nu} d\nu}$} & \multicolumn{1}{c}{$v$} & \multicolumn{5}{c}{$\sigma_\textup{v}$} & \multicolumn{1}{c}{log $L_\textup{SiO}$} & \multicolumn{1}{c}{$\sigma$} \\

& \multicolumn{1}{c}{(J2000)} & \multicolumn{1}{c}{(J2000)} & \multicolumn{1}{c}{(pc)} & \multicolumn{1}{c}{(Jy km s$^{-1}$)} & \multicolumn{1}{c}{(km s$^{-1}$)} & \multicolumn{5}{c}{(km s$^{-1}$)} & \multicolumn{1}{c}{($L_{\odot}$)} & \multicolumn{1}{c}{(Jy beam$^{-1}$)} \\ \cline{7-11} 
&  &  &  &  &  
        & \multicolumn{1}{c|}{$D_\textup{min}$} 
        & \multicolumn{1}{c|}{$D_\textup{max}$} 
        & \multicolumn{1}{c|}{$D_\textup{mean}$} 
        & \multicolumn{1}{c|}{$D_\textup{median}$} 
        & \multicolumn{1}{c|}{$D_\textup{std}$} 
        & &
  \\ \cline{7-11}
\multicolumn{1}{l}{I08076-3556} & 08:09:33.21 & -36:05:14.2 & 0.03$^\textup{a}$ & 3.47±0.26 & 39.4±2.47 & 0.96 & 8.6 & 3.71 & 3.35 & 3.26 & -12.76±0.031 & 0.099 \\
\multicolumn{1}{l}{I08303-4303} & 08:32:10.05 & -43:13:47.71  & 0.05±0.066 & 2.3±0.61 & 14.61±0.23 & 0.17 & 1.69 & 0.67 & 0.59 & 0.17 & -11.42±0.102 & 0.123 \\
I08448-4343   C1 & 08:46:31.64 & -43:54:29.1 & 0.07±0.016 & 24.7±4.5 & 3.6±1.94 & 0.35 & 2.6 & 1.29 & 1.3 & 0.29 & -11.42±0.073 & 0.124 \\
\multicolumn{1}{r}{C2} & 08:46:36.78 & -43:54:26.2 & 0.05$^\textup{a}$ & 4.96±0.36 & 4.11±0.92 & 0.32 & 2.07 & 1.08 & 1.01 & 0.22 & -12.12±0.03 & 0.124 \\
\multicolumn{1}{l}{I08470-4243}  & 08:48:47.86 & -42:54:28.9 & 0.15$^\textup{a}$ & 12.4±1.2 & 14.66±4.17 & 0.32 & 5.16 & 2.21 & 1.85 & 1.65 & -10.76±0.04 & 0.121 \\
I09002-4732   C1 & 09:01:52.18 & -47:43:42.7  & 0.06±0.019 & 10.46±0.84 & 2.78±0.08 & 0.29 & 1.27 & 0.8 & 0.81 & 0.05 & -11.32±0.034 & 0.128 \\
\multicolumn{1}{r}{C2} & 09:01:51.71 & -47:44:14.1 & 0.09$^\textup{a}$ & 2.41±0.3 & 1.86±0.02 & 0.17 & 0.56 & 0.38 & 0.38 & 0.01 & -11.96±0.051 & 0.128 \\
\multicolumn{1}{r}{C3} & 09:01:55.28 & -47:44:21.7  & 0.09$^\textup{a}$ & 0.55±0.05 & 4.08±0.06 & 0.17 & 0.59 & 0.36 & 0.37 & 0.02 & -12.6±0.038 & 0.128 \\
I09018-4816 C1 & 09:03:33.61 & -48:27:59.3  & 0.09±0.047 & 11.11±0.9 & 9.64±0.31 & 0.3 & 1.9 & 1.04 & 1.03 & 0.21 & -10.63±0.034 & 0.115 \\
\multicolumn{1}{r}{C2} & 09:03:33.4 & -48:27:21.6 & 0.19$^\textup{a}$ & 1.21±0.16 & 7.92±0.95 & 0.39 & 5.24 & 1.49 & 1.2 & 0.71 & -11.59±0.054 & 0.115 \\
\multicolumn{1}{l}{I09094-4803} & 09:11:08.42 & -48:15:43.0  & 0.71$^\textup{a}$ & 0.76±0.07 & 74.19±0.04 & 0.17 & 0.8 & 0.54 & 0.62 & 0.04 & -10.66±0.038 & 0.128 \\
\multicolumn{1}{l}{I10365-5803} & 10:38:33.17 & -58:19:14.4 & 0.19$^\textup{a}$ & 3.02±0.34 & -22.24±6.94 & 0.36 & 2.53 & 1.73 & 1.91 & 0.38 & -11.26±0.046 & 0.093 \\
\multicolumn{1}{l}{I11298-6155} & 11:32:05.36 & -62:12:26.3 & 0.78$^\textup{a}$ & 1.55±0.1 & 33.13±0.32 & 0.72 & 1.87 & 1.36 & 1.31 & 0.14 & -10.31±0.027 & 0.124 \\
\multicolumn{1}{l}{I11332-6258} & 11:35:32.00 & -63:14:45.1 & 0.15$^\textup{a}$ & 0.95±0.04 & -15.68±0.43 & 0.17 & 1.16 & 0.79 & 0.74 & 0.04 & -11.97±0.018 & 0.125 \\
\multicolumn{1}{l}{I11590-6452}    & 12:01:39.09 & -65:09:03.7 & 0.03$^\textup{a}$ & 9.82±0.96 & 1.68±0.36 & 0.17 & 2.37 & 0.87 & 0.93 & 0.12 & -12.31±0.041 & 0.123 \\
\multicolumn{1}{l}{I12320-6122} & 12:34:52.43 & -61:39:52.0  & 0.14 & 4.48±0.71 & -43.35±1.25 & 0.29 & 3.32 & 1.42 & 1.45 & 0.58 & -10.78±0.064 & 0.130 \\
\multicolumn{1}{l}{I12326-6245} & 12:35:34.85 & -63:02:29.4 & 0.23±0.039 & 92.9±4.5 & -39.9±2.41 & 0.53 & 6.8 & 2.5 & 2.18 & 2.35 & -9.21±0.021 & 0.123 \\
\multicolumn{1}{l}{I12572-6316} & 13:00:24.41 & -63:32:31.8 & 0.91$^\textup{a}$ & 0.91±0.01 & 29.85±0.21 & 0.17 & 1.18 & 0.82 & 0.94 & 0.09 & -10.42±0.005 & 0.131 \\
\multicolumn{1}{l}{I13079-6218} & 13:11:14.11 & -62:34:45.4 & 0.12±0.026 & 114.3±3 & -40.24±1.13 & 0.75 & 7.14 & 3.86 & 3.86 & 3.59 & -9.29±0.011 & 0.138 \\
\multicolumn{1}{l}{I13111-6228} & 13:14:26.50 & -62:44:26.9 & 0.3$^\textup{a}$ & 0.62±0.06 & -38.83±0.48 & 0.32 & 0.93 & 0.59 & 0.58 & 0.02 & -11.55±0.04 & 0.123 \\
\multicolumn{1}{l}{I13134-6242} & 13:16:43.88 & -62:58:32.5  & 0.3$^\textup{a}$ & 45±2.3 & -31.78±2.66 & 0.59 & 5.2 & 2.97 & 3.05 & 1.78 & -9.69±0.022 & 0.125 \\
\multicolumn{1}{l}{I13140-6226} & 13:17:15.68 & -62:42:27.1 & 0.3$^\textup{a}$ & 7.94±0.56 & -34.56±0.65 & 0.17 & 2.86 & 1.48 & 1.45 & 0.6 & -10.44±0.03 & 0.138 \\
\multicolumn{1}{l}{I13291-6249} & 13:32:32.05 & -63:05:08.3 & 0.61$^\textup{a}$ & 17.7±2.9 & -32.62±2.78 & 0.29 & 7.84 & 2.16 & 1.3 & 3.14 & -9.49±0.066 & 0.137 \\
\multicolumn{1}{l}{I13295-6152} & 13:32:52.90 & -62:07:49.2  & 0.31$^\textup{a}$ & 0.55±0.06 & -45.08±0.03 & 0.17 & 0.45 & 0.32 & 0.34 & 0.0 & -11.58±0.045 & 0.129 \\
\multicolumn{1}{l}{I13471-6120} & 13:50:44.03 & -61:35:07.5  & 0.44$^\textup{a}$ & 3.82±0.47 & -59.11±1.01 & 0.23 & 2.22 & 1.15 & 1.06 & 0.31 & -10.45±0.05 & 0.119 \\
\multicolumn{1}{l}{I13484-6100} & 13:51:58.44 & -61:15:36.5 & 0.24±0.086 & 25.6±2.6 & -54.99±0.26 & 0.47 & 4.14 & 1.52 & 1.34 & 0.62 & -9.63±0.042 & 0.127 \\
\multicolumn{1}{l}{I14013-6105} & 14:04:54.67 & -61:20:07.6 & 0.33$^\textup{a}$ & 1.88±0.19 & -55.31±0.11 & 0.42 & 1.0 & 0.81 & 0.84 & 0.03 & -11±0.042 & 0.126 \\
\multicolumn{1}{l}{I14164-6028} & 14:20:08.47 & -60:42:01.6 & 0.26 $^\textup{a}$& 5.58±0.27 & -46.94±0.05 & 0.35 & 2.01 & 1.15 & 1.12 & 0.23 & -10.75±0.021 & 0.120 \\
\multicolumn{1}{l}{I14212-6131}  & 14:25:04.65 & -61:44:55.8 & 0.24±0.045 & 33.7±3.1 & -49.99±1.08 & 0.49 & 2.96 & 1.72 & 1.66 & 0.44 & -9.9±0.038 & 0.127 \\
\multicolumn{1}{l}{I14453-5912} & 14:49:06.37 & -59:24:36.3 & 0.13 & 9.8±1.4 & -40.79±0.26 & 0.23 & 1.58 & 0.94 & 1.01 & 0.12 & -10.61±0.058 & 0.148 \\
I14498-5856 C1 & 14:53:43.74 & -59:08:54.0 & 0.23$^\textup{a}$ & 2.88±0.34 & -50.6±0.12 & 0.21 & 1.37 & 0.79 & 0.64 & 0.15 & -11.04±0.048 & 0.141 \\
\multicolumn{1}{r}{C2}  & 14:53:44.58 & -59:09:29.6 & 0.23$^\textup{a}$ & 0.75±0.04 & -51.97±0.66 & 0.22 & 1.79 & 1.11 & 1.06 & 0.26 & -11.63±0.023 & 0.141 \\
\multicolumn{1}{l}{I15290-5546}  & 15:32:52.77 & -55:56:04.3 & 0.47$^\textup{a}$ & 15.36±0.97 & -85.76±2.44 & 0.23 & 6.24 & 3.73 & 3.73 & 3.21 & -9.66±0.027 & 0.148 \\
\multicolumn{1}{l}{I15394-5358}  & 15:43:16.66 & -54:07:14.7 & 0.2±0.023 & 80.5±6.3 & -41.15±8.28 & 0.41 & 3.51 & 1.96 & 2.01 & 0.71 & -10.08±0.033 & 0.158 \\
\multicolumn{1}{l}{I15408-5356}  & 15:44:43.32 & -54:05:20.8 & 0.2±0.037 & 62.5±7.9 & -36.54±1.28 & 0.55 & 4.54 & 2.27 & 2.15 & 0.82 & -10.19±0.052 & 0.149 \\
I15411-5352 C1  & 15:44:59.27 & -54:02:25.5 & 0.13$^\textup{a}$ & 1.22±0.1 & -43.09±0.05 & 0.21 & 1.77 & 0.79 & 0.75 & 0.14 & -11.9±0.034 & 0.148 \\
\multicolumn{1}{r}{C2} & 15:45:00.13 & -54:01:39.4 & 0.13$^\textup{a}$ & 0.91±0.12 & -43.25±2.08 & 0.17 & 1.76 & 0.48 & 0.45 & 0.07 & -12.02±0.054 & 0.148 \\
I15439-5449 C1 & 15:47:46.18 & -54:58:16.6  & 0.24$^\textup{a}$ & 0.46±0.05 & -55.71±0.45 & 0.32 & 0.9 & 0.54 & 0.48 & 0.02 & -11.81±0.045 & 0.148 \\
\multicolumn{1}{r}{C2} & 15:47:48.78 & -54:58:10.2 & 0.24$^\textup{a}$ & 0.16±0.08 & -54.44±0.14 & 0.3 & 0.92 & 0.55 & 0.55 & 0.03 & -12.26±0.176 & 0.148 \\
I15520-5234 C1 & 15:55:50.17 & -52:43:13.8 & 0.24±0.027 & 202±16 & -41.66±2.55 & 0.58 & 5.73 & 3.26 & 3.55 & 1.52 & -9.35±0.033 & 0.156 \\
\multicolumn{1}{r}{C2} & 15:55:48.04 & -52:43:09.4  & 0.2±0.042 & 118±11 & -45.68±0.66 & 0.91 & 2.85 & 2.1 & 2.17 & 0.27 & -9.58±0.039 & 0.156 \\
\multicolumn{1}{l}{I15522-5411}& 15:56:07.13 & -54:20:02.6 & 0.2$^\textup{a}$ & 0.95±0.1 & -47.44±0.07 & 0.3 & 1.17 & 0.66 & 0.63 & 0.06 & -11.65±0.043 & 0.157 \\
\multicolumn{1}{l}{I15557-5215} & 15:59:40.96 & -52:23:27.6 & 0.3$^\textup{a}$ & 36.8±2.6 & -68.02±13.86 & 0.3 & 6.58 & 2.74 & 2.66 & 2.23 & -9.73±0.03 & 0.149 \\
\multicolumn{1}{l}{I15567-5236} & 16:00:33.56 & -52:44:39.1 & 0.44$^\textup{a}$ & 5.98±0.74 & -107.71±1.59 & 0.17 & 5.39 & 1.75 & 1.06 & 2.44 & -10.17±0.051 & 0.155 \\
\multicolumn{1}{l}{I15584-5247} & 16:02:20.54 &- 52:55:13.0  & 0.32$^\textup{a}$ & 2.15±0.16 & -77.13±1.35 & 0.3 & 2.22 & 1.08 & 1.23 & 0.24 & -10.88±0.031 & 0.158 \\
\multicolumn{1}{l}{I15596-5301} & 16:03:32.55 & -53:09:29.8  & 0.74$^\textup{a}$ & 13.1±1.2 & -73.1±0.88 & 0.3 & 1.66 & 1.04 & 1.06 & 0.11 & -9.38±0.038 & 0.159 \\
\multicolumn{1}{l}{I16026-5035} & 16:06:24.78 &- 50:43:06.2 & 0.33$^\textup{a}$ & 0.59±0.03 & -79.97±0.05 & 0.24 & 0.82 & 0.61 & 0.64 & 0.03 & -11.42±0.022 & 0.166 \\
\multicolumn{1}{l}{I16037-5223} & 16:07:38.08 & -52:31:00.9 & 0.72$^\textup{a}$ & 1.152±0.07 & -80.1±0.91 & 0.41 & 1.76 & 1.17 & 1.22 & 0.16 & -10.46±0.026 & 0.156 \\
\multicolumn{1}{l}{I16060-5146} & 16:09:52.40 & -51:54:56.2 & 0.49±0.067 & 109±8.9 & -90.96±3.62 & 0.61 & 4.37 & 2.37 & 2.28 & 1.18 & -9.02±0.034 & 0.160 \\
\multicolumn{1}{l}{I16065-5158} & 16:10:20.09 & -52:06:11.0  & 0.14 & 49.6±3 & -64.04±0.5 & 0.46 & 5.72 & 2.67 & 2.49 & 1.67 & -9.61±0.026 & 0.155 \\
\multicolumn{1}{l}{I16071-5142} & 16:10:59.84 & -51:50:21.2  & 0.18±0.031 & 51.4±1.7 & -86.48±0.47 & 0.3 & 5.01 & 2.56 & 2.35 & 1.66 & -9.34±0.014 & 0.153 \\
\multicolumn{1}{l}{I16076-5134} & 16:11:26.66 & -51:41:56.3  & 0.39$^\textup{a}$ & 56.1±1.6 & -86.74±0.41 & 0.69 & 6.67 & 3.66 & 3.64 & 2.98 & -9.31±0.012 & 0.158 \\
\multicolumn{1}{l}{I16119-5048} & 16:15:45.35 & -50:55:52.5 & 0.23$^\textup{a}$ & 10.17±0.58 & -51.19±0.54 & 0.55 & 4.18 & 2.32 & 2.07 & 0.89 & -10.51±0.024 & 0.158 \\
\multicolumn{1}{l}{I16164-5046} & 16:20:11.30 &- 50:53:16.7& 0.09 & 33.2±1.9 & -55.12±1.21 & 0.48 & 3.3 & 1.72 & 1.61 & 0.56 & -9.88±0.024 & 0.164 \\
I16172-5028 C1& 16:21:02.64 & -50:35:00.4  & 0.25±0.065 & 95.6±9.7 & -52.32±9.85 & 0.72 & 4.68 & 2.6 & 2.14 & 1.74 & -9.42±0.042 & 0.159 \\
\multicolumn{1}{r}{C2} & 16:21:03.14 & -50:35:49.6 & 0.26$^\textup{a}$ & 14.3±1.9 & -49.82±31.3 & 0.36 & 8.79 & 1.47 & 1.3 & 0.98 & -10.24±0.054 & 0.159 \\
\multicolumn{1}{r}{C3} & 16:20:59.42 & -50:35:53.1 & 0.26$^\textup{a}$ & 6.78±0.94 & -50.55±82.11 & 0.76 & 22.34 & 6.1 & 5.43 & 22.08 & -10.57±0.056 & 0.159 \\
\multicolumn{1}{l}{I16272-4837} & 16:30:58.27 & -48:43:45.2 & 0.32±0.056 & 66.6±8.7 & -46.32±7.23 & 0.53 & 5.95 & 2.48 & 2.37 & 1.54 & -9.75±0.053 & 0.163 \\
\multicolumn{1}{l}{I16297-4757} & 16:33:29.23 & -48:03:45.6  & 0.36$^\textup{a}$ & 3.85±0.26 & -80.13±0.06 & 0.43 & 1.84 & 1.16 & 1.15 & 0.16 & -10.51±0.028 & 0.168 \\
\multicolumn{1}{l}{I16318-4724} & 16:35:33.72 & -47:31:10.1  & 0.54$^\textup{a}$ & 12.9±0.64 & -121.48±0.41 & 0.43 & 2.98 & 1.68 & 1.65 & 0.48 & -9.62±0.021 & 0.174 \\
\multicolumn{1}{l}{I16344-4658} & 16:38:09.36 & -47:04:59.4 & 0.86$^\textup{a}$ & 3.23±0.16 & -49.88±0.6 & 0.64 & 2.25 & 1.57 & 1.65 & 0.2 & -9.83±0.021 & 0.172 \\
\multicolumn{1}{l}{I16348-4654} & 16:38:29.27 & -47:00:39.0 & 0.86$^\textup{a}$ & 33.7±3.5 & -47.53±2.12 & 0.45 & 3.6 & 2.27 & 2.37 & 0.6 & -8.81±0.043 & 0.157\\
\end{tabular}
\end{table*}

\begin{table*}
\centering
	\renewcommand\tabcolsep{2.8pt}
    \begin{tabular}{lcclllccccclc}
	    \hline
	    \hline
\multicolumn{1}{c}{IRAS} & \multicolumn{1}{c}{RA} & \multicolumn{1}{c}{DEC} & \multicolumn{1}{c}{$R_\textup{a}$} & \multicolumn{1}{c}{$\int{F_{\nu} d\nu}$} & \multicolumn{1}{c}{$v$} & \multicolumn{5}{c}{$\sigma_\textup{v}$} & \multicolumn{1}{c}{log $L_\textup{SiO}$} & \multicolumn{1}{c}{$\sigma$} \\
& \multicolumn{1}{c}{(J2000)} & \multicolumn{1}{c}{(J2000)} & \multicolumn{1}{c}{(pc)} & \multicolumn{1}{c}{(Jy km s$^{-1}$)} & \multicolumn{1}{c}{(km s$^{-1}$)} & \multicolumn{5}{c}{(km s$^{-1}$)} & \multicolumn{1}{c}{($L_{\odot}$)} & \multicolumn{1}{c}{(Jy beam$^{-1}$)} \\ \cline{7-11} 
&  &  &  &  &  
        & \multicolumn{1}{c|}{$D_\textup{min}$} 
        & \multicolumn{1}{c|}{$D_\textup{max}$} 
        & \multicolumn{1}{c|}{$D_\textup{mean}$} 
        & \multicolumn{1}{c|}{$D_\textup{median}$} 
        & \multicolumn{1}{c|}{$D_\textup{std}$} 
        & &
  \\ \cline{7-11}
\multicolumn{1}{l}{I16351-4722} & 16:38:50.40 & -47:27:58.3  & 0.18±0.069 & 79.4±8 & -40.4±4.29 & 0.23 & 6.61 & 2.42 & 2.1 & 2.46 & -9.64±0.042 & 0.177 \\
\multicolumn{1}{l}{I16385-4619} & 16:42:13.74 & -46:25:29.8 & 0.51$^\textup{a}$ & 9.84±0.59 & -119.57±0.64 & 0.53 & 2.96 & 1.76 & 1.66 & 0.49 & -9.81±0.025 & 0.176 \\
\multicolumn{1}{l}{I16424-4531} & 16:46:06.26 & -45:36:39.0 & 0.18$^\textup{a}$ & 1.78±0.14 & -31.75±30.16 & 0.17 & 6.37 & 2.6 & 1.37 & 5.17 & -11.41±0.033 & 0.173 \\
\multicolumn{1}{l}{I16445-4459} & 16:48:04.54 & -45:05:03.9 & 0.55$^\textup{a}$ & 1.3±0.08 & -121.08±0.23 & 0.24 & 1.09 & 0.8 & 0.85 & 0.05 & -10.59±0.026 & 0.164 \\
I16458-4512 C1 & 16:49:29.22 & -45:18:09.3 & 0.25$^\textup{a}$ & 11.4±1.8 & -50.77±1.97 & 0.36 & 4.93 & 1.63 & 1.53 & 0.89 & -10.34±0.064 & 0.167 \\
\multicolumn{1}{r}{C2} & 16:49:32.31 & -45:17:30.0 & 0.25$^\textup{a}$ & 0.78±0.08 & -49.19±0.15 & 0.24 & 1.88 & 0.56 & 0.53 & 0.07 & -11.51±0.042 & 0.167 \\
\multicolumn{1}{l}{I16484-4603} & 16:52:02.10 & -46:08:16.6 & 0.15$^\textup{a}$ & 17.1±1.5 & -33.15±1.62 & 0.65 & 3.53 & 1.57 & 1.4 & 0.48 & -10.63±0.037 & 0.165 \\
\multicolumn{1}{l}{I16487-4423} & 16:52:23.92 & -44:27:47.9 & 0.23$^\textup{a}$ & 3.36±0.3 & -42.69±0.23 & 0.17 & 1.22 & 0.74 & 0.84 & 0.1 & -10.95±0.037 & 0.169 \\
\multicolumn{1}{l}{I16489-4431} & 16:52:34.13 & -44:36:26.0  & 0.22$^\textup{a}$ & 6.98±0.6 & -40.45±1.61 & 0.3 & 2.67 & 1.69 & 1.77 & 0.35 & -10.63±0.036 & 0.162 \\
I16524-4300 C1 & 16:56:03.80 & -43:04:59.7  & 0.24±0.083 & 12.4±2.2 & -41.45±1.8 & 0.35 & 2.17 & 1.24 & 1.36 & 0.22 & -10.34±0.071 & 0.106 \\
\multicolumn{1}{r}{C2} & 16:56:02.36 & -43:04:18.9& 0.15±0.089 & 7.7±1.2 & -40.73±2.38 & 0.17 & 2.33 & 1.09 & 0.95 & 0.38 & -10.55±0.063 & 0.106 \\
\multicolumn{1}{r}{C3} & 16:56:06.55 & -43:04:12.5  & 0.25$^\textup{a}$ & 0.81±0.1 & -42.1±0.2 & 0.23 & 1.09 & 0.63 & 0.62 & 0.05 & -11.52±0.051 & 0.106 \\
I16547-4247 C1 & 16:58:16.79 & -42:52:03.0 & 0.17±0.057 & 50.7±5.2 & -29.37±1.77 & 0.51 & 4.58 & 2.26 & 2.32 & 1.25 & -9.92±0.042 & 0.119 \\
\multicolumn{1}{r}{C2} & 16:58:18.11 & -42:52:37.2 & 0.2$^\textup{a}$ & 3.81±0.92 & -36.64±2.34 & 0.57 & 5.02 & 2.12 & 1.75 & 0.95 & -11.05±0.094 & 0.119 \\
I16562-3959 C1 & 16:59:41.33 & -40:03:37.6 & 0.14 & 36.8±3.9 & -11.74±5.39 & 0.46 & 4.95 & 2.36 & 2.3 & 1.65 & -10.18±0.044 & 0.115 \\
\multicolumn{1}{r}{C2} & 16:59:38.93 & -40:04:11.0  & 0.1±0.033 & 20.9±1.6 & -11.4±0.44 & 0.3 & 1.53 & 1.01 & 1.05 & 0.11 & -10.43±0.032 & 0.115 \\
\multicolumn{1}{r}{C3} & 16:59:43.59 & -40:03:08.5 & 0.16 & 22.2±3 & -10.53±2.6 & 0.37 & 2.55 & 1.44 & 1.54 & 0.26 & -10.4±0.055 & 0.115 \\
\multicolumn{1}{l}{I16571-4029} & 17:00:32.34 & -40:34:08.0 & 0.17$^\textup{a}$ & 15.2±0.68 & -13.78±1.32 & 0.48 & 4.71 & 2.55 & 2.44 & 1.22 & -10.57±0.019 & 0.112 \\
\multicolumn{1}{l}{I17006-4215} & 17:04:11.88 & -42:19:50.0 & 0.16$^\textup{a}$ & 2.39±0.22 & -21.78±0.2 & 0.23 & 2.23 & 1.21 & 1.19 & 0.35 & -11.44±0.038 & 0.117 \\
I17008-4040 C1 & 17:04:22.96 & -40:44:19.6 & 0.16±0.056 & 15.6±2.8 & -18.15±2.77 & 0.23 & 2.54 & 1.35 & 1.36 & 0.4 & -10.56±0.072 & 0.120 \\
\multicolumn{1}{r}{C2} & 17:04:25.22 & -40:44:28.4 & 0.17$^\textup{a}$ & 5.44±0.74 & -12.02±5.12 & 0.68 & 2.82 & 1.72 & 1.72 & 0.35 & -11.01±0.055 & 0.120 \\
I17016-4124  C1 & 17:05:10.61 & -41:29:18.9  & 0.14±0.037 & 122±18 & -26.25±33.11 & 0.65 & 6.52 & 3.5 & 3.25 & 2.81 & -10.14±0.06 & 0.116 \\
\multicolumn{1}{r}{C2}  & 17:05:11.25 & -41:28:52.7 & 0.09 & 96.1±9.2 & -31.02±14.83 & 0.63 & 7.13 & 3.97 & 3.81 & 3.41 & -10.25±0.04 & 0.116 \\
\multicolumn{1}{l}{I17143-3700} & 17:17:45.50 & -37:03:12.5 & 0.87$^\textup{a}$ & 1.58±0.13 & -31.18±0.24 & 0.36 & 1.38 & 0.87 & 0.92 & 0.04 & -10.1±0.034 & 0.140 \\
\multicolumn{1}{l}{I17158-3901} & 17:19:20.41 & -39:03:49.5 & 0.12±0.062 & 19.7±2.1 & -17.54±2.0 & 0.34 & 3.11 & 1.65 & 1.59 & 0.6 & -10.15±0.044 & 0.159 \\
I17160-3707 C1 & 17:19:26.61 & -37:10:21.4 & 0.66 & 14±1.9 & -69.0±7.56 & 0.42 & 2.89 & 1.69 & 1.72 & 0.42 & -9.31±0.055 & 0.151 \\
\multicolumn{1}{r}{C2} & 17:19:25.83 & -37:10:02.8  & 0.72$^\textup{a}$ & 6.9±1.4 & -69.92±26.24 & 0.62 & 18.32 & 6.3 & 4.71 & 25.68 & -9.62±0.08 & 0.151 \\
\multicolumn{1}{r}{C3} & 17:19:26.60 & -37:11:12.6  & 0.72$^\textup{a}$ & 1.11±0.09 & -69.17±0.21 & 0.35 & 1.38 & 0.88 & 0.87 & 0.13 & -10.41±0.034 & 0.151 \\
I17175-3544 C1 & 17:20:52.71 & -35:47:04.9 & 0.09$^\textup{a}$ & 65.5±9.8 & -9.18±13.26 & 0.35 & 9.7 & 3.8 & 3.01 & 9.74 & -10.43±0.061 & 0.155 \\
\multicolumn{1}{r}{C2} & 17:20:54.55 & -35:46:46.8 & 0.09$^\textup{a}$ & 7.7±2.1 & 1.17±0.5 & 0.91 & 4.44 & 2.93 & 3.14 & 1.17 & -11.36±0.105 & 0.155 \\
\multicolumn{1}{r}{C3} & 17:20:50.26 & -35:46:40.7 & 0.09$^\textup{a}$ & 6.48±0.42 & 0.88±10.1 & 0.84 & 6.37 & 3.71 & 3.62 & 1.58 & -11.44±0.027 & 0.155 \\
\multicolumn{1}{l}{I17204-3636} & 17:23:50.18 & -36:38:56.7 & 0.23$^\textup{a}$ & 3.21±0.29 & -19.79±0.68 & 0.36 & 2.23 & 1.29 & 1.48 & 0.24 & -10.95±0.038 & 0.132 \\
\multicolumn{1}{l}{I17220-3609} & 17:25:25.09 & -36:12:40.6  & 0.48±0.222 & 25.6±3.4 & -94.88±2.66 & 0.46 & 3.94 & 1.72 & 1.6 & 0.74 & -9.29±0.054 & 0.154 \\
I17233-3606 C1 & 17:26:40.91  & -36:09:26.8 & 0.14±0.008 & 213.8±9.6 & 1.68±0.65 & 0.31 & 8.77 & 4.75 & 5.28 & 3.71 & -9.92±0.019 & 0.169 \\
\multicolumn{1}{r}{C2} & 17:26:42.53 & -36:09:17.3 & 0.07 & 159±19 & 2.47±0.51 & 0.83 & 10.5 & 4.77 & 4.35 & 7.03 & -10.05±0.049 & 0.169 \\
\multicolumn{1}{r}{C3} & 17:26:45.55 & -36:09:20.4 & 0.15±0.031 & 53.3±8.9 & 0.99±0.51 & 0.67 & 5.03 & 2.83 & 2.54 & 1.81 & -10.52±0.067 & 0.169 \\
\multicolumn{1}{l}{I17244-3536} & 17:27:50.28 & -35:38:57.8 & 0.09$^\textup{a}$ & 0.59±0.04 & -10.43±0.28 & 0.46 & 1.44 & 1.15 & 1.15 & 0.04 & -12.46±0.028 & 0.148 \\
I17258-3637 C1 & 17:29:18.35 & -36:40:27.8 & 0.15±0.051 & 12.1±1.6 & -13.37±2.06 & 0.36 & 1.86 & 1.1 & 1.06 & 0.16 & -10.59±0.054 & 0.161 \\
\multicolumn{1}{r}{C2} & 17:29:18.65 & -36:40:50.0  & 0.18$^\textup{a}$ & 3.13±0.49 & -10.86±5.07 & 0.26 & 1.45 & 0.89 & 0.85 & 0.09 & -11.18±0.063 & 0.161 \\
\multicolumn{1}{l}{I17269-3312} & 17:30:14.47 & -33:15:01.1 & 0.3$^\textup{a}$ & 10.73±0.69 & -24.11±2.5 & 0.35 & 3.08 & 1.59 & 1.62 & 0.64 & -10.19±0.027 & 0.149 \\
\multicolumn{1}{l}{I17271-3439} & 17:30:26.8 & -34:41:48.9  & 0.37±0.091 & 51.6±8 & 4.81±0.0 & 0.24 & 3.71 & 1.66 & 1.77 & 0.72 & -9.81±0.063 & 0.148 \\
I17278-3541 C1 & 17:31:15.88 & -35:44:15.3  & 0.09$^\textup{a}$ & 6.45±0.46 & 4.54±2.61 & 0.37 & 3.29 & 1.97 & 2.22 & 0.84 & -11.45±0.03 & 0.151 \\
\multicolumn{1}{r}{C2} & 17:31:13.08 & -35:43:41.8  & 0.09$^\textup{a}$ & 3.55±0.17 & -8.39±4.01 & 0.58 & 4.03 & 2.57 & 2.78 & 1.14 & -11.7±0.02 & 0.151 \\
\multicolumn{1}{r}{C3} & 17:31:16.02 & -35:43:27.2 & 0.09$^\textup{a}$ & 1.08±0.12 & 0.94±0.25 & 0.17 & 6.92 & 0.99 & 0.63 & 1.32 & -12.22±0.046 & 0.151 \\
\multicolumn{1}{r}{C4} & 17:31:16.52 & -35:43:46.0 & 0.09$^\textup{a}$ & 2.54±0.49 & 9.51±27.86 & 0.57 & 4.06 & 2.03 & 1.66 & 0.75 & -11.85±0.077 & 0.151 \\
\multicolumn{1}{r}{C5} & 17:31:13.26 & -35:44:07.9 & 0.09$^\textup{a}$ & 0.66±0.03 & -8.82±0.74 & 0.35 & 2.19 & 1.58 & 1.79 & 0.32 & -12.44±0.019 & 0.151 \\
\multicolumn{1}{r}{C6} & 17:31:14.42 & -35:43:58.0 & 0.09$^\textup{a}$ & 0.43±0.05 & 0.44±0.03 & 0.17 & 0.81 & 0.51 & 0.61 & 0.05 & -12.62±0.048 & 0.151 \\
\multicolumn{1}{l}{I17439-2845} & 17:47:10.18 & -28:45:44.2 & 0.53$^\textup{a}$ & 0.44±0.01 & 22.74±1.63 & 0.74 & 3.6 & 2.29 & 2.63 & 0.73 & -11.05±0.01 & 0.105 \\
I17441-2822 C1 & 17:47:22.78 & -28:23:15.1 & 0.54$^\textup{a}$ & 71±14 & 60.87±22.29 & 0.45 & 9.32 & 3.82 & 3.08 & 8.19 & -8.83±0.078 & 0.526 \\
\multicolumn{1}{r}{C2} & 17:47:18.57 & -28:23:15.8  & 0.54$^\textup{a}$ & 53±11 & 58.9±50.44 & 0.48 & 10.36 & 4.02 & 3.68 & 9.57 & -8.96±0.082 & 0.526 \\
\multicolumn{1}{r}{C3} & 17:47:22.06 & -28:22:49.4 & 0.54$^\textup{a}$ & 28.1±4.9 & 64.93±40.02 & 0.74 & 5.85 & 3.16 & 2.65 & 2.7 & -9.24±0.07 & 0.526 \\
\multicolumn{1}{r}{C4} & 17:47:18.30 & -28:22:46.6 & 0.54$^\textup{a}$ & 16.3±3.4 & 59.47±14.94 & 0.42 & 4.28 & 1.62 & 1.21 & 1.07 & -9.47±0.082 & 0.526 \\
\multicolumn{1}{r}{C5} & 17:47:20.82 & -28:22:59.5 & 0.54$^\textup{a}$ & 4.75±0.69 & 45.77±0.4 & 0.54 & 2.07 & 1.33 & 1.33 & 0.18 & -10.01±0.059 & 0.526 \\
\multicolumn{1}{l}{I17455-2800} & 17:48:41.59 & -28:01:56.2 & 0.67$^\textup{a}$ & 6.33±0.53 & -15.74±0.48 & 0.24 & 2.02 & 1.11 & 1.07 & 0.27 & -9.7±0.035 & 0.129 \\
\multicolumn{1}{l}{I17545-2357} & 17:57:34.59 & -23:57:32.5 & 0.2$^\textup{a}$ & 1.43±0.38 & 7.56±0.3 & 0.17 & 1.28 & 0.42 & 0.3 & 0.05 & -11.41±0.102 & 0.100 \\
I17589-2312 C1 & 18:01:59.12 & -23:12:56.7  & 0.21$^\textup{a}$ & 3.83±0.65 & 23.84±3.49 & 0.32 & 1.96 & 1.03 & 0.92 & 0.15 & -10.97±0.068 & 0.098 \\
\multicolumn{1}{r}{C2} & 18:01:58.39 & -23:12:21.1 & 0.21$^\textup{a}$ & 3.78±0.45 & 19.88±1.68 & 0.35 & 2.31 & 1.01 & 0.97 & 0.23 & -10.98±0.049 & 0.098 \\
\multicolumn{1}{l}{I18032-2032} & 18:06:15.05 & -20:31:37.7  & 0.18 & 64.8±3.1 & 5.08±5.54 & 0.38 & 10.76 & 4.26 & 3.77 & 8.97 & -9.27±0.02 & 0.088 \\
I18056-1952 C1 & 18:08:38.16 & -19:51:50.1  & 0.21 & 36.9±1.7 & 66.88±0.82 & 0.54 & 4.08 & 2.45 & 2.48 & 0.96 & -9.07±0.02 & 0.099 \\
\multicolumn{1}{r}{C2} & 18:08:38.00 & -19:51:12.5 & 0.6$^\textup{a}$ & 11.32±0.56 & 65.47±5.17 & 0.3 & 4.86 & 2.5 & 2.67 & 1.95 & -9.59±0.021 & 0.099 \\
\multicolumn{1}{r}{C3} & 18:08:36.37 & -19:52:14.3 & 0.6$^\textup{a}$ & 2.75±0.2 & 70.98±4.31 & 0.37 & 1.51 & 1.08 & 1.09 & 0.09 & -10.2±0.03 & 0.099 \\
I18079-1756 C1 & 18:10:51.58 & -17:55:46.5 & 0.13$^\textup{a}$ & 3.31±0.43 & 18.02±1.27 & 0.17 & 1.09 & 0.63 & 0.67 & 0.08 & -11.46±0.053 & 0.139 \\
\multicolumn{1}{r}{C2} & 18:10:49.38 & -17:56:01.7 & 0.13$^\textup{a}$ & 1.37±0.1 & 11.65±0.35 & 0.29 & 1.43 & 0.93 & 1.01 & 0.06 & -11.84±0.031 & 0.139 \\
\multicolumn{1}{l}{I18089-1732} & 18:11:51.58 & -17:31:25.6 & 0.18$^\textup{a}$ & 5.83±0.28 & 33.25±0.97 & 0.3 & 2.74 & 1.54 & 1.46 & 0.43 & -10.94±0.02 & 0.136\\
\end{tabular}
\end{table*}

\begin{table*}
\centering
	\begin{threeparttable}
	\renewcommand\tabcolsep{2.7pt}
    \begin{tabular}{lcclllccccclc}
	    \hline
	    \hline
\multicolumn{1}{c}{IRAS} & \multicolumn{1}{c}{RA} & \multicolumn{1}{c}{DEC} & \multicolumn{1}{c}{$R_\textup{a}$} & \multicolumn{1}{c}{$\int{F_{\nu} d\nu}$} & \multicolumn{1}{c}{$v$} & \multicolumn{5}{c}{$\sigma_\textup{v}$} & \multicolumn{1}{c}{log $L_\textup{SiO}$} & \multicolumn{1}{c}{$\sigma$} \\
& \multicolumn{1}{c}{(J2000)} & \multicolumn{1}{c}{(J2000)} & \multicolumn{1}{c}{(pc)} & \multicolumn{1}{c}{(Jy km s$^{-1}$)} & \multicolumn{1}{c}{(km s$^{-1}$)} & \multicolumn{5}{c}{(km s$^{-1}$)} & \multicolumn{1}{c}{($L_{\odot}$)} & \multicolumn{1}{c}{(Jy beam$^{-1}$)} \\ \cline{7-11} 
&  &  &  &  &  
        & \multicolumn{1}{c|}{$D_\textup{min}$} 
        & \multicolumn{1}{c|}{$D_\textup{max}$} 
        & \multicolumn{1}{c|}{$D_\textup{mean}$} 
        & \multicolumn{1}{c|}{$D_\textup{median}$} 
        & \multicolumn{1}{c|}{$D_\textup{std}$} 
        & &
  \\ \cline{7-11}
\multicolumn{1}{l}{I18117-1753} & 18:14:39.21 & -17:52:08.6 & 0.21±0.066 & 31.6±3.9 & 35.6±9.08 & 0.42 & 4.11 & 1.94 & 1.64 & 1.03 & -10.18±0.051 & 0.138 \\
\multicolumn{1}{l}{I18139-1842} & 18:16:51.47 & -18:41:39.5  & 0.22$^\textup{a}$ & 3.15±0.17 & 39.36±0.27 & 0.17 & 1.8 & 1.0 & 1.01 & 0.27 & -11.04±0.023 & 0.134 \\
\multicolumn{1}{l}{I18159-1648} & 18:18:54.84 & -16:47:50.4 & 0.09 & 47.1±3.5 & 20.13±4.86 & 0.36 & 5.89 & 2.36 & 2.24 & 1.92 & -10.49±0.031 & 0.138 \\
\multicolumn{1}{l}{I18182-1433} & 18:21:09.03 & -14:31:46.0 & 0.34$^\textup{a}$ & 12.24±0.78 & 59.76±0.63 & 0.4 & 2.82 & 1.61 & 1.64 & 0.52 & -10.07±0.027 & 0.130 \\
\multicolumn{1}{l}{I18236-1205} & 18:26:26.50 & -12:03:53.4  & 0.16$^\textup{a}$ & 1.84±0.16 & 27.13±6.9 & 0.17 & 2.97 & 1.15 & 1.27 & 0.44 & -11.57±0.036 & 0.144 \\
\multicolumn{1}{l}{I18264-1152} & 18:29:14.91 & -11:50:25.6 & 0.17±0.061 & 19.4±2.7 & 43.35±4.27 & 0.55 & 4.88 & 1.78 & 1.35 & 1.04 & -10.17±0.057 & 0.137 \\
\multicolumn{1}{l}{I18290-0924} & 18:31:43.13 & -09:22:32.9  & 0.39$^\textup{a}$ & 1.87±0.24 & 84.15±0.46 & 0.3 & 1.38 & 0.81 & 0.8 & 0.08 & -10.78±0.052 & 0.101 \\
\multicolumn{1}{l}{I18311-0809} & 18:33:53.54 & -08:07:12.2  & 0.45$^\textup{a}$ & 3.35±0.18 & 113.7±0.19 & 0.32 & 2.3 & 1.44 & 1.47 & 0.33 & -10.41±0.023 & 0.096 \\
I18316-0602 C1 & 18:34:21.17 & -06:00:14.5  & 0.16$^\textup{a}$ & 28.3±1.3 & 35.61±12.71 & 0.29 & 8.83 & 4.23 & 3.88 & 7.48 & -10.41±0.02 & 0.093 \\
\multicolumn{1}{r}{C2} & 18:34:20.52 & -05:59:33.6  & 0.16$^\textup{a}$ & 25.4±2.5 & 43.5±33.06 & 0.42 & 6.5 & 2.74 & 2.3 & 2.76 & -10.46±0.041 & 0.093 \\
\multicolumn{1}{r}{C3} & 18:34:20.52 & -05:59:33.6  & 0.05±0.131 & 6.6±2.6 & 29.93±5.4 & 1.05 & 7.78 & 2.48 & 1.91 & 2.37 & -11.04±0.144 & 0.093 \\
\multicolumn{1}{l}{I18341-0727} & 18:36:49.87 & -07:24:53.6 & 0.45$^\textup{a}$ & 9.6±1.1 & 112.5±3.89 & 0.29 & 2.97 & 1.52 & 1.45 & 0.52 & -9.96±0.047 & 0.096 \\
\multicolumn{1}{l}{I18411-0338} & 18:43:46.11 & -03:35:31.2 & 0.57$^\textup{a}$ & 5.23±0.33 & 103.62±0.06 & 0.23 & 1.82 & 1.08 & 1.18 & 0.22 & -10.04±0.027 & 0.115 \\
\multicolumn{1}{l}{I18434-0242} & 18:46:03.58 & -02:39:24.7  & 0.39$^\textup{a}$ & 19.81±0.49 & 97.65±0.11 & 0.42 & 4.24 & 2.19 & 2.06 & 1.29 & -9.78±0.011 & 0.094 \\
\multicolumn{1}{l}{I18461-0113} & 18:48:41.74 & -01:10:01.3 & 0.39$^\textup{a}$ & 14.65±0.59 & 95.48±0.34 & 0.46 & 2.53 & 1.58 & 1.63 & 0.35 & -9.91±0.017 & 0.094 \\
\multicolumn{1}{l}{I18469-0132} & 18:49:33.05 & -01:29:02.1  & 0.39$^\textup{a}$ & 4.57±0.24 & 86.96±0.12 & 0.17 & 1.72 & 1.04 & 0.92 & 0.18 & -10.42±0.022 & 0.123 \\
\multicolumn{1}{l}{I18479-0005} & 18:50:30.63 & -00:01:58.9 & 0.99$^\textup{a}$ & 8.42±0.57 & 14.64±3.11 & 0.45 & 2.84 & 1.83 & 1.85 & 0.46 & -9.35±0.028 & 0.097 \\
I18507p0110  C1& 18:53:18.50 & +01:15:01.2 & 0.08±0.039 & 35.1±6.1 & 58.9±1.88 & 0.32 & 5.8 & 2.01 & 1.95 & 1.8 & -10.57±0.07 & 0.126 \\
\multicolumn{1}{r}{C2} & 18:53:16.67 & +01:15:11.1 & 0.08±0.029 & 32.4±3.9 & 61.56±5.32 & 0.5 & 6.48 & 2.9 & 2.88 & 1.78 & -10.61±0.049 & 0.126 \\
\multicolumn{1}{r}{C3} & 18:53:15.62 & +01:14:20.1 & 0.12$^\textup{a}$ & 4.6±1.4 & 54.97±2.24 & 0.44 & 3.3 & 1.12 & 1.11 & 0.2 & -11.45±0.115 & 0.126 \\
I18507p0121 C1 & 18:53:18.02 & +01:25:24.8 & 0.12$^\textup{a}$ & 48.7±4.8 & 59.29±8.86 & 0.23 & 4.78 & 2.29 & 2.19 & 2.27 & -10.43±0.041 & 0.110 \\
\multicolumn{1}{r}{C2} & 18:53:17.90 & +01:24:48.1 & 0.09±0.03 & 29±3.3 & 57.25±7.38 & 0.67 & 7.04 & 2.71 & 2.44 & 2.13 & -10.65±0.047 & 0.110 \\
\multicolumn{1}{l}{I18517p0437} & 18:54:14.50 & +04:41:39.3  & 0.18$^\textup{a}$ & 8±1 & 43.85±0.23 & 0.23 & 1.43 & 0.72 & 0.7 & 0.09 & -10.85±0.051 & 0.110 \\
\multicolumn{1}{l}{I18530p0215} & 18:55:33.42 & +02:19:01.6 & 0.36$^\textup{a}$ & 0.41±0.06 & 77.43±3.21 & 0.17 & 3.22 & 0.97 & 0.67 & 0.62 & -11.55±0.059 & 0.099 \\
\multicolumn{1}{l}{I19078p0901} & 19:10:13.24 & +09:06:14.0 & 0.97±0.099 & 278±17 & 6.46±2.0 & 0.88 & 6.91 & 3.81 & 4.0 & 2.27 & -7.97±0.026 & 0.122 \\
\multicolumn{1}{l}{I19095p0930} & 19:11:53.85 & +09:35:50.7 & 0.43$^\textup{a}$ & 18.7±1 & 43.51±0.18 & 0.95 & 3.32 & 2.16 & 2.22 & 0.4 & -9.67±0.023 & 0.116 \\
\multicolumn{1}{l}{I19097p0847} & 19:12:09.04 & +08:52:12.2 & 0.6$^\textup{a}$ & 6.51±0.45 & 57.27±0.38 & 0.31 & 2.83 & 1.6 & 1.76 & 0.56 & -9.83±0.029 & 0.139\\
\hline
\end{tabular}
 \begin{tablenotes}
        \footnotesize
        \item[]Note. $R_\textup{a}$: the linear radius of SiO clump. $\int{F_{\nu} d\nu}$: the integrated intensity of SiO clump. $v$: the velocity of SiO clump. $\sigma_\textup{v}$: the velocity dispersion of SiO clump. $L_\textup{sio}$: the luminosity of the SiO clump. $\sigma$: rms in one source.
        \item[a] symbol after radius of SiO clump indicates that this SiO clump is unresolved by CASA.
      \end{tablenotes}
      
    \end{threeparttable}
\end{table*}

\begin{table*}
   \centering
   \caption{ SiO (2-1) Line Parameters and Clump Properties in the Region Without Clear Detections}
   \label{tab:TableA2}
\begin{threeparttable}
\begin{tabular}{ccclllclc}
	\hline
	\hline
	IRAS & RA & DEC & \multicolumn{1}{c}{$R_\textup{a}$} & \multicolumn{1}{c}{$\int{F_{\nu} d\nu}$} & $v$  & $\sigma_\textup{v}$ &\multicolumn{1}{c}{log $L_\textup{SiO}$} & \multicolumn{1}{c}{$\sigma$} \\ 
	& J2000 & J2000 & \multicolumn{1}{c}{(pc)} & (Jy km s$^{-1}$ ) &(km s$^{-1}$)& (km s$^{-1}$) &\multicolumn{1}{c}{($L_{\odot}$)} & \multicolumn{1}{c}{(Jy beam$^{-1}$)}\\
\multicolumn{1}{l}{I12383-6128} & 12:41:18.18 & -61:44:17.4 & 0.25$^\textup{a}$
& 1.08 ± 0.38 & -39 ± 2 & 1.3 ± 1.7 & -11.44 ± 0.13 & 0.14 \\
\multicolumn{1}{l}{I13080-6229} & 13:11:12.81 & -62:45:02.1 & 0.58 ± 0.4 & 3.6 ± 1.4 & -35.7 ± 8.1  & 2.6 ± 6.6 & -10.79± 0.143 & 0.122 \\
\multicolumn{1}{l}{I13291-6229} & 13:32:32.54 & -62:45:22.6 & 0.35 ± 0.18 & 2.35 ± 0.5 & -37.6 ± 3.8  & 0.99 ± 3.1 & -11.21 ± 0.08 & 0.125 \\
\multicolumn{1}{l}{I14050-6056}& 14:08:40.89 & -61:11:18.4 & 0.28$^\textup{a}$ & 3.81 ± 0.99 & -53 ± 3 & 1.8 ± 2.4 & -10.85 ± 0.1 & 0.154 \\
\multicolumn{1}{l}{I14382-6017}&14:42:03.76 & -60:30:30.9 & 0.39 ± 0.59 & 3.6 ± 1.7 & -60.7 ± 2.6 & 1.2 ± 2.1 & -10.17 ± 0.095 & 0.153 \\
\multicolumn{1}{l}{I15254-5621}& 15:29:18.84 & -56:31:38.4 & 0.28$^\textup{a}$ & 4.1 ± 1 & -68.5 ± 2.3 & 2.1 ± 1.8 & -10.69 ± 0.095 & 0.16 \\
\multicolumn{1}{l}{I15437-5343}& 15:47:33.06 & -53:52:39.4 & 0.46 & 4.9 ± 1.4 & -83.1 ± 2.8 & 1.8 ± 2.3 & -10.69 ± 0.109 & 0.163 \\
\multicolumn{1}{l}{I15502-5302}& 15:54:03.32 & -53:11:37.9 & 0.43 $^\textup{a}$ & 3.91 ± 0.74 & -92.4 ± 1.7 & 2 ± 1.4& -10.38 ± 0.08 & 0.176 \\
\multicolumn{1}{l}{I16304-4710}	& 16:34:05.19 & -47:16:33.0 & 0.8 $^\textup{a}$& 1.42 ± 0.48 & -62.3 ±2.3 & 0.93 ± 1.9 & -10.24 ± 0.126 & 0.211 \\
\multicolumn{1}{l}{I16313-4729}& 16:34:54.38 & -47:35:33.6 & 0.33 $^\textup{a}$ & 5.4 ± 1.2 & -74.3 ±2.2 & 3 ±1.8 & -10.42 ± 0.087 & 0.188 \\
\multicolumn{1}{l}{I16330-4725}& 16:36:43.25 & -47:31:25.9 & 0.78 $^\textup{a}$& 4.1 ± 1.2 & -71.4 ± 2.7 & 2.5 ± 2.2 & -9.81 ± 0.111 & 0.205 \\
\multicolumn{1}{l}{I16372-4545}& 16:40:54.07 & -45:50:52.3 & 0.3 $^\textup{a}$& 2.8 ± 0.5 & -58.2 ± 2.4 & 2 ± 2 & -10.82 ± 0.071 & 0.204 \\

\multicolumn{1}{l}{I16506-4512}& 16:54:13.35 & -45:17:54.7 & 0.07 ± 0.05 & 5.74 ± 0.88 & -35.3 ± 1.9 & 3.8 ± 1.6 & -10.98 ± 0.062 & 0.181 \\
\multicolumn{1}{l}{I17136-3617}& 17:17:02.31 & -36:20:49.4 & 0.09 $^\textup{a}$ & 2.64 ± 0.48 & -10.5 ± 2.7 & 2.8 ± 2.2 & -11.83 ± 0.073 & 0.176 \\
\multicolumn{1}{l}{I18110-1854}& 18:14:02.41 & -18:53:14.1 & 0.39 ± 0.157 & 13.5 ± 4.1 & 38.5 ± 2.5 & 3 ± 2 & -10.32 ± 0.115 & 0.144 \\
\multicolumn{1}{l}{I18116-1646}& 18:14:36.17 & -16:45:45.7 & 0.29 $^\textup{a}$ & 1.43 ± 0.52 & 48.6 ± 2.5 & 1.3 ± 2 & -11.16 ± 0.135 & 0.145 \\
\multicolumn{1}{l}{I18223-1243}& 18:25:11.12 & -12:42:21.1 & 0.24$^\textup{a}$ & 1.31 ± 0.33 &  48.3 ± 2.4 & 2 ± 2  & -10.95 ± 0.076 & 0.168 \\
\multicolumn{1}{l}{I18228-1312}& 18:25:41.66 & -13:10:11.6 & 0.29 & 4.24 ± 0.93 & 31.7 ± 2.7 & 0.77 ± 2.2 & -10.86 ± 0.086 & 0.148 \\
\hline
\end{tabular}
 \begin{tablenotes}
        \footnotesize
        \item[]Note. $R_\textup{a}$: The linear radius of SiO clump. $\int{F_{\nu} d\nu}$: The integrated intensity of SiO clump. $v$: SiO line central velocity. $\sigma_\textup{v}$: The velocity dispersion of SiO clump. $L_\textup{sio}$: The luminosity of the SiO clump. $\sigma$: rms in one source.
        \item[a] symbol after radius of SiO clump indicates that this SiO clump is unresolved by CASA.
      \end{tablenotes}
    \end{threeparttable}
\end{table*}

\begin{table*}
	\caption{SiO (2-1) Line Parameters and Clump Properties}
	\label{tabA3}
	\renewcommand\tabcolsep{6pt}
     \centering
     \begin{tabular}{lcclccccccc}
       \hline
	   \hline 
IRAS &  $L_\textup{bol}/M$ & log$n$ & [SiO]/[H$^{13}$CO$^+$] & $D_\textup{d}$ & H40$\alpha$ & \multicolumn{3}{c}{Wing} & outflow & type  \\
\cline{7-9}
 & & &  & & & SiO & HCO$^+$ & CS  \\
\cline{7-9}
 & ($L{_\odot} M{_\odot}^{-1}$) & (cm$^{-3}$) & & (pc) & & & & & \\
 
I08076-3556 & 3.16& 5.51 & - & 0.01 &  & \checkmark &  &  &\checkmark & C \\
I08303-4303 & 25.12&5.4 & 0.41±0.01 & 0.16 &  & \checkmark & \checkmark & \checkmark & \checkmark & D \\
I08448-4343 C1& 25.12 & 5.68 & 1.86±0.07& 0.06 &  & \checkmark &  &  & \checkmark & D \\
\multicolumn{1}{r}{C2} & 25.12 & 5.68 & 3.35±0.37 & 0.08 &  & \checkmark &  &  & \checkmark & D \\
I08470-4243 & 39.81 & 5.4 & 2.98±0.1 & 0.02 &  & \checkmark & \checkmark & \checkmark & \checkmark & C \\
I09002-4732 C1& 158.49 & 5.78 & 0.74±- & 0.2 & \checkmark &  &  &  &  & B \\
\multicolumn{1}{r}{C2} & 158.49 & 5.78& - & 0.15 & \checkmark &  &  &  &  & B \\
\multicolumn{1}{r}{C3} & 158.49 & 5.78 & - & 0.08 & \checkmark &  &  &  &  & B \\
I09018-4816 C1 & 50.12& 5.59 & 0.9±- & 0.05 &  &  &  &  &  & C \\
\multicolumn{1}{r}{C2} & 50.12 & 5.59 & - & 0.48 &  &  &  &  &  & D \\
I09094-4803 & 31.62 & 4.18 & 1.39±0.16 & 0.93 &  &  &  &  &  & D \\
I10365-5803 & 39.81 & 5.29 &0.32±- & 0.11 &  & \checkmark & \checkmark &  & \checkmark & D \\
I11298-6155 & 63.1 & 4.52 & 3.97±0.19 & 0.37 & \checkmark &  & \checkmark & \checkmark & \checkmark & A \\
I11332-6258 & 39.81& 5.28 & 0.22±0.05 & 0.03 &  &  & \checkmark &  & \checkmark & C \\
I11590-6452 & 3.98 & 6.29 & - & 0.04 &  &  &  &  &  & D \\
I12320-6122 & 398.11 & 3.61 & 0.7±0.25 & 0.25 & \checkmark & \checkmark & \checkmark & \checkmark & \checkmark & B \\
I12326-6245 & 79.43 & 4.36 & 6.16±0.12 & 0.07 & \checkmark & \checkmark & \checkmark & \checkmark & \checkmark & A \\
I12383-6128 & 5.01& 4.32 & 0.99±0.25 & 0.37 & \checkmark &  &  &  &  & B \\
I12572-6316 & 5.01 & 3.88 & 0.42±0.05 & 0.12 & \checkmark &  &  &  &  & A \\
I13079-6218 & 39.81 & 4 & 4.77±0.05 & 0.11 & \checkmark & \checkmark & \checkmark & \checkmark & \checkmark & A \\
I13080-6229 & 79.43& 3.78 & 0.15±0.01 & 0.17 & \checkmark & \checkmark &  &  & \checkmark & B \\
I13111-6228 & 50.12 & 3.77 & 0.09±- & 0.04 & \checkmark &  & \checkmark &  & \checkmark & A \\ 
I13134-6242 & 31.62 & 4.45 & 3.04±0.14 & 0.08 &  & \checkmark & \checkmark & \checkmark & \checkmark & C \\
I13140-6226 & 7.94  & 4.42 & 1±- & 0.05 &  & \checkmark & \checkmark & \checkmark & \checkmark & C \\
I13291-6229 & 39.81 & 4.09 & 0.13±0.01 & 0.38 & \checkmark &  &  &  &  & B \\
I13291-6249 & 31.62  & 3.6 & 1.85±0.05 & 0.5 & \checkmark & \checkmark & \checkmark & \checkmark & \checkmark & B \\
I13295-6152 & 3.98 & 4.64 & 0.09±- & 0.14 &  &  &  &  &  & C \\
I13471-6120 & 79.43  & 4 & 0.62±0.09 & 0.43 & \checkmark & \checkmark & \checkmark & \checkmark & \checkmark & B \\
I13484-6100 & 50.12 & 3.94 & 1.24±0.05 & 0.14 &  & \checkmark & \checkmark & \checkmark & \checkmark & C \\
I14013-6105 & 63.1 & 4.1 & - & 0.08 & \checkmark &  &  &  &  & A \\
I14050-6056 & 79.43  & 3.95 & - & 0.61 & \checkmark &  &  &  &  & B \\
I14164-6028 & 31.62  & 5.12 & 0.94±0.04 & 0.04 &  &  &  &  &  & C \\
I14212-6131 & 10.0 & 4.35 & 1.75±0.12 & 0.36 &  & \checkmark & \checkmark & \checkmark & \checkmark & D \\
I14382-6017 & 39.81 & 3.54 & - & 0.58 & \checkmark &  &  &  &  & B \\
I14453-5912 & 19.95 & 3.82 & 1.65±0.45 & 0.33 & \checkmark &  &  &  &  & B \\
I14498-5856 C1 & 25.12& 4.01 & 0.52±0.01 & 0.12 &  &  & \checkmark &  & \checkmark & D \\
\multicolumn{1}{r}{C2} & 25.12 & 4.01 & - & 0.59 &  &  &  &  &  & D \\
I15254-5621 & 100.0 & 3.87 & - & 0.35 & \checkmark & \checkmark &  & \checkmark & \checkmark & B \\
I15290-5546 & 79.43 & 3.65 & 2.31±0.03 & 0.18 & \checkmark & \checkmark & \checkmark & \checkmark & \checkmark & A \\
I15394-5358 & 6.31 & 4.53 & 1.75±0.03 & 0.02 &  & \checkmark & \checkmark & \checkmark & \checkmark & C \\
I15408-5356 & 100.0 & 4.16 & 3.85±0.54 & 0.26 & \checkmark & \checkmark & \checkmark & \checkmark & \checkmark & B \\
I15411-5352 C1 & 63.1& 4.05 & 0.22±- & 0.08 & \checkmark & \checkmark &  & \checkmark & \checkmark & B \\
\multicolumn{1}{r}{C2}  & 63.1 & 4.05 & 0.33±- & 0.34 & \checkmark &  &  &  &  & B \\
I15437-5343 & 39.81 & 3.95 & 0.42±0.11 & 0.11 &  &  &  &  &  & C \\
I15439-5449 C1 & 25.12 & 4.1 & 0.77±0.001 & 0.57 & \checkmark &  &  &  &  & B \\
\multicolumn{1}{r}{C2} & 25.12 & 4.1 & 0.04±0.003 & 0.41 & \checkmark &  &  &  &  & B \\
I15502-5302 & 125.89 & 3.65 & 1.04±0.02 & 0.81 & \checkmark &  &  & \checkmark & \checkmark & B \\
I15520-5234 C1 & 79.43 & 4.34 & 16.45±4.935 & 0.18 & \checkmark & \checkmark & \checkmark & \checkmark & \checkmark & B \\
\multicolumn{1}{r}{C2}  & 79.43 & 4.34 & 4.06±0.005 & 0.06 & \checkmark & \checkmark &  & \checkmark & \checkmark & A \\
I15522-5411 & 7.94 & 4.09 & 0.33±0.002 & 0.04 &  & \checkmark &  & \checkmark & \checkmark & C \\
I15557-5215 & 5.01& 4.07 & 4.81±0.144 & 0.04 &  & \checkmark & \checkmark & \checkmark & \checkmark & C \\
I15567-5236 & 158.49 & 3.76 & 0.38±0.051 & 0.29 & \checkmark & \checkmark & \checkmark & \checkmark & \checkmark & B \\
I15584-5247 & 12.59 & 3.91 & 0.62±0.043 & 0.4 &  & \checkmark & \checkmark & \checkmark & \checkmark & D \\
I15596-5301 & 39.81  & 3.74 & 0.87±0.033 & 0.34 &  &  &  &  &  & C \\
I16026-5035 & 79.43 & 3.89 & 0.36±0.07 & 0.31 &  &  &  &  &  & D \\
I16037-5223 & 63.1 & 3.42 & 0.23±0.028 & 0.1 & \checkmark & \checkmark & \checkmark & \checkmark & \checkmark & A \\
I16060-5146 & 79.43 & 4.23 & - & 0.08 & \checkmark & \checkmark & \checkmark & \checkmark & \checkmark & A \\
I16065-5158 & 50.12 & 3.87 & 1.5±0.061 & 0.08 & \checkmark & \checkmark & \checkmark & \checkmark & \checkmark & A \\
I16071-5142 & 12.59 & 3.07 & 2.77±0.022& 0.08 &  & \checkmark & \checkmark & \checkmark & \checkmark & C \\
I16076-5134 & 50.12 & 3.63 & 5.61±0.016 & 0 &  & \checkmark & \checkmark & \checkmark & \checkmark & C\\
I16119-5048 & 12.59 & 3.91 & 2.08±0.141 & 0.07 &  & \checkmark & \checkmark & \checkmark & \checkmark & C \\
I16164-5046 & 63.1 & 3.9 & 0.92±0.091 & 0.05 & \checkmark & \checkmark &  & \checkmark & \checkmark & A \\
\end{tabular}
\end{table*}

\begin{table*}
	\renewcommand\tabcolsep{6pt}
     \centering
     \begin{tabular}{lcclccccccc}
       \hline
	   \hline 
IRAS &  $L_\textup{bol}/M$ & log$n$ & [SiO]/[H$^{13}$CO$^+$] & $D_\textup{d}$ & H40$\alpha$ & \multicolumn{3}{c}{Wing} & outflow & type  \\
\cline{7-9}
 & & &  & & & SiO & HCO$^+$ & CS  \\
\cline{7-9}
 & ($L{_\odot} M{_\odot}^{-1}$) & (cm$^{-3}$) & & (pc) & & & & & \\
I16172-5028 C1 & 63.1 & 4.08 & 1.27±0.03 & 0.23 & \checkmark & \checkmark & \checkmark & \checkmark & \checkmark & B \\
\multicolumn{1}{r}{C2} & 63.1 & 4.08 & - & 0.65 & \checkmark &  &  & \checkmark & \checkmark & B \\
\multicolumn{1}{r}{C3} & 63.1 & 4.08 & - & 0.89 & \checkmark & \checkmark &  & \checkmark & \checkmark & B \\
I16272-4837 & 12.59 & 4.04 & 2.53±0.861 & 0.14 &  & \checkmark &  & \checkmark & \checkmark & D \\
I16297-4757 & 31.62 & 3.11 & 0.73±0.055 & 0.14 & \checkmark &  &  &  &  & A \\
I16304-4710 & 25.12 & 3.7 & 0.26±0.01 & 0.22 & \checkmark &  &  &  &  & A \\
I16313-4729 & 100.0 & 4.37 & 0.7±0.04 & 0.13 &  &  &  &  &  & C \\
I16318-4724 & 31.62 & 3.83 & 1.02±0.042 & 0.09 &  & \checkmark & \checkmark & \checkmark & \checkmark & C \\
I16330-4725 & 100.0 & 3.14 & 0.23±- & 0.19 & \checkmark &  &  &  &  & A \\
I16344-4658 & 19.95 & 3.42 & 0.63±0.028 & 0.24 &  & \checkmark &  & \checkmark & \checkmark & C \\
I16348-4654 & 10.0 & 3.87 & 2.14±0.008 & 0.3 & \checkmark & \checkmark &  & \checkmark & \checkmark & A \\
I16351-4722 & 50.12 & 4.3 & 2.27±0.114 & 0.05 &  & \checkmark &  & \checkmark & \checkmark & C \\
I16372-4545 & 19.95 & 3.92 & 0.51±0.02 & 0.2 &  &  &  &  &  & D \\
I16385-4619 & 79.43 & 3.43 & 1.51±0.126 & 0.06 & \checkmark & \checkmark & \checkmark & \checkmark & \checkmark & A \\
I16424-4531 & 15.85 & 4.12 & 0.74±0.143 & 0.12 &  & \checkmark &  & \checkmark & \checkmark & D \\
I16445-4459 & 12.59 & 3.3 & 0.33±0.07 & 0.39 &  &  &  &  &  & D \\
I16458-4512 C1 & 7.94 & 3.76 & 4.39±0.054 & 0.46 & \checkmark & \checkmark &  & \checkmark & \checkmark & B \\
\multicolumn{1}{r}{C2} & 7.94 & 3.76 & - & 0.47 & \checkmark &  &  &  &  & B \\
I16484-4603 & 100.0 & 4.39 & 2.65±0.346 & 0.33 &  & \checkmark &  &  & \checkmark & D \\
I16487-4423 & 25.12 & 3.6 & 1.31±0.071 & 0.4 &  & \checkmark &  &  & \checkmark & D \\
I16489-4431 & 7.94 & 3.96 & 1.09±0.007 & 0.24 &  & \checkmark &  & \checkmark & \checkmark & D \\
I16506-4512 & 79.43 & 3.91 & 3.58±0.67 & 0.55 & \checkmark &  &  & \checkmark & \checkmark & B \\
I16524-4300 C1 & 10.0 & 3.94 & 1.48±0.13 & 0.3 &  &  &  &  &  & D \\
\multicolumn{1}{r}{C2} & 10.0 & 3.94 & 3.11±0.262 & 0.47 &  &  &  &  &  & D \\
\multicolumn{1}{r}{C3} & 10.0 & 3.94 & - & 0.87 &  &  &  &  &  & D \\
I16547-4247 C1 & 39.81 & 4.3 & 0.73±0.06 & 0.08 &  & \checkmark & \checkmark & \checkmark & \checkmark & C \\
\multicolumn{1}{r}{C2} & 39.81 & 4.3 & 16.68±7.892 & 0.43 &  & \checkmark & \checkmark & \checkmark & \checkmark & D \\
I16562-3959 C1 & 316.23 & 4.24 & 0.8±0.197 & 0.05 &  & \checkmark & \checkmark &  & \checkmark & C \\
\multicolumn{1}{r}{C2} & 316.23 & 4.24 & - & 0.5 &  &  &  &  &  & D \\
\multicolumn{1}{r}{C3} & 316.23 & 4.24 & - & 0.47 &  &  &  &  &  & D \\
I16571-4029 & 25.12 & 4.77 & 2.59±0.083 & 0.02 &  & \checkmark & \checkmark & \checkmark & \checkmark & C \\
I17006-4215 & 39.81 & 4.32 & 0.61±0.087 & 0.14 & \checkmark & \checkmark &  &  & \checkmark & B \\
I17008-4040 & 39.81 & 3.31 & 0.57±0.025 & 0.05 &  & \checkmark &  &  & \checkmark & C \\
\multicolumn{1}{r}{C2} & 39.81 & 3.31 & 42.78±31.127 & 0.3 &  &  &  &  &  & D \\
I17016-4124 C1 & 31.62 & 4.79 & 6.61±0.19 & 0.09 &  & \checkmark & \checkmark & \checkmark & \checkmark & D \\
\multicolumn{1}{r}{C2} & 31.62 & 4.79 & 8.95±1.579 & 0.08 &  & \checkmark & \checkmark & \checkmark & \checkmark & D \\
I17136-3617 & 125.89 & 4.15 & - & 0.14 & \checkmark & \checkmark &  & \checkmark & \checkmark & B \\
I17143-3700 & 63.1 & 3.0 & 0.19±0.007 & 0.11 & \checkmark &  &  & \checkmark & \checkmark & A \\
I17158-3901 & 25.12 & 3.8 & 0.88±0.02 & 0.04 &  & \checkmark &  &  & \checkmark & C \\
I17160-3707 C1 & 79.43 & 4.03 & 1.16±0.11 & 2.15 & \checkmark & \checkmark &  &  & \checkmark & B \\
\multicolumn{1}{r}{C2} & 79.43 & 4.03 & - & 1.4 & \checkmark &  &  & \checkmark & \checkmark & B \\
\multicolumn{1}{r}{C3} & 79.43 & 4.03 & 0.21±0.024 & 3.35 & \checkmark & \checkmark &  & \checkmark & \checkmark & B \\
I17175-3544 C1 & 50.12 & 5.08 & 1.9±0.014 & 0.07 & \checkmark & \checkmark & \checkmark & \checkmark & \checkmark & B \\
\multicolumn{1}{r}{C2} & 50.12 & 5.08 & - & 0.12 & \checkmark & \checkmark &  &  & \checkmark & B \\
\multicolumn{1}{r}{C3} & 50.12 & 5.08 & 1.66±0.274 & 0.29 & \checkmark & \checkmark & \checkmark & \checkmark & \checkmark & B \\
I17204-3636 & 19.95 & 4.18 & 0.28±0.009 & 0.06 &  & \checkmark &  & \checkmark & \checkmark & C \\
I17220-3609 & 25.12 & 3.77 & - & 0.22 & \checkmark & \checkmark &  & \checkmark & \checkmark & A \\
I17233-3606 C1 & 39.81 & 4.91 & 8.93±0.712 & 0.15 & \checkmark & \checkmark & \checkmark & \checkmark & \checkmark & B \\
\multicolumn{1}{r}{C2} & 39.81 & 4.91 & 2.91±0.241 & 0.01 & \checkmark & \checkmark & \checkmark & \checkmark & \checkmark & A \\
\multicolumn{1}{r}{C3} & 39.81 & 4.91 & 1.87±0.235 & 0.22 & \checkmark & \checkmark &  & \checkmark & \checkmark & B \\
I17244-3536 & 39.81 & 4.28 & 0.79±0.182 & 0.16 & \checkmark &  &  &  &  & B \\
I17258-3637 C1 & 398.11 & 4.02 & - & 0.3 & \checkmark & \checkmark &  &  & \checkmark & B \\
\multicolumn{1}{r}{C2} & 398.11 & 4.02 & - & 0.57 & \checkmark &  &  &  &  & B \\
I17269-3312 & 12.59 & 3.66 & 25.06±2.55 & 0.04 &  & \checkmark & \checkmark & \checkmark & \checkmark & C \\
I17271-3439 & 39.81 & 4.23 & 1.16±0.094 & 0.1 & \checkmark & \checkmark &  & \checkmark & \checkmark & A \\
I17278-3541 C1 & 19.95 & 4.64 & - & 0.16 &  & \checkmark &  & \checkmark & \checkmark & D \\
\multicolumn{1}{r}{C2} & 19.95 & 4.64 & - & 0.15 &  & \checkmark &  & \checkmark & \checkmark & D \\
\multicolumn{1}{r}{C3} & 19.95 & 4.64 & 1.32±0.118 & 0.25 &  &  &  &  &  & D \\
\multicolumn{1}{r}{C4} & 19.95 & 4.64 & - & 0.21 &  & \checkmark &  &  & \checkmark & D \\
\multicolumn{1}{r}{C5} & 19.95 & 4.64 & - & 0.08 &  &  & \checkmark & \checkmark & \checkmark & D \\
\multicolumn{1}{r}{C6} & 19.95 & 4.64 & - & 0.02 &  & \checkmark & \checkmark &  & \checkmark & C \\
I17439-2845 & 79.43 & 3.66 & - & 1.39 & \checkmark &  &  & \checkmark & \checkmark & B \\
I17441-2822 C1 & 15.85 & 4.81 & 0.6±0.063 & 1.37 & \checkmark & - & - & - &  & B \\
\multicolumn{1}{r}{C2} & 15.85 & 4.81 & 0.52±0.046 & 0.96 & \checkmark & - & - & - &  & B \\
\end{tabular}
\end{table*}

\begin{table*}
	\renewcommand\tabcolsep{6pt}
     \centering
     \begin{threeparttable}
     \begin{tabular}{lcclccccccc}
       \hline
	   \hline 
IRAS &  $L_\textup{bol}/M$ & log$n$ & [SiO]/[H$^{13}$CO$^+$] & $D_\textup{d}$ & H40$\alpha$ & \multicolumn{3}{c}{Wing} & outflow & Type  \\
\cline{7-9}
 & & &  & & & SiO & HCO$^+$ & CS  \\
\cline{7-9}
 & ($L{_\odot} M{_\odot}^{-1}$) & (cm$^{-3}$) & & (pc) & & & & & \\
\multicolumn{1}{r}{C3} & 15.85 & 4.81 & 0.28±0.057 & 1.13 & \checkmark & - & - & - &  & B \\
\multicolumn{1}{r}{C4} & 15.85 & 4.81 & 0.16±0.017 & 1.26 & \checkmark & - & - & - &  & B \\
\multicolumn{1}{r}{C5} & 15.85 & 4.81 & - & 0.36 & \checkmark & - & - & - &  & B\\
I17455-2800 & 63.1 & 3.7 & 3±0.401 & 0.78 & \checkmark & \checkmark &  &  & \checkmark & B \\
I17545-2357 & 10.0 & 3.89 & - & 0.49 & \checkmark &  &  &  &  & B \\
I17589-2312 C1 & 10.0 & 4.17 & - & 0.45 &  &  &  &  &  & D \\
\multicolumn{1}{r}{C2}  & 10.0 & 4.17 & 1.16±0.105 & 0.23 &  &  &  &  &  & D \\
I18032-2032 & 79.43 & 3.8 & 2.45±0.016 & 0.24 & \checkmark & \checkmark & \checkmark & \checkmark & \checkmark & B \\
I18056-1952 C1 & 19.95 & 3.92 & 1.02±0.017 & 0 & \checkmark & \checkmark & \checkmark & \checkmark & \checkmark & A \\
\multicolumn{1}{r}{C2} & 19.95 & 3.92 & 2.08±0.012 & 1.55 & \checkmark & \checkmark & \checkmark & \checkmark & \checkmark & B \\
\multicolumn{1}{r}{C3} & 19.95 & 3.92 & 0.39±0.026 & 0.73 & \checkmark & \checkmark &  &  & \checkmark & B \\
I18079-1756 C1 & 19.95 & 4.28 & 0.34±0.026 & 0.05 &  &  &  &  &  & C \\
\multicolumn{1}{r}{C2} & 19.95 & 4.28 & 0.47±0.099 & 0.26 &  &  &  &  &  & D \\
I18089-1732 & 15.85 & 3.85 & 0.62±0.069 & 0.03 &  & \checkmark & \checkmark & \checkmark & \checkmark & C \\
I18110-1854 & 39.81 & 3.99 & 0.87±0.13 & 0.35 & \checkmark & \checkmark &  &  & \checkmark & B \\
I18116-1646 & 100.0 & 3.73 & 0.09±- & 0.18 & \checkmark &  &  &  &  & B \\
I18117-1753 & 12.59 & 4.09 & 1.52±0.171 & 0.11 &  & \checkmark & \checkmark & \checkmark & \checkmark & D \\
I18139-1842 & 251.19 & 4.15 & 0.6±0.103 & 0.14 & \checkmark & \checkmark &  &  & \checkmark & B \\
I18159-1648 & 10.0 & 4.37 & 2.31±0.259 & 0.03 &  & \checkmark & \checkmark & \checkmark & \checkmark & C \\
I18182-1433 & 15.85 & 4.16 & 0.92±0.08 & 0.04 &  & \checkmark &  &  & \checkmark & C \\
I18223-1243 & 19.95 & 3.76 & 2.47±0.68 & 0.81 &  &  & \checkmark &  & \checkmark & D \\
I18228-1312 & 50.12 & 3.89 & 0.57±0.24 & 0.19 & \checkmark &  &  &  &  & B \\
I18236-1205 & 3.98 & 4.14 & 0.71±0.153 & 0.11 &  & \checkmark &  & \checkmark & \checkmark & D \\
I18264-1152 & 5.01 & 4.09 & 1.24±0.015 & 0.14 &  & \checkmark & \checkmark & \checkmark & \checkmark & D \\
I18290-0924 & 6.31 & 3.7 & 1.59±0.245 & 0.24 &  &  &  &  &  & D \\
I18311-0809 & 19.95 & 3.33 & 1.05±0.091 & 0.12 & \checkmark & \checkmark & \checkmark & \checkmark & \checkmark & A \\
I18316-0602  C1 & 12.59 & 4.27 & 42.04±1.271 & 0.09 &  & \checkmark &  & \checkmark & \checkmark & D \\
\multicolumn{1}{r}{C2} & 12.59 & 4.27 & 1.53±0.103 & 0.17 &  & \checkmark & \checkmark & \checkmark & \checkmark & D \\
\multicolumn{1}{r}{C3} & 12.59 & 4.27 & - & 0.07 &  &  &  & \checkmark & \checkmark & C \\
I18341-0727 & 25.12 & 3.4 & 1.49±0.084 & 0.45 & \checkmark & \checkmark & \checkmark & \checkmark & \checkmark & B \\
I18411-0338 & 25.12 & 3.42 & 0.42±0.002 & 0.07 &  & \checkmark &  &  & \checkmark & C \\
I18434-0242 & 125.89 & 3.7 & 1.04±0.026 & 0.2 & \checkmark & \checkmark &  & \checkmark & \checkmark & A \\
I18461-0113 & 19.95 & 4.27 & 0.73±0.004 & 0.1 &  & \checkmark & \checkmark & \checkmark & \checkmark & C \\
I18469-0132 & 63.1 & 4.12 & 0.34±0.002 & 0.44 &  & \checkmark &  &  & \checkmark & D \\
I18479-0005 & 79.43 & 3.65 & 0.56±0.027& 0.12 & \checkmark &  &  &  &  & A \\
I18507p0110 C1 & 39.81 & 4.88 & 0.75±0.008 & 0.03 & \checkmark & \checkmark &  & \checkmark & \checkmark & A \\
\multicolumn{1}{r}{C2} & 39.81 & 4.88 & - & 0.24 & \checkmark & \checkmark & \checkmark & \checkmark & \checkmark & B \\
\multicolumn{1}{r}{C3} & 39.81 & 4.88 & - & 0.18 & \checkmark &  &  &  &  & B \\
I18507p0121 C1 & 7.94 & 4.83 & 3.42±0.067 & 0.01 &  & \checkmark & \checkmark & \checkmark & \checkmark & C \\
\multicolumn{1}{r}{C2} & 7.94 & 4.83 & 3.42±0.399 & 0.12 &  & \checkmark &  & \checkmark & \checkmark & D \\
I18517p0437 & 50.12 & 5.43 & 0.5±0.022 & 0.05 &  &  &  &  &  & C \\
I18530p0215 & 25.12 & 3.84 & 0.13±0.029 & 0.23 & \checkmark & \checkmark &  &  & \checkmark & B \\
I19078p0901 & 79.43 & 3.73 & 2.12±0.02 & 0.16 & \checkmark & \checkmark & \checkmark & \checkmark & \checkmark & A \\
I19095p0930 & 100.0 & 4.29 & 0.8±0.014 & 0.06 &  & \checkmark & \checkmark & \checkmark & \checkmark & C \\
I19097p0847 & 15.85 & 3.5 & 1.18±0.166 & 0.31 &  & \checkmark & \checkmark & \checkmark & \checkmark & D\\

\hline
\end{tabular}
\begin{tablenotes}
        \footnotesize
        \item[]Note. $L_\textup{bol}/M$: The bolometric luminosity to mass ratio. 
        $n$: The particle number density. [SiO]/[H$^{13}$CO$^+$]: The integrated intensity ratio of SiO emission and H$^{13}$CO$^+$ emssion, and the - symbol indicates that the source has absorption features in their spectra. 
        $D_\textup{d}$: The separated linear distance between the central position of SiO clumps and the peak position of the 3 mm continuum emission. 
        H40$\alpha$ column: the \checkmark symbol indicates that the SiO clump is associated with H40$\alpha$ emission. 
        Wing columns: The \checkmark symbol represents that the SiO clump shows line wing emission; the sources with - symbol are excluded in analysis.
        outflow column: The \checkmark symbol indicates that the SiO clump is with outflow.
        Type column: A is the SiO clumps associated with both H40$\alpha$ emission and 3 mm continuum emission. B is the SiO clumps separated from both H40$\alpha$ emission and 3 mm continuum emission. C is the SiO clumps associated with 3mm continuum emission undetected H40$\alpha$ emission. D is the SiO clumps partly separated from 3mm continuum emission undetected H40$\alpha$ emission.
      \end{tablenotes}
    \end{threeparttable}
\end{table*}

————————————————
\bsp	
\label{lastpage}
\end{document}